\documentclass[11pt]{article}
\usepackage[top=2.5cm,bottom=2.5cm,left=2.55cm,right=2.55cm]{geometry}

\usepackage{enumitem}

\usepackage{bbm}
\usepackage{overpic}
\usepackage{caption}
\usepackage{mathtools}
\usepackage{latexsym} 
\usepackage{verbatim}  
\usepackage{tikz}
\usetikzlibrary{matrix}
\usepackage{color}
\usepackage{graphicx,amssymb,amsfonts,amsmath,amssymb,amscd,amstext, mathrsfs}
\usepackage{graphicx}
\usepackage{dsfont}
\usepackage{xcolor}
\usepackage{amsmath}
\usepackage{empheq}
\usepackage{cite}
\usepackage{tikz}
\usetikzlibrary{shapes.misc, positioning,decorations.pathreplacing,angles,quotes}
\usetikzlibrary{decorations.pathmorphing}

\numberwithin{equation}{section}

\newlength\dlf

\newcommand{\bw}{\begin{widetext}}
\newcommand{\ew}{\end{widetext}}
\newcommand{\bea}{\begin{eqnarray}}
\newcommand{\eea}{\end{eqnarray}}

\renewcommand{\bar}[1]{\overline{#1}}
\renewcommand{\tilde}[1]{\widetilde{#1}}
\renewcommand{\hat}[1]{\widehat{#1}}

\newcommand{\<}{\langle}
\renewcommand{\>}{\rangle}

\renewcommand{\cal}{\mathcal}

 \definecolor{themecolor}{RGB}{0, 64, 168}
\definecolor{myRed}{RGB}{150,22,22}
\definecolor{myRedL}{RGB}{255,239,237}
\definecolor{myGrayL}{RGB}{220,220,220}
\definecolor{blul}{RGB}{51, 153,255}
\definecolor{greenl}{RGB}{76,153,0}

\DeclareFontShape{OT1}{cmr}{mx}{n}{<->cmr10}{}
\newcommand{\titlefont}{\fontseries{mx}\selectfont}

\def\frac#1#2{{#1\over #2}}

\usepackage{subcaption}
\usepackage{graphicx}
\usepackage{jheppub}
\usepackage{mymacroMOD}
\usepackage{braket}
\usetikzlibrary{decorations.pathmorphing}
\usetikzlibrary{decorations.markings}
\tikzset{snake it/.style={decorate, decoration=snake}}
\usetikzlibrary{arrows.meta}
\makeatletter
\newlength\lrvec@height
\newlength\lrvec@width
\newif\iflrvec@same@height
\def\lrvec{\@ifstar\slrvec@\lrvec@}
\newcommand{\slrvec@}[2][.4ex]{
  \lrvec@same@heighttrue
  \mathpalette\lrvec@@{{#1}{#2}}
}
\newcommand{\lrvec@}[2][.4ex]{
  \lrvec@same@heightfalse
  \mathpalette\lrvec@@{{#1}{#2}}
}
\def\lrvec@@#1#2{\lrvec@@@#1#2}
\def\lrvec@@@#1#2#3{%
  \iflrvec@same@height
    \settoheight{\lrvec@height}{$\m@th#1 \mathbf{T}#3$}
  \else
    \settoheight{\lrvec@height}{$\m@th#1#3$}
  \fi
  \settowidth{\lrvec@width}{$\m@th#1#3$}
  \kern.08em
  \raisebox{#2}{\raisebox{\lrvec@height}{\rlap{%
    \kern-.05em
    \begin{tikzpicture}[<-> /.tip={To[width=.4em, length=.2em]}]
      \draw [<->] (-.05em,0)--(\lrvec@width+.05em,0);
    \end{tikzpicture}%
  }}}%
  #3
  \kern.08em
}
\makeatother

\usepackage{simpler-wick}
\graphicspath{ {./figure/} }
\newcommand\blfootnote[1]{%
  \begingroup
  \renewcommand\thefootnote{}\footnote{#1}%
  \addtocounter{footnote}{-1}%
  \endgroup
}
\makeatletter
\newcommand*{\Xbar}{}%
\DeclareRobustCommand*{\Xbar}{%
  \mathpalette\@Xbar{}%
}
\newcommand*{\@Xbar}[2]{%
  \sbox0{$#1\mathrm{X}\m@th$}%
  \sbox2{$#1X\m@th$}%
  \rlap{%
    \hbox to\wd2{%
      \hfill
      $\overline{%
        \vrule width 0pt height\ht0 %
        \kern\wd0 %
      }$%
    }%
  }%
  \copy2 %
}
\makeatother

\newcommand{\exch}{%
  \raisebox{-0.4ex}{%
    \tikz[scale=0.7, line width=0.4pt]{
      \draw (0,0.1)--(0, -0.1);
   \draw (0,0.1) -- (0.1,0.2);
   \draw (0,0.1) -- (-0.1,0.2);
   \draw (0,-0.1) -- (0.1,-0.2);
   \draw (0,-0.1) -- (-0.1,-0.2);
    }%
  }%
}

\newcommand{\three}{%
  \raisebox{-0.4ex}{%
    \tikz[scale=0.7, line width=0.4pt]{
      \draw (0,0.15)--(0, -0.15);
   \draw (0,0.15) -- (0.12,0.2);
   \draw (0,0.15) -- (-0.12,0.2);
      \draw (0,0.15) -- (-0.12,0.1);
   \draw (0,-0.15) -- (0.12,-0.2);
      \draw (0,-0.15) -- (0.12,-0.1);
   \draw (0,-0.15) -- (-0.12,-0.2);
    }%
  }%
}

\newcommand{\contact}{%
{%
    \tikz[scale=0.7, line width=0.4pt]{
\draw (-0.1,0.1)--(0.1, -0.1);
\draw (0.1,0.1)--(-0.1, -0.1);
\filldraw (0,0) circle (0.75pt);
    }%
  }%
}

\newcommand{\bubble}{%
  \raisebox{-0.4ex}{%
    \tikz[scale=0.7, line width=0.4pt]{
    \draw (0,0) circle (0.1);
   \draw (0,0.1) -- (0.1,0.2);
   \draw (0,0.1) -- (-0.1,0.2);
   \draw (0,-0.1) -- (0.1,-0.2);
   \draw (0,-0.1) -- (-0.1,-0.2);
    }%
  }%
}

\newcommand{\contactell}[1]{%
  {%
    \tikz[scale=0.7, line width=0.4pt, every node/.style={scale=0.6}]{
      \draw (-0.1,0.1)--(0.1, -0.1);
      \draw (0.1,0.1)--(-0.1, -0.1);
      \filldraw (0,0) circle (0.75pt);
      \node[above=0cm] at (0,0) {\scriptsize $#1$};
    }%
  }%
}

\def\headline#1{\hbox to \hsize{\hrulefill\quad\lower.3em\hbox{#1}\quad\hrulefill}}

\begin{document}
\pagestyle{myplain}
\begin{titlepage}

\begin{flushright} 
\end{flushright}

\begin{center} 

\vspace{0.35cm}

{\fontsize{20.5pt}{25pt}
{\titlefont  Towards  Large-Spin Effective  Theory I:\\
  Three-Particle States in AdS $\phi^4$ Theory}}

\vspace{1.6cm}  

{{Giulia Fardelli$^{\contact}$\blfootnote{${}^{\contact}$\href{mailto:fardelli@bu.edu}{\tt fardelli@bu.edu}}, A. Liam Fitzpatrick$^{\bubble}$\blfootnote{${}^{\bubble}$\href{mailto:fitzpatr@bu.edu}{\tt fitzpatr@bu.edu}},  Wei Li$^{\three}$\blfootnote{${}^{\three}$\href{mailto:weili17@bu.edu}{\tt weili17@bu.edu}}
}}

\vspace{1cm} 

{{\it
Department of Physics, Boston University, 
Boston, MA  02215, USA
}}\\
\end{center}
\vspace{1.5cm}

{\noindent 
We describe how to construct an effective Hamiltonian for leading twist states in $d\ge 3$ CFTs based on the separation of scales that emerges at large spin $J$ between the AdS radius $\ell_{\rm AdS}$ and the characteristic distance $\sim \ell_{\rm AdS} \log J$ between particles rotating in AdS with angular momentum $J$.  As a controlled example, we work specifically with the toy model  of a bulk complex scalar field $\phi$ with a $\lambda |\phi|^4$ coupling in AdS, up to $O(\lambda^2)$. For a given choice of twist cutoff $\Lambda_\tau$ in the effective theory, interactions are separated into long-distance nonlocal potential terms, arising from $t$-channel exchange of states with twist $\le \Lambda_\tau$, and short-distance local terms fixed by matching to low spin CFT data.  At $O(\lambda^2)$, the effective Hamiltonian for the toy model has two-body nonlocal potential terms from one-loop bulk diagrams as well as three-body nonlocal potential terms from tree-level exchange of $\phi$.  We describe in detail how these contributions are evaluated and how they are related to the CFT data entering in the large spin expansion.  We discuss how to apply the construction of such effective Hamiltonians  for models which do not have a large central charge or a sparse spectrum and are not typically considered holographic.
}

\end{titlepage}

\tableofcontents

\newpage

\section{Introduction and Summary} 

What Conformal Field Theories (CFTs) can be described holographically? This is an enduring question that has been addressed from myriad perspectives since the discovery of the AdS/CFT correspondence. Typically, to make quantitative comparisons with the putative holographic dual, it requires some kind of weakly coupled expansion in which to do perturbative computations in the bulk description.  Most commonly, the small bulk parameter is provided from the CFT by a large $N$ expansion.  However, one might hope to formulate a much more general bulk description based on expanding in large angular momentum, or spin, $J$, which shares many key features with a large $N$ expansion.  The basic idea is that, due to the kinematics of AdS, a large angular momentum $J$  implies a long-distance scale.  Moreover, because of the unitarity bound on twists, for $d>2$ the force between two objects due to exchanging a particle, even a massless one, will decay exponentially at long distances.  Thus we expect that the lowest energy states at a fixed large value of $J$ will be described by two distant objects in AdS interacting mainly due to a weak force between them. The chief limitation of this approach is that the objects themselves, and their couplings to the low-twist exchanged fields, must be taken as inputs from the CFT (in the form of scaling dimensions and OPE coefficients), similarly to how one could compute the gravitational force between two distant protons with only the proton mass as an input.  This kind of separation of physics into short-distance effects, which must be parameterized, and long-distances effects, which can be computed perturbatively, is most elegantly and systematically handled in the language of Effective Field Theory (EFT).  

Our motivation in this paper is to formulate a precise holographic EFT description for low-twist CFT states based upon this separation of scales that emerges at large $J$.  Concretely, the low-twist states that we would like such an EFT to describe  can be built up starting from the lowest-dimension nontrivial operator in the theory, ``$\Phi$'' with dimension $\Delta$, and recursively looking at lowest-twist trajectories in the OPE as one adds more factors of $\Phi$.  Two-particle, or `double-twist', states ``$[\Phi, \Phi]_J$'' can be defined by applying the Lorentzian inversion formula in order to connect the large $J$ accumulation point, where the twist of $[\Phi, \Phi]_J$ approaches $2\Delta$, to finite $J$.  Three-twist states are obtained by then considering the OPE of $\Phi$ and $[\Phi, \Phi]_J$, and so on for multi-twist states with more factors of $\Phi$.  Ultimately, we would like to construct an EFT that is local in AdS, with a well-defined cutoff, in which one can perform systematically improvable computations of the anomalous dimensions of such operators even in CFTs such as the $3d$ Ising model and $3d$ $O(2)$ model, which have a small central charge and a small gap in dimensions, and therefore are not usually thought of as being holographic.  

However, in order to better understand the rules of such EFTs in general, in this paper we will have the more modest goal of working out the details of the EFT in a case which is manifestly holographic, namely the theory of a single bulk complex scalar\footnote{Working with a scalar of definite $U(1)$ charge eliminates mixing with operators of different particle number, although we are confident that our analysis can be generalized without much difficulty. }  field $\phi$, dual to the CFT operator $\Phi$, with a perturbative bulk quartic local interaction $ \lambda (\phi \phi^*)^2$ in AdS$_5$:
\begin{equation}
S = \int d^{d+1} x \sqrt{g} \left(  |\partial \phi|^2 -  m^2 |\phi|^2 - \frac{\lambda}{4} |\phi|^4 \right).
\label{eq:BareLag}
\end{equation}
For $d\ne 3$, the bulk coupling $\lambda$ is dimensionful and at times it will be helpful to make this explicit by defining
\begin{equation}
\lambda \equiv M^{3-d}
\label{eq:MDef}
\end{equation}
so that $M$ is the corresponding energy scale.   

 We will mostly focus on the case of three-particle states with twist equal to $3\Delta + O(\lambda)$ and $U(1)$ charge equal to $3$, and in this paper we will only consider states in the symmetric-traceless representations of the rotation group. The triple-twist states at spin $J$ have an overcomplete basis of states of the form $[\Phi, [\Phi, \Phi]_\ell]_{J-\ell}$ given by recursively adding $\Phi$s as described above.  Despite the simplicity of the model, its large $J$ EFT is sufficiently rich that it will illuminate many aspects of large $J$ EFT that should apply more generally.

In principle, all computations of dimensions of multi-particle states in this theory can be computed perturbatively in $\lambda$  using standard techniques based on evaluating correlation functions with Witten diagrams, after regularization and renormalization of UV divergences.  However, such computations quickly become extremely difficult beyond the simplest multiparticle states.  It is significantly easier to instead obtain operator dimensions by diagonalizing the Hamiltonian of the theory, which is the strategy we will adopt.  In particular, at any spin $J$, there are a finite number  of lowest-twist three-particle states, so the effective Hamiltonian acting on this sector is just a finite-dimensional matrix.  

Crucially, in order to obtain an effective Hamiltonian acting only within the lowest-twist sector of three-particle states, we must `integrate out' the higher-twist states. In perturbation theory, their effect first appears at $O(\lambda^2)$, because it takes at least two factors of the bare Hamiltonian in order to mix out of and back into the lowest twist sector.  Most of the technical work we do will be to obtain the resulting effective Hamiltonian for triple-twist states at $O(\lambda^2)$.  We will see exactly how this effective Hamiltonian can be separated into short-distance, local counter-terms and long-distance, non-local potential terms due to bulk exchange of low-twist states. The exact division between these effects depends on our choice of a cutoff $\Lambda_\tau$ in the twist.   Since we only consider up to three-particle states, we only need to consider the two-body and three-body interactions in the Hamiltonian,\footnote{At $Q=1$, the only lowest-twist states are $\Phi$ and its lowest-twist descendants.  The effective Hamiltonian is just the  mass and kinetic term, so any contributions are just absorbed into the physical dimension $\Delta$ and bulk wavefunction renormalization.  }  
 \begin{equation}
 H= H_{\rm GFF} + H_2 + H_3,
 \end{equation}
where $H_{\rm GFF}$ is the Hamiltonian for a generalized free field (GFF) theory (dual to a free theory in AdS), and $H_2$ $(H_3)$ contains terms that are quartic (sextic) in the bulk field $\phi$. We furthermore break up $H_2$ and $H_3$ into short-range local terms and long-range potential terms:
\begin{equation}
H_n = H_{n, \rm loc} + H_{n, \rm potl}.
\end{equation}
 Schematically, $H_{n, \rm loc}$ is a sum over local terms of the form
 \begin{equation}
 H_{n, \rm loc} = \int d^d x \sqrt{g} \sum_m c_{n,m}  [ \phi^n \phi^{*n}]_m,
 \end{equation}
where the subscript $m$ on the local counterterms `$[ \phi^n \phi^{*n}]_m$'  is meant to indicate that they also involve derivatives of the bulk $\phi$ fields, and can be organized into linear combinations of derivatives in order to pick out specific spin structures.  The coefficients $c_{n,m}$ depend on the twist cutoff $\Lambda_\tau$, and in our explicit calculations we will see that this dependence on $\Lambda_\tau$ is given by dimensional analysis in the standard way.  Because of the simplicity of the model, in some cases we will resum an infinite number of such counterterms contributing to some observable, but in practice this is not necessary and at a fixed level of accuracy only a finite number of such counterterms must be included. 

By contrast, the nonlocal potential terms in $H_{n, \rm potl}$ are long-range interactions due to exchange of bulk fields in AdS below the twist cutoff. Schematically,
\begin{equation}
\begin{aligned}
H_{2, \rm potl} &= 
 \int d^d x \sqrt{g} \sum_{\tau}^{\Lambda_\tau} \sum_\ell g^{(2)}_{\tau, \ell}\int d^{d+1} y \sqrt{g} \phi \phi^*(x) G^{(\rm reg)}_{\tau, \ell}(x,y) \phi \phi^*(y), \\
H_{3, \rm potl} &= 
 \int d^d x \sqrt{g} \sum_{\tau}^{\Lambda_\tau} \sum_\ell g^{(3)}_{\tau, \ell}\int d^{d+1} y \sqrt{g} \phi^2 \phi^*(x) G^{(\rm reg)}_{\tau, \ell}(x,y) \phi \phi^{*2}(y).
\end{aligned}
\end{equation}
The superscript `(reg)' indicates that the propagators $G_{\tau,\ell}$ for a twist-$\tau$ spin-$\ell$ field must be regulated to remove any poles associated with redundant contributions to the Hamiltonian, in a manner we will explain in detail.  As usual, as one decreases the twist cutoff, there is a tradeoff in complexity in that a lower cutoff means one has fewer bulk exchanges in the EFT but more local counterterms, for a fixed level of accuracy.

 Why should a holographic Hamiltonian of the form written above apply to CFTs without any large $N$ limit? If we put no restrictions on the size of the coefficients of the Wilson terms in the Hamiltonian, it is almost a tautology that such a construction is possible.  The basic reason is that at some fixed $Q$ and sufficiently large value of $J$, we can analyze all four-point functions of the form $\< \phi \phi^* [\phi^{Q-1}] [\phi^{* Q-1}]\>$, where $ [\phi^{Q-1}] $ are the charge $(Q-1)$ operators in the theory. Assuming our charge $Q$ spin $J$ operator of interest $[\phi^Q]$ lies on the leading large-spin trajectory of operators in the $\phi \times [\phi^{Q-1}]$ OPE, then  the large spin expansion \cite{Alday:2007mf,Fitzpatrick:2012yx,Komargodski:2012ek} applied to this four-point function tells us that the dimension of the state can be well-approximated by the dimension of $\phi$ plus the dimension of $[\phi^{Q-1}]$ plus small corrections that are fixed by the leading low-twist conformal blocks in the $t$-channel and $u$-channel of the correlator.  As long as our holographic Hamiltonian includes nonlocal potential terms for these low-twist exchanges, with coefficients designed to match the OPE coefficients of the corresponding conformal blocks, then it will correctly capture these leading anomalous dimensions at large $J$.  All that remains at that point is to consider the finite set of spins $J$ that are not large enough for the large spin expansion to be a good approximation.  All possible corrections to anomalous dimensions that truncate at some finite value of $J$ can be captured by local terms in AdS~\cite{Heemskerk:2009pn}.  This argument implies that we can implement a recursive strategy, where we work our way up from smaller to larger values of $Q$.  That is, at each value of $Q$, we assume our effective Hamiltonian already approximately reproduces all the leading-twist charge $(Q-1)$ states.  To reproduce the leading-twist charge $Q$ states, we only need to add additional bulk exchanges with coefficients tuned to match the OPE coefficients for the corresponding conformal blocks in order to capture the large $J$ behavior, plus a finite set of local terms to capture the small $J$ behavior.  

In practice, of course, we would like to put restrictions on the size of the Wilson coefficients in the Hamiltonian.  For one, if they are not small in any sense, then there is no way to use them to compute higher order corrections.  Moreover, the naive procedure outlined above may break down when one encounters accumulation points in twist (which must be present on general grounds).  In this case, the sum over contributions to the anomalous dimensions from individual conformal blocks may diverge, and it can become necessary to sum over them before applying the large spin expansion or the Lorentzian inversion formula.  An important feature of the construction we use is that the two-body Hamiltonian $H_2$ contributes also to all states with $Q>2$ as well, because of the nature of second-quantization.  Consequently, in the recursive construction, some piece of the nonlocal exchanges at charge $Q$ are already accounted for by the construction at $(Q-1)$ or smaller.  Since the accumulation points in twist can often be understood themselves as multi-particle states at large spin, they are automatically inherited from the Hamiltonian at lower $Q$ without needing to add them explicitly, which in turn leads to better convergence.   In our $\lambda |\phi|^4$ example, the corresponding statement is that at $O(\lambda^2)$, we will not need to add any nonlocal potentials to account for the $t$-channel exchange of neutral two-particle states with large spin (in fact, at this order only spin-$0$ exchanges will be necessary), despite the presence of an accumulation point in spin at twist $2 \Delta$.

In more detail, the matching procedure for constructing the effective Hamiltonian in the $\lambda |\phi|^4$ holographic model proceeds as follows. At linear order in $\lambda$, the two-body effective Hamiltonian $H_2$ is simply the tree-level Hamiltonian $H= \frac{\lambda}{4} \int d^d x \sqrt{g} |\phi|^4$ acting on the charged two-particle states, and is nonzero only for the $J=0$ primary $\Phi^2$ and its descendants.  At second order in $\lambda$, however, there are contributions from intermediate two-particle, four-particle and six-particle states, since the bare interaction includes $n\rightarrow n$, $n\rightarrow n+2$, and $n\rightarrow n+4$ couplings.  Diagrammatically, these correspond to $s$-, $t$-, and $u$- channel one-loop diagrams, depicted in Figure \ref{Fig:Q2bubble}.

Each of these one-loop diagrams can be represented as a sum over a tower of two-particle states in the corresponding channel with spin 0 and twists $2\Delta+2n$ for $n=0,1,2,\dots$.  Let the twist cutoff  be $\Lambda_\tau = 2\Delta+2N$, so that the two-particle states with $n\le N$ are at or below the cutoff.\footnote{In this perturbative model, it is straightforward to do such a decomposition at the level of the covariant 1-loop Witten diagrams for the four-point function $\< \Phi \Phi \Phi^* \Phi^*\>$, but more generally we would consider the (nonperturbative) conformal block decompositions of the four-point function. } When we integrate out primary operators with twist below the cutoff $\Lambda_\tau$,  we use the full bulk-to-bulk propagator for a bulk field with the corresponding dimension and spin,   roughly analogous to integrating out a particle in flat space  but keeping its entire propagator $\sim \frac{1}{p^2+m^2}$.  Keeping the entire propagator is necessary in order to capture the analytic properties of its contributions as a function of spin $J$, especially the large $J$ behavior.  By contrast, the contribution from exchanges of primaries above the twist cutoff are absorbed into local counterterms.  
Then the effective interactions in the lowest-twist $Q=2$ Hamiltonian $H_2$ is
 \begin{align*}
 H_2 \!\supset \!\! \int\! d^d x \sqrt{g}\! \left[ \sum_m\! c_m (\phi \phi^*) \Box^m (\phi \phi^*)  +\!\sum_{n=0}^N g^{(2)}_{2\Delta+2n,0}\! \! \int\! d^{d+1} y \sqrt{g} (\phi \phi^*)(x) G_{2\Delta+2n}(x,y) (\phi \phi^*)(y) \right]\!,
 \end{align*}
 where $G_{2\Delta+2n}(x,y)$ is the bulk-to-bulk propagator for a scalar with dimension $2\Delta+2n$.   The bulk couplings $ g^{(2)}_{2\Delta+2n,0}$ can be derived from the spectral decomposition of the loop, or more generally to match the $t$-channel conformal block decomposition.\footnote{An interesting aspect of the  procedure where the $g^{(2)}_{\tau,\ell}$ coefficients are obtained from the CFT data is that these bulk coefficients vanish in the limit that the exchanged operators have double-trace twists $\tau=2\Delta+2n$, and therefore their leading order values are sensitive to the anomalous dimensions of the double-trace operators.  From the bulk perspective, the equivalent statement is that the couplings $g^{(2)}_{2\Delta+2n,0}$ determine the double-trace anomalous dimensions in the $t$-channel. } The Wilson coefficients $c_j$ for the local counterterms arise from all the two-particle states above the twist cutoff, and can be derived by doing a spectral decomposition of the loop and then series expanding the bulk-to-bulk propagator as a sum over local terms, $\frac{1}{\Box + m^2} = \frac{1}{m^2} \sum_{m=0}^\infty (-\Box/m^2)^m$.  More generally, for more realistic models, the Wilson coefficients should be chosen by first including the bulk exchanges and comparing to physical observables (anomalous dimensions and OPE coefficients) as a function of $J$. The lightcone bootstrap guarantees that the resulting mismatch will decay like $J^{-\Lambda_\tau}$ at large $J$, and so to good approximation we can include only a finite number of local counterterms in order to match a finite set of small $J$ observables.\footnote{Moreover, analyticity in spin~\cite{Caron-Huot:2017vep} guarantees that for  $J  > 1$,  the Wilson coefficients will shrink to zero as $\Lambda_\tau$ approaches $\infty$; it is important to keep in mind that this is an asymptotic statement, and that the Wilson coefficients can, and often do, grow with $\Lambda_\tau$ in intermediate regimes.
 } For more general models, or for this model at higher-loop order, the local counterterms will involve additional spin structures beyond the ones shown above.
 
Things become more interesting once we pass to $Q=3$ states.  At this point, all the interactions in the two-particle Hamiltonian $H_2$  still contribute, because any of the three particles in the state can simply be a `spectator' which does not see the interaction.  However, there are also additional terms in the effective Hamiltonian that involve all three particles.  At $O(\lambda^2)$, the new three-body interaction is a tree-level ``three-to-three''  interaction due to $\phi$ exchange; the corresponding diagram
is shown in Figure~\ref{fig:3to3}.     When $\Delta$ is below our twist cutoff, we must include this diagram by using the full bulk-to-bulk propagator for $\phi$.  
There is an interesting complication with this diagram, however, due to the fact that,  in its $s$-channel decomposition,  it includes intermediate states that are lowest-twist $Q=3$ states.  It is important {\it not} to include such states in higher order diagrams.  The reason is that these states are already part of our lowest-twist $Q=3$ space of states, and therefore when we diagonalize the $O(\lambda)$ Hamiltonian we are already including them at all orders in $\lambda$.  Therefore, including them again in the diagram in Figure~\ref{fig:3to3} would be double-counting.  When we use the bulk-to-bulk propagator for the $\phi$ exchange in Figure~\ref{fig:3to3}, we must therefore be careful to subtract out the contribution from the lowest-twist $Q=3$ states. The correct treatment of this and similar diagrams is one of the key points of this paper.

 In the actual computation, this redundancy shows up as an unphysical divergence.  The divergence is more transparent if we consider the bulk exchange of a field with dimension $\Delta_\chi$ slightly shifted away from $\Delta$.  In this case, the divergence is turned into a pole at the point $\Delta_\chi = \Delta$, which is removed when we subtract out the redundant contribution from the lower-order Hamiltonian.  The resulting three-body interaction takes the form of a regulated (by removing the pole) $\phi$ bulk exchange:
 \begin{equation}
 \begin{aligned}
 H&= H_{\rm GFF} + H_2 + H_3 + \dots, \\
  H_3 &=    g^{(3)}_{\Delta,0} \int d^d x \sqrt{g(x)}\int d^{d+1}y \sqrt{g(y)}
    (\phi^2 \phi^*)(x)  G^{(\rm reg)}_{\Delta}(x,y)  (\phi^{*2} \phi)(y) .
  \end{aligned}
 \end{equation}
When evaluated in the $Q=3$ sector,  this Hamiltonian gives the anomalous dimension of $[\Phi, \Phi^2]_J$ at large $J$ summarized in~\eqref{resQ3Full}. Notably,  $[\Phi, \Phi^2]_J$ is the only $Q=3$ state that receives a correction at order $\lambda$.

Finally, at the end we return to the question of how these methods would apply to a wider class of CFTs,  where we are given some set of anomalous dimensions and OPE coefficients, rather than a holographic Lagrangian as the starting point.  In that case, we would start by constructing a holographic Lagrangian designed to reproduce the available CFT data for the $Q=2$ states.  In practice, the structure of the EFT will be the same, involving bulk propagators for low-twist fields exchanged in the $t$-channel, which capture the leading $Q=2$ anomalous dimensions and OPE coefficients at large $J$, and local contact terms obtained by matching the CFT data at small $J$.  An important aspect of this matching is that one must add enough local contact terms to do the small $J$ matching for  both the $Q=2$ anomalous dimensions and their OPE coefficients with $\Phi$, i.e. the OPE coefficients in the three-point functions $\< \Phi^*(x) \Phi^*(y) [\Phi, \Phi]_J(z)\>$.

The paper is organized as follows.  In section~\ref{sec:LSCFT}, we begin by reviewing the large-spin results derived from CFT four-point functions and their interpretation in terms of AdS physics.  We then introduce our bulk setup, define the relevant states, and explain how to compute corrections to their binding energies from the Hamiltonian. Section~\ref{sec:LO} contains explicit computations of anomalous dimensions due to a $\lambda |\phi|^4$ interaction at leading order for $Q=2$ and $Q=3$ states and a new, direct method for how to get OPE coefficients in the Hamiltonian setup at $Q=2$. In section~\ref{sec:NLOBubbles}, we analyze the $O(\lambda^2)$ corrections from one-loop diagrams,  illustrating the computation explicitly in the case $\Delta=2$ and $d=4$.  In section~\ref{sec:NLO3To3} we discuss the genuine three-body interaction and the proper way to compute them. 
Section~\ref{sec:Results} summarizes the main results obtained for the $[\Phi, \Phi^2]_J$ anomalous dimensions and compares them with the CFT predictions obtained from  large spin and the Lorentzian inversion formula. 
We conclude in section~\ref{sec:Generalization} with a discussion of possible extensions to non-holographic models. Several appendices collect additional technical details.

\section{Large Spin CFT Techniques}\label{sec:LSCFT}

\subsection{Large Spin Expansion from Conformal Block Decomposition}

Given any two operators $\Phi_1$ and $\Phi_2$ in a $d>2$ CFT with twists $\tau_1$ and $\tau_2$ respectively, the OPE of $\Phi_1 \times \Phi_2$ contains an infinite tower of operators $[\Phi_1, \Phi_2]_J$ with increasing spin $J$ and an accumulation point in twist $\tau_J$ at $\lim_{J\rightarrow \infty} \tau_J = \tau_1+\tau_2$~\cite{Fitzpatrick:2012yx,Komargodski:2012ek}.  In the case where $\Phi_1$ and $\Phi_2$ are scalars, this result has been proven rigorously from the conformal bootstrap \cite{Pal:2022vqc, vanRees:2024xkb}, and moreover the leading corrections to the twist of the operators  $[\Phi_1, \Phi_2]_J$ at large $J$ is given in terms of the low-twist operators appearing in the `$s$-channel' $\Phi_1 \Phi_1^* \rightarrow \CO_s \rightarrow \Phi_2 \Phi_2^*$ and $t$-channel $\Phi_1 \Phi_2^* \rightarrow \CO_t \rightarrow \Phi_1 \Phi_2^*$ conformal block decompositions of the $\<\Phi_1 \Phi_2 \Phi_1^* \Phi_2^*\>$ four-point function.   For a single low-twist operator $\CO_s$, with twist $\tau_s$ and spin $\ell_s$,  and $\CO_t$, with twist $\tau_t$ and spin $\ell_t$,  (besides the identity operator), the leading correction to $\tau_J$ at large $J$ is~\cite{Simmons-Duffin:2016wlq}
{\allowdisplaybreaks
\threeseqn{
&\qquad \qquad \quad \, \, \delta{\tau_J}=\frac{\delta(c_{\Phi_1\Phi_2[\Phi_1,\Phi_2]}\gamma)}{\delta c_{\Phi_1\Phi_2[\Phi_1,\Phi_2]}}\, , 
}[]
{
&\begin{aligned}
\delta(c_{\Phi_1\Phi_2[\Phi_1,\Phi_2]}\gamma)&=-\frac{c_{\Phi_1\Phi_1^*\CO_s}c_{\Phi_2\Phi_2^*\CO_s}}{\Gamma\mleft(\Delta_1-\frac{\tau_s}{2}\mright)\Gamma\mleft(\Delta_2-\frac{\tau_s}{2}\mright)}\frac{\Gamma(\tau_s+2\ell_s)}{\Gamma\mleft(\frac{\tau_s+2\ell_s}{2} \mright)^2}\frac{1}{(\mathcal{J}^2)^\frac{\tau_s}{2}}\\
&\quad\,-(-1)^J\frac{(-1)^{\ell_t}c_{\Phi_1\Phi_2\CO_t}^2}{\Gamma\mleft( \frac{\Delta_1+\Delta_2-\tau_t}{2}\mright)^2}\frac{\Gamma(\tau_t+2\ell_t)}{\Gamma\mleft(\frac{\tau_t+2\ell_t+\Delta_{12}}{2}\mright)\Gamma\mleft(\frac{\tau_t+2\ell_t-\Delta_{12}}{2}\mright)}\frac{1}{(\CJ^2)^\frac{\tau_t}{2}}\, ,
\end{aligned}
}[]
{
&\begin{aligned}
\phantom{(\gamma\,\, }\delta c_{\Phi_1\Phi_2[\Phi_1,\Phi_2]}&=\frac{1}{2\Gamma(\Delta_1)\Gamma(\Delta_2)}-\frac{c_{\Phi_1\Phi_1^*\CO_s}c_{\Phi_2\Phi_2^*\CO_s}}{\Gamma\mleft(\Delta_1-\frac{\tau_s}{2}\mright)\Gamma\mleft(\Delta_2-\frac{\tau_s}{2}\mright)}\frac{\Gamma(\tau_s+2\ell_s)H_{\frac{\tau_s}{2}+\ell_s-1}}{\Gamma\mleft(\frac{\tau_s+2\ell_s}{2} \mright)^2(\mathcal{J}^2)^\frac{\tau_s}{2}}\\
&\quad\,-(-1)^J\frac{(-1)^{\ell_t}c_{\Phi_1\Phi_2\CO_t}^2}{\Gamma\mleft( \frac{\Delta_1+\Delta_2-\tau_t}{2}\mright)^2}\frac{\Gamma(\tau_t+2\ell_t)(H_{\frac{\tau_t+2\ell_t+\Delta_{12}-2}{2}}+H_{\frac{\tau_t+2\ell_t-\Delta_{12}-2}{2}})}{2\Gamma\mleft(\frac{\tau_t+2\ell_t+\Delta_{12}}{2}\mright)\Gamma\mleft(\frac{\tau_t+2\ell_t-\Delta_{12}}{2}\mright)(\CJ^2)^\frac{\tau_t}{2}}\, ,
\end{aligned}
}[][largeSpinDT]
}
$\!\!\!$where $H_x$ is the Harmonic number, $\Delta_{ij}\equiv \Delta_i-\Delta_j$, $c_{ijk}$ are OPE coefficients  and we have introduced
the ``conformal spin''  $\CJ$~\cite{Alday:2016njk}:
\begin{equation}
\CJ^2 = \left(J+ \frac{\Delta_1 + \Delta_2}{2}\right) \left(J -1 + \frac{\Delta_1 + \Delta_2}{2}\right).
\end{equation}
An advantage of using the conformal spin $\CJ$ rather than $J$ is that the contribution from an individual conformal block is $\CJ^{-\tau/2}$ times a Taylor series in $1/\CJ^2$ \cite{Alday:2016njk,Caron-Huot:2017vep}. 

 More generally, an exact definition of $\tau_J$ as an analytic function of $J$ for finite $J$ (Re$(J)\ge 2$) is provided by the Lorentzian inversion formula~\cite{Caron-Huot:2017vep}, as an integral over the four-point function $\<\Phi_1 \Phi_2 \Phi_1^* \Phi_2^*\>$.  Applied to an individual term in the $s$- or $t$-channel conformal block expansions, its large spin expansion reproduces the large spin expansions described above.  In  principle, given the CFT data (in this context, the conformal block dimensions and spins and OPE coefficients) for all operators up to some maximum twist $\tau_{\rm max}$ allows one to determine the function $\tau_J$ to increasing accuracy as $\tau_{\rm max}$ is increased.\footnote{Because of accumulation points in twist,  at some finite $\tau_{\rm max}$ the number of operators with twist below $\tau_{\rm max}$ becomes infinite, and it may be necessary to sum their contributions to the four-point function before applying the inversion formula.}
 
\subsection{Connection to  AdS Physics}

A simple way to see the connection between CFT behavior at large spin $J$ and locality in AdS is through the behavior of bulk exchanges, specifically their bulk-to-bulk propagator.  For general spin, the bulk-to-bulk propagator of a field with twist $\tau$ decays exponentially in the limit of large geodesic separation $\sigma(x,y)$ between its endpoints $x,y$, like  \cite{Costa:2014kfa}\footnote{Consider the bulk-to-bulk propagator in embedding space as given in~\cite{Costa:2014kfa}
\eqna{
G(X_1,X_2,W_1,W_2)=\sum_{k=0}^J (W_1\cdot W_2)^J [(W_1\cdot X_2) (W_2\cdot X_1)]^k g_k(u)\, .
}
At large geodesic separation $\sigma$, or equivalently  large $u\equiv \cosh\sigma-1\equiv -1-X_1\cdot X_2$,  the authors showed 
\eqna{
g_k(u)\approx \frac{J!}{k!(J-k)!}\frac{J+\Delta-1}{\Delta-1}u^{-\Delta-k}.
}[gkLargeU1]
However, to fully capture the large-$u$ behavior of the  propagator, it is necessary to extract the embedding space indices. Due to the action of the projector $K_A$ --- see (12) in~\cite{Costa:2014kfa} --- powers of $W_i$ can be converted into powers of $X_i$.  Since we are interested only in the leading large-$u$ behavior, we can focus on the highest powers of $(X_1\cdot X_2)$ generated by the action of $K$ on $G(X_1,X_2,W_1,W_2)$.  It is possible to show that these appear as
\eqna{
g_k(u)(X_1\cdot X_2)^{J-k}\alpha_{A_1, \ldots A_J;B_1, \ldots B_J}\,,
}[gkLargeU2]
where $\alpha_{A_1, \ldots A_J;B_1, \ldots B_J}$ is some tensor structure that depends on $J$ but not on $k$.  For intuition on how such powers arise, consider the case $k=0$:  $K_{A_1} \cdots K_{A_J}K_{B_1} \cdots K_{B_J}(W_1\cdot W_2)^J$.  Since each $K_{A}$ contains a term $X_i \cdot \partial_{W_i} $,   acting on $(W_1\cdot W_2)^J$ produces terms proportional to $(X_1\cdot X_2)^J$.  Finally, combining~\eqref{gkLargeU1} and~\eqref{gkLargeU2},  we find that the propagator behaves as:
\eqna{
G(X_1,X_2)\sim u^{-\Delta+J}\sim e^{-\tau \sigma}\, .
}[]
}  
\begin{equation}
G_{\tau}(x,y) \sim e^{-\tau \sigma(x,y)}.
\end{equation}
Moreover, two particles with total spin $J$ in their center-of-mass frame in AdS are separated by a distance that grows logarithmically with $J$ in the large $J$ limit:
\begin{equation}
\sigma \sim \log J,
\end{equation}
so that the shift in their energy from exchange of a bulk field with twist $\tau$ is approximately $ e^{-\tau \log J}=  J^{-\tau}$.  

The dimensions of operators in a CFT are eigenvalues of the dilatation generator, and the CFT dilatation operator $D$ is equivalent to the  Hamiltonian for AdS global coordinates.  Consequently, working in AdS global coordinates maps the problem of finding dimensions of operators in the CFT to the problem of finding the eigenvalues of a many-body Hamiltonian in AdS \cite{Fitzpatrick:2010zm}.  When a CFT is close to a generalized free field (GFF) theory, we can approximate it by a free theory in AdS, and use the free bulk modes as a starting point for perturbation theory.  A free scalar field with bulk action
\begin{equation}
 S = \int d^{d+1} x \sqrt{g} (|\partial \phi|^2 - m^2 |\phi|^2)
 \end{equation}
 has the following mode expansion:
 \begin{equation}
 \begin{aligned}
\phi(x) &= \sum_{n \ell L} \phi_{n \ell L}(x) a_{n \ell L} + \phi^*_{n \ell L}(x) b^\dagger_{n \ell L}, \\
\phi_{n \ell L} &= \frac{1}{N_{\Delta,n,\ell}}  e^{i E_{n,\ell} t} Y_{\ell, L}(\Omega) \sin^\ell \rho \cos^\Delta \rho \  {}_2F_1(-n,\Delta+\ell+n, \ell+\frac{d}{2}, \sin^2 \rho), \\
N_{\Delta, n, \ell} &= (-1)^n \sqrt{ \frac{n! \Gamma^2(\ell+\frac{d}{2}) \Gamma(\Delta+n - \frac{d-2}{2}) }{\Gamma(n+\ell + \frac{d}{2}) \Gamma(\Delta+n + \ell)}}, \qquad N_{\Delta, \ell} \equiv N_{\Delta, 0, \ell},
\label{eq:BulkFieldDecomp}
\end{aligned}
\end{equation}
where $\ell$ labels the total spin of the state, $L$ labels the specific state within the spin-$\ell$ representation (all descendants of a scalar are in the symmetric traceless representation), $E_{n,\ell} = \Delta+2n + \ell$,  $m^2 = \Delta(\Delta-d)$.  
The  corresponding boundary CFT operator is a limit of the bulk operator,
\begin{equation}
\Phi(x) = \lim_{\rho \rightarrow \frac{\pi}{2}} \frac{\sqrt{\textrm{vol}(S^{d-1})}N_{\Delta, 0}}{\cos^\Delta\rho} \phi(x),
\label{eq:BdPhiDecomp}
\end{equation}
normalized to have two-point function $\<\Phi(x) \Phi^*(y)\> = |x-y|^{-2\Delta}$.  We will mostly only be interested in the lowest-twist $n=0$, highest-weight (i.e.~highest-weight within a spin-$\ell$ representation) modes of the bulk field,  in order to build our lowest-twist effective Hamiltonian. These lowest-twist, highest-weight wavefunctions are particularly simple when written in embedding space coordinates,
\begin{equation} 
X_0 = \frac{\cos t}{\cos \rho}\, , \qquad \ X_{d+1} =\frac{\sin t}{\cos \rho}\, , \qquad \ X_i = \tan \rho\lsp  \Omega_i\, ,
\end{equation}
and their null coordinates in the $0,d+1$ and $1,2$ directions,
\begin{equation}
\begin{aligned}
X_\pm &\equiv X_0 \pm X_{d+1} =  \frac{e^{\pm i t}}{\cos \rho}, \\
\bar{X}_{\pm} & \equiv X_1 \pm i X_2 =   \tan \rho \sin \theta_1 \sin \theta_2 \dots \sin \theta_{d-2} e^{\pm i \varphi} = \tan\! \rho \lsp  x_{\pm}.
\end{aligned}
\end{equation}
In terms of these variables, the lowest-twist, highest-weight AdS wavefunctions are\footnote{In the notation of~\cite{Fardelli:2024heb}, $\tilde{N}_{\Delta, \ell_1, \ell_2} = (\tilde{N}_{\Delta, \ell_1} \tilde{N}_{\Delta, \ell_2})^{-1}$.  Integrals over spatial slices of global AdS are facilitated by the identity 
\begin{equation}
\int d^d X_{\rm AdS} \frac{(\bar{X}_+ \bar{X}_-)^A}{ (X_+ X_-)^{B}} = \int d \Omega (x_+ x_-)^A \int_0^{\frac{\pi}{2}} d \rho \frac{\sin^{d-1} \rho}{\cos^{d+1} \rho} \tan^{2A} \rho \cos^{2B} \rho =  \frac{\pi^{\frac{d}{2}} \Gamma(A+1) \Gamma(B-A-\frac{d}{2})}{\Gamma(B)}\, .
\label{eq:intAdSXs}
\end{equation}
}
\begin{equation}\label{eq:phiell}
\phi_\ell(x) = \frac{1}{\tilde{N}_{\Delta,  \ell}} \frac{\bar{X}^\ell_+}{X_-^{\Delta+\ell}}, \quad \phi^*_\ell(x) = \frac{1}{\tilde{N}_{\Delta,  \ell}} \frac{\bar{X}^\ell_-}{X_+^{\Delta+\ell}},  \quad \tilde{N}_{\Delta, \ell} \equiv  \sqrt{\frac{2\pi^{\frac{d}{2}}  \Gamma(\ell+1)\Gamma(\Delta-\frac{d-2}{2})}{\Gamma(\Delta+\ell)}}. 
\end{equation}
A general lowest-twist, highest-weight charge $Q$ state is a linear combination of `monomial' states of the form
\begin{equation}
|\ell_1, \ell_2, \dots, \ell_n \> \equiv b_{\ell_1}^\dagger  b_{\ell_2}^\dagger \dots b_{\ell_n}^\dagger | {\rm vac} \>.
\label{eq:monomialstate}
\end{equation}
Primary states are specific linear combinations of these monomial states, and can be generated efficiently using recursion relations.\footnote{Local interactions in AdS do not mix primary states with any descendant states of the same dimension \cite{Fitzpatrick:2010zm}, so as long as our effective Hamiltonian acts only within a sector of fixed twist then we can restrict it to the space of primaries.}  The two-particle primary at spin $J$ is~\cite{Fitzpatrick:2011dm,Penedones:2010ue,Mikhailov:2002bp}
\begin{equation}
\begin{aligned}
|\Psi\>_{2,J} &= \frac{1}{\sqrt{{\cal N}_{J;2}}} \sum_{\ell=0}^J \frac{(-1)^\ell}{\sqrt{(J-\ell)!\ell! \Gamma(\Delta+J-\ell)\Gamma(\Delta+\ell)}} | \ell, J-\ell\> , \\
{\cal N}_{J;2} & = (1+(-1)^J) \frac{\Gamma(2\Delta+2J-1)}{\Gamma(J+1)\Gamma(\Delta+J)^2\Gamma(2\Delta+J-1)}. 
\end{aligned}
\end{equation}
The Hamiltonian treatment has many technical advantages over the covariant treatment, especially for computing anomalous dimensions.  
However, by choosing a specific frame and breaking up the bulk fields into modes, it obscures some important consequences of conformal symmetry, since a generic conformal transformation changes the reference frame.  For our purposes, the chief one of these consequences is the large-distance decay of bulk propagators.  In terms of the individual modes of bulk fields, a bulk propagator for $t$-channel exchange of a field $\chi$ is recovered by summing over intermediate states with the primary and descendant modes of $\chi$, including time-reversed diagrams where $\chi$ is emitted and absorbed in the opposite order.  This is the well-known fact that the spectral decomposition of the time-ordered propagator is a sum over free particle wavefunctions,
\begin{equation}
G_\chi(x,x',t,t') = i \sum_i  \chi_i(x) \chi_i(x') \left( e^{i E_i (t-t')} \theta(t-t') + e^{-i E_i(t-t')} \theta(t'-t)\right).
\end{equation}
For example, consider how an interaction $g \phi_1 \phi_2 \chi$ corrects the energy of the two-particle state  $|\phi_1 \phi_2\>$ at second order in $g$, through bulk $t$-channel exchange. 
 In time-independent perturbation theory (TIPT), there is a  contribution from 3-particle intermediate states $|\phi_1 \phi_1 \chi_n\>$ and $| \phi_2 \phi_2 \chi_n\>$.   The second-order TIPT formula for the energy is ($\Delta_{ij}\equiv \Delta_i-\Delta_j$)
\begin{equation}
\begin{aligned}
\delta E &= \sum_n \frac{|V_n|^2}{\Delta E_n} = \sum_{n}  \left( \frac{\left| \int d^d x \sqrt{g} \phi_{1}(x)\phi_2(x)\chi_n(x)\right|^2}{\Delta_1+\Delta_2 - (2\Delta_1+ \Delta_\chi+2n)} + \frac{\left| \int d^d x \sqrt{g} \phi_{1}(x)\phi_2(x)\chi_n(x)\right|^2}{\Delta_1+\Delta_2 - (2\Delta_2+ \Delta_\chi+2n)}\right)\\
& = i\! \!\int\! d^d x \sqrt{g} d^d y \sqrt{g} (\phi_1 \phi_2 \chi_n)(x) (\phi_1 \phi_2 \chi_n)(y)\!\! \int_{-\infty}^\infty\! \!\!\!dt\! \left( e^{ i (\Delta_\chi+2n)t} \theta(t) + e^{-i (\Delta_\chi + 2n)t} \theta(-t)\!\right)\!e^{i \Delta_{21} t} \\
& = i\! \!\int\! d^{d+1} x \sqrt{g} \phi^*_1(x,t)\phi_2(x,t) G_\chi(x,y,t,0) \phi_1(y) \phi_2(y),
\label{eq:BulkToBulkFromModes}
\end{aligned}
\end{equation}
thereby reproducing the bulk $\chi$ propagator.  Consequently, when summing over intermediate states in a Hamiltonian treatment, in order to see the correct large $J$ behavior from an intermediate state, we must sum over all the corresponding descendants as well as their time-reversed contributions.  

In practice, rather than doing these explicit sums, it is much more efficient to directly use the bulk propagator that arises from the sum.  However, the interpretation of the bulk propagator as coming from a sum over states will still be crucial, because of the fact that often one of the intermediate states in the sum is identical to the external state, and therefore in TIPT we are instructed {\it not to include this state in the sum}.  Therefore, the bulk propagator must be modified to remove the contribution coming from the term where the intermediate state and the external state are the same.  In practice, this term will be a pole where the energy denominator $\Delta E_n$ vanishes, and so in the propagator calculation will show up as a divergence that goes away once the $\Delta E_n =0$ term is subtracted out.
\section{Analysis at $O(\lambda)$}\label{sec:LO}

\subsection{2-Body Terms}

At $\CO(\lambda)$, the effective Hamiltonian for the lowest-twist states does not get any contributions from integrating out the higher-twist states, because two factors of $\lambda$ are required in order to mix out of and back into the lowest twist sector.  Consequently, the $\CO(\lambda)$ effective Hamiltonian is just the bare Hamiltonian, which follows from the bulk Lagrangian in the standard way:
\begin{equation}
H_2 =  \frac{\lambda}{4} \int d^d x \sqrt{g}  |\phi(x)|^4 \, ,
\end{equation}
where the product of fields is implicitly normal-ordered. Restricting to the 2-body (i.e.  2-to-2 interactions) acting on the lowest-twist and highest-weight (under rotations) states, 
\begin{equation}
\begin{aligned}
H_2 &\supset H_{2\rightarrow 2}= \frac{1}{4}\sum_{\ell_1, \ell_2, \ell_3, \ell_4} V(\ell_1, \ell_2 ,\ell_3 ,\ell_4) a_{\ell_1}^\dagger a_{\ell_2}^\dagger a_{\ell_3} a_{\ell_4}, \\
 V(\ell_1, \ell_2 ,\ell_3 ,\ell_4)  & = \lambda
  \int d^d X_{\rm AdS}\phi_{\ell_1}(x) \phi_{\ell_2}(x) \phi^*_{\ell_3}(x) \phi^*_{\ell_4}(x) \\
   & = \lambda \left(\prod_{i=1}^4 \frac{1}{\tilde{N}_{\Delta, \ell_i}}\right)\frac{\pi^{\frac{d}{2}} \Gamma(\ell_1+\ell_2 +1) \Gamma\mleft(2\Delta-\frac{d}{2}\mright)}{\Gamma(2\Delta+\ell_1+\ell_2)} \delta_{\ell_1 + \ell_2, \ell_3 + \ell_4},
   \label{eq:PhiFourV2}
\end{aligned}
\end{equation}
where we have used (\ref{eq:intAdSXs}).

\subsection{$Q=2$ States}
\label{sec:Q2states}

\subsubsection*{\hspace{-3pt}$Q=2$ Spectrum}
Due to the interaction,   the $Q=2$ states acquire an anomalous dimension $\gamma_{[\Phi, \Phi]_J}$, so that their twist becomes
\eqna{
\tau_{[\Phi, \Phi]_J}=2\Delta+\gamma_{[\Phi, \Phi]_J}\, .
}[]
To compute these anomalous dimensions, we simply diagonalize the Hamiltonian within the space of primaries.  In this case, a short calculation shows that $H_{2 \rightarrow 2}$ vanishes on any primaries with positive spin $J>0$.  In fact, for later reference, consider a more general interaction of the form (\ref{eq:PhiFourV2}) with
\begin{equation}\label{genSpin0potential}
V(\ell_1, \ell_2, \ell_3, \ell_4) =  \left(\prod_{i=1}^4 \frac{1}{\tilde{N}_{\Delta, \ell_i}}\right) f(\ell_1 + \ell_2)  \delta_{\ell_1 + \ell_2, \ell_3 + \ell_4}\, .
\end{equation}
Between two $Q=2$ primaries, the resulting matrix elements are
\begin{equation}
\begin{aligned}
\gamma_{[\Phi,\Phi]_J}\delta_{J,J'} &= {}_{2, J}\< \Psi | H_2 |\Psi\>_{2,J'}\\ &=   \delta_{J,J'}\frac{f(J)}{{\cal N}_{J;2} } \left( \sum_{\ell=0}^J \frac{1}{\tilde{N}_{\Delta, \ell}\tilde{N}_{\Delta, J-\ell }}  \frac{ (-1)^\ell}{\sqrt{(J-\ell)! \ell! \Gamma(\Delta+J-\ell) \Gamma(\Delta+\ell)} }\right)^2 \\
& = \delta_{J',0}  \delta_{J,0} f(0)\frac{ \Gamma(\Delta)^2}{8 \pi^d\Gamma(\Delta-\frac{d-2}{2})^2}\, .
\end{aligned}
\end{equation}
Note that, because this result depends only on the value of $f$ at $J=\ell_1+\ell_2=0$, many different interactions cannot be distinguished based only on the $Q=2$ anomalous dimensions.  For the specific case of our $|\phi|^4$ interaction, 
\begin{equation}
\gamma_{[\Phi,\Phi]_J} = \lambda \delta_{J,0} \frac{ \Gamma(\Delta)^2 \Gamma(2\Delta-\frac{d}{2})}{8 \pi^{\frac{d}{2}} \Gamma(2\Delta) \Gamma( \Delta - \frac{d-2}{2})^2},
\end{equation}
in agreement with previous results in the literature \cite{Heemskerk:2009pn,Fitzpatrick:2010zm}.

\subsubsection*{\hspace{-3pt}$Q=2$ OPE Coefficients}\label{sec:OPEcoeff}

Next, we would like to compute the corrections to the OPE coefficients for $Q=2$ operators in the  $\Phi \times \Phi$ OPE.  Several methods exist already for doing such computations, but we would like to see how compute them within the Hamiltonian framework, using only the matrix elements of $H_2$ as input.  As a practical matter, some of the techniques we develop here will be used again later in more complicated situations, so this computation will be a useful warm-up.  

More precisely, we will directly compute the three-point function $\< \Phi \Phi [\Phi, \Phi]_J\>$ after conformally mapping one of the $\Phi$s to conformal $\infty$, the other $\Phi$ to a point $\hat{x}$ on the unit sphere (i.e. at $t=0$ in radial quantization), and the $[\Phi, \Phi]_J$ to the origin, where it is related to the OPE coefficient $c_{\Phi\Phi[\Phi, \Phi]_J}$ in a conventional normalization by
\begin{equation}
\< \Phi | \Phi(0,\hat{x}) | [ \Phi, \Phi]_J \> = c_{\Phi\Phi[\Phi, \Phi]_J} \hat{x}_-^J, \qquad \hat{x}_\pm \equiv \hat{x}_1 \pm i \hat{x}_2.
\end{equation}
An advantage of computing the OPE coefficients this way is that it makes direct reference to the CFT state $ | [ \Phi, \Phi]_J \>$, and therefore we can use the eigenstates computed from diagonalizing the interacting Hamiltonian.  The presence of the operator $\Phi(x)$, however, involves higher-twist descendants of $\Phi$, as can be seen explicitly in the mode decomposition  (\ref{eq:BdPhiDecomp}). As a result, even at $O(\lambda)$ we need to look at Hamiltonian matrix elements that mix the lowest-twist and higher-twist states.

More precisely, consider the $O(\lambda)$ corrections to the bra and ket state above for $J=0$, $\Phi^2\equiv [\Phi, \Phi]_0$, in TIPT:
\begin{equation}
\begin{aligned}
\ket{\Phi}&=\ket{\Phi}^{(0)}+\sum_{m}\frac{\langle m|H_2| \Phi\rangle^{(0)}}{E_{\Phi}-E_m}\ket{m}\, ,\\
\ket{\Phi^2}&=\ket{\Phi^2}^{(0)}+\sum_{m^\prime}\frac{\langle m'|H_2| \Phi^2\rangle^{(0)}}{E_{\Phi^2}-E_{m^\prime}}\ket{m^\prime}\, .
\end{aligned}
\end{equation}
The resulting correction to the OPE coefficient is
\begin{equation}
\delta c_{\Phi \Phi \Phi^2} = \sum_m \frac{{}^{(0)}\langle \Phi |H_2| m\rangle}{E_{\Phi}-E_m} \< m | \Phi(0,\hat{x}) | \Phi^2\>^{(0)} + \sum_{m'} \frac{\< m' | H_2 | \Phi^2 \>^{(0)} }{E_{\Phi^2}-E_{m^\prime}} {}^{(0)}\<\Phi | \Phi(0, \hat{x}) | m'\>.
\end{equation}
Since $H_2$ is a normal-ordered bulk quartic potential,  $\ket{m}$ is a 3-particle state, while $\ket{m^\prime}$ is a 2-particle state. By inspection,  three of the $\phi$s in the interaction $H_2$ must contract with lowest-twist external states and one $\phi$ must contract with a state created by the operator $\Phi(0,\hat{x})$, so there will always be exactly one mode that is summed over higher-twist descendants. So the sum over $m$ and $m'$ reduces to a sum over the twist label $n$ of the $\Phi$ descendants:
\begin{equation}
\delta c_{\Phi \Phi \Phi^2} = \sum_{n\ge 0} \frac{V_{000n} \< 0 0 n | \Phi(0,\hat{x})| 0 0  \> }{-2\Delta - 2n} + \sum_{n \ge 1}  \frac{V_{000n}}{-2n} \< 0 | \Phi(0, \hat{x} )| 0 n\>,
\end{equation}
where $|0 0 n\> \propto (b_0^\dagger)^2 b_n^\dagger |{ \rm vac}\>, |0 n \> \propto b_0^\dagger b_n^\dagger  | {\rm vac}\>$, $|0\> = b_0^\dagger |{\rm vac}\>$, and 
\begin{equation}
V_{000n} = \frac{\lambda}{N_{\Delta, 0}^3 N_{\Delta, n,0}}  \int d^d x \sqrt{g} \phi^3_{0}(0,x) \phi_{n,0}(0,x)  .
\end{equation}
At this point, one could proceed by brute force evaluation of the integrals and the sum on $n$.  However, there is a significantly more efficient way to evaluate this expression for $\delta c_{\Phi \Phi \Phi^2}$, using the fact that the sum on $n$ is effectively reproducing the bulk-to-boundary propagator of $\phi$, exactly along the lines of the bulk-to-bulk propagator in (\ref{eq:BulkToBulkFromModes}).  The key subtlety is that the term $n=0$ is missing from the second sum above, because of the familiar fact that TIPT at first order does not include the external state in the sum over intermediate states.  It would be convenient if we could simply add this term back in to get the bulk-to-boundary propagator, but to do this literally would mean adding in a divergence of the form $1/0$. 
 To avoid such a singularity, we can first separately parameterize the dimension of the various bulk fields $\phi$. 
 At the level of the action, this corresponds to changing the quartic interaction to
\begin{equation}
S \supset -\frac{\lambda}{4}\int d^{d+1} x \sqrt{g} \phi^{(1)} \phi^{(2)} \phi^{(\Delta_\chi/2)} \phi^{(\Delta_\chi/2)}\, ,
\label{eq:LagContinueDelta}
\end{equation}
and considering the OPE coefficient $\< \Phi^{(1)} | \Phi^{(2)}(0,\hat{x}) | [\Phi^{(\Delta_\chi/2)}, \Phi^{(\Delta_\chi/2)}]_0\>$, where $\Phi^{(1)}$,  $\Phi^{(2)}$, $\Phi^{(\Delta_\chi/2)}$ have dimensions $\Delta_1, \Delta_2, \Delta_\chi/2$, respectively.\footnote{Strictly speaking, analytically continuing in $\Delta_2$ and $\Delta_\chi$ is not quite the same as using the Lagrangian (\ref{eq:LagContinueDelta}), because the symmetry factors of various diagrams change depending on whether or not all the fields in the Lagrangian are identical or distinct.  In order for the limit $\Delta_1 \rightarrow \Delta, \Delta_2 \rightarrow \Delta, \Delta_\chi \rightarrow 2\Delta$ to reproduce the matrix elements of the original interaction, we use the symmetry factors associated with identical rather than distinct fields.}
  Now, the change in the OPE coefficient takes the form
\begin{equation}
\begin{aligned}
V_{000n} & = \frac{\lambda}{N_{\Delta_\chi/2, 0}^2 N_{\Delta_1,0} N_{\Delta_2, n,0}}  \int d^d x \sqrt{g} (\phi_{0}^{(\Delta_\chi/2)})^2(0,x) \phi^{(1)}_{0}(0,x) \phi^{(2)}_{n,0}(0,x)  , \\
\delta c_{\Phi \Phi \Phi^2} &= \sum_{n\ge 0} \frac{V_{000n} \< 0 0 n | \Phi^{(2)}(0,\hat{x})| 0 0  \> }{\Delta_1-(\Delta_\chi+\Delta_2+ 2n)} + \sum_{n \ge 1}  \frac{V_{000n}}{\Delta_\chi-(\Delta_1+\Delta_2+2n)} \< 0  | \Phi^{(2)}(0, \hat{x} )| 0 n\> \\
 & =\frac{1}{\sqrt{2}} \left( - \frac{V_{0000}}{\Delta_\chi-(\Delta_1+\Delta_2)}   - \lambda \int d^{d+1} X (\phi^{(\Delta_\chi/2)}_0)^2(X) \phi^{(1)*}_0(X) K_{\Delta_2}(X,\hat{x})\right).
\end{aligned}
\label{eq:OPEpoleSubtraction}
\end{equation}
The first term in the second line subtracts out the `missing' $n=0$ term, which accounts for the fact that the second sum in the first line is missing the $n=0$ term, and therefore it must be subtracted out when we rewrite the sum in terms of  the bulk-to-boundary propagator. The factor of $1/\sqrt{2}$ is from the normalization factor for the states containing two identical bosons, and equals the undeformed OPE coefficient $c_{\Phi \Phi \Phi^2}$ at $\lambda=0$.   The primary bulk wavefunctions $\phi^*_0(X)$ and $\phi_0(X)$ are just the bulk-to-boundary propagators with the boundary point at conformal infinity $P_\infty$ or the origin $P_0$, respectively:
\begin{equation}
\begin{aligned}
\phi^*_0(X) &=  K_\Delta(X,P_\infty) = \CC_{\Delta} \frac{1}{(-2P_\infty \cdot X)^\Delta}  = \CC_{\Delta}(e^{-i t} \cos \rho)^{\Delta} ,
\\
\phi_0(X) &=  K_\Delta(X,P_0) = \CC_{\Delta} \frac{1}{(-2P_0 \cdot X)^\Delta}  = \CC_{\Delta}(e^{i t} \cos \rho)^{\Delta},
\\
\CC_{\Delta}&\equiv\frac{\Gamma(\Delta)}{2\pi^{\frac{d}{2}}\Gamma\mleft(\Delta-\frac{d-2}{2}\mright)} = \frac{1}{N_{\Delta, 0}^2\textrm{vol}(S^{d-1})}\, ,
\end{aligned}
\end{equation}
where (see e.g.~\cite{Costa:2014kfa}) $P^A=(1,x^2, x^a), P_0 = (1,0,0), P_\infty=(0,1,0)$ and $-2 X \cdot Y = X^+ Y^- + X^- Y^+ -2 X^a Y_a$.  Now the full sum over the descendants is taken care of by the covariant integral in the second line, which is a standard result in AdS/CFT for three-point functions~\cite{Freedman:1998tz,Paulos:2011ie}
\begin{equation}
\begin{aligned}
&W_{\Delta_1\Delta_2\Delta_\chi}^{\rm( 0-con)} \equiv \sqrt{\CC_{\Delta_1}\lsp \CC_{\Delta_2}\lsp \CC_{
\frac{\Delta_\chi}{2}}^2}\int d^{d+1}X  \frac{-\lambda}{(-2P_1\cdot X)^{\Delta_1}(-2P_1\cdot X)^{\Delta_2}(-2P_3\cdot X)^{\Delta_\chi}}\,, \\
&\quad =\frac{-\lambda \pi^{\frac{d}{2}}\sqrt{\CC_{\Delta_1}\lsp \CC_{\Delta_2}\lsp \CC_{
\frac{\Delta_\chi}{2}}^2}\Gamma\mleft(\frac{\Delta_1+\Delta_2+\Delta_\chi-d}{2}\mright)}{2\Gamma(\Delta_1)\Gamma(\Delta_2)\Gamma(\Delta_\chi)}
\frac{\Gamma\mleft(\frac{\Delta_1+\Delta_2-\Delta_\chi}{2}\mright)\Gamma\mleft(\frac{\Delta_1-\Delta_2+\Delta_\chi}{2}\mright)\Gamma\mleft(\frac{-\Delta_1+\Delta_2+\Delta_\chi}{2}\mright)}{P_{12}^{\frac{\Delta_1+\Delta_2-\Delta_\chi}{2}}P_{13}^{\frac{\Delta_1-\Delta_2+\Delta_\chi}{2}}P_{23}^{\frac{-\Delta_1+\Delta_2+\Delta_\chi}{2}}}\, , 
\end{aligned}
\end{equation}
for the tree-level Witten diagram from a spin-0 contact term (``0-con''). One can immediately see that because of  $\Gamma\mleft(\frac{\Delta_1+\Delta_2-\Delta_\chi}{2}\mright)$  this expression diverges when $\Delta_\chi=2\Delta_1=2\Delta_2$.  In the context of our computation, this divergence is due to the fact that the covariant calculation includes the pole arising from the $n=0$ term, which should be subtracted out as shown above. \\
If we denote the pole piece as
\begin{equation}
f^{\rm( 0-con)}_{\Delta_1 \Delta_2\Delta_\chi} \equiv  \frac{V_{0000}}{\Delta_\chi-\Delta_1-\Delta_2} 
= -\lambda\frac{\sqrt{\CC_{\Delta_1}\CC_{\Delta_2}\CC^2_{\frac{\Delta_\chi}{2}}}}{(\Delta_1+\Delta_2-\Delta_\chi)}\frac{\pi^\frac{d}{2}\Gamma\mleft(\frac{\Delta_1+\Delta_2+\Delta_\chi-d}{2}\mright)}{\Gamma\mleft(\frac{\Delta_1+\Delta_2+\Delta_\chi}{2}\mright)}
\end{equation}
and set $P_{ij} =1$ per our choice of boundary insertions, we find the shift in the OPE coefficient-squared is
\begin{equation}
\begin{aligned}
\delta c^2_{\Phi \Phi \Phi^2} = & 2 c_{\Phi \Phi \Phi^2} \delta c_{\Phi \Phi \Phi^2} = 2 \sqrt{2} \delta c_{\Phi \Phi \Phi^2} \\
 = &2 \lim_{\Delta_1\to\Delta} W^{(\rm 0-con)}_{\Delta_1\Delta(2\Delta)}-f^{(\rm 0-con)}_{\Delta_1\Delta(2\Delta)}=2\lim_{\Delta_\chi\to 2 \Delta} W^{(\rm 0-con)}_{\Delta \Delta \Delta_\chi}-f^{(\rm 0-con)}_{\Delta \Delta \Delta_\chi}\\
&\quad =\lambda \frac{ \Gamma (\Delta )^2 \left(2 H_{\Delta -1}-H_{2 \Delta -1}\right) \Gamma
   \mleft(2 \Delta -\frac{\mathit{d}}{2}\mright)}{4 \pi ^{\frac{\mathit{d}}{2}}\Gamma (2 \Delta ) \Gamma
   \mleft(\Delta -\frac{\mathit{d}-2}{2}\mright)^2} . 
   \end{aligned}
\end{equation}
As we go to higher orders in $\lambda$ and start generating effective terms from integrating out higher-twist fields, we will generically have to include additional quartic terms with derivatives.  In a matching computation, the anomalous dimension $\gamma_{\Phi^2}$ and OPE coefficient $c_{ \Phi \Phi \Phi^2}$ for the scalar double-trace operator $\Phi^2$ would be independent pieces of data, which can be used to fix two (linearly independent combinations of) parameters in the effective Lagrangian. If we want to reproduce both the anomalous dimension and the OPE coefficient, we will therefore generically need at least one additional quartic coupling in the effective Hamiltonian, in order to provide a second free parameter in addition to the coefficient $\phi$ of $|\phi|^4$.   
A key point is that, if we only work in the leading twist sector, then for each spin $J$ there are an infinite number of different (linear combinations of) local terms that at tree-level only affect the $Q=2$  primary anomalous dimension for spin $J$.  Therefore, for each spin $J$ {\it one can always choose two local terms out of this infinite set to use to match both the anomalous dimension of $[\Phi, \Phi]_J$ and the OPE coefficient $c_{\Phi \Phi [\Phi, \Phi]_J}$}.

\subsection{$Q=3$ States} \label{Sec:Q3states}

Starting at $Q=3$, for $J=6$ or $J\ge 8$,  there are multiple primaries that are degenerate with each other in the GFF limit.  In the recursive construction of primaries, this degeneracy shows up as multiple ways to take a $Q=2$ state with spin $\ell$ and add another $\Phi$ to obtain total spin $J$:
\begin{equation}
[\Phi, [\Phi, \Phi]_\ell]_{J-\ell}, \qquad  0 \le \ell \le J.
\end{equation}
At finite $J$, these states are not orthogonal to each other, and one must compute their inner product  to construct an orthonormal basis.  At large $J$, their overlap is suppressed as
\eqna{
\langle [\Phi,[\Phi, \Phi]_{\ell^\prime}]_{J-\ell^\prime}| [\Phi,[\Phi, \Phi]_\ell]_{J-\ell}\rangle &=\begin{cases}
1 &\quad \ell^\prime =\ell\, ,\\
 \frac{2}{J^\Delta}\sqrt{\frac{(2\Delta+2\ell-1)(2\Delta+2\ell^\prime-1)\Gamma(2\Delta+\ell-1)\Gamma(2\Delta+\ell^\prime-1)}{\Gamma(\ell+1)\Gamma(\ell^\prime+1)\Gamma(\Delta)^2}} & \quad \ell^\prime \neq \ell\, ,
\end{cases}
}[ovs]
and in particular for $\ell^\prime=0$, 
\eqna{
&\langle [\Phi,\Phi^2]_{J}| [\Phi,[\Phi, \Phi]_\ell]_{J-\ell}\rangle \stackrel{J\gg 1}{\approx}  \frac{1}{J^\Delta} \sqrt{\frac{4 \Gamma (2 \Delta ) (2 \Delta +2 \ell -1) \Gamma (\ell +2 \Delta -1)}{\Gamma (\Delta )^2 \Gamma
   (\ell +1)}}  \\
   &\qquad\qquad\qquad \,\times\left(1  -\frac{(-1)^J}{J^\Delta \Gamma(\Delta)}\left(\Gamma(2\Delta)+(2\Delta+2\ell-1)\frac{\Gamma(2\Delta+\ell-1)}{\Gamma(\ell+1)}\right)\right)+\cdots \, .
}[overlapWithZero]
Because our $O(\lambda)$ interaction is $|\phi|^4$, which at $Q=2$ affects only the spin 0 primary, a special role will be played by the $Q=3$ state recursive built on top of $\Phi^2$.  In the GFF theory this state can be written explicitly in terms of monomial states as follows:
\eqna{
[\Phi, \Phi^2]_J&=\frac{1}{\sqrt{\CN}}\sum_{m=0}^J\sum_{i=0}^{J-m}\sqrt{\frac{2^J\Gamma (i+\Delta ) \Gamma (J-m-i+\Delta )}{i! m! \Gamma (\Delta )^7 \Gamma (m+\Delta ) (J-m-i)!}} \frac{(-1)^m \ket{i, m,J-m-i}}{\Gamma(2\Delta+J-m)}\, , \\
\CN&=\frac{2^{J+1} \left(2 (-1)^J \Gamma (2 \Delta ) \Gamma (J+\Delta )+\Gamma (\Delta ) \Gamma (J+2 \Delta
   )\right) \Gamma (2 J+3 \Delta -1)}{\Gamma (\Delta )^6 \Gamma (2 \Delta ) \Gamma (J+1) \Gamma (J+\Delta
   ) \Gamma (J+2 \Delta )^2 \Gamma (J+3 \Delta -1)}\,.
}[tTrace]

\subsubsection*{\hspace{-3pt}Expectations at Large Spin}
Before moving on to compute the spectrum of the Hamiltonian at $Q=3$, we can gain some intuition about the expected large spin behavior by viewing the three-particle state as a double-trace operator built from $\Phi_1=\Phi$ and $\Phi_2= \Phi^2$, and by applying~\eqref{largeSpinDT}.   Apart from the identity,  at order $\lambda$ and at leading large $J$, the only contributions we need to include are the exchange of $\Phi$ itself in the $t$-channel OPE and of $[\Phi, \Phi^*]_0$ in the $s$-channel OPE.  Higher-twist operators would produce higher powers of $J$, while higher-spin neutral lowest-twist double-trace operators would start contribute at higher order in $\lambda$, since their anomalous dimensions start at order $\lambda^2$.  At order $\lambda$, the large spin expansion predicts 
\eqna{
\tau_{[\Phi, \Phi^2]_J} &\approx \Delta+\Delta_{\Phi^2}+\lambda \frac{ (-1)^J}{J^\Delta} \frac{\Gamma(2\Delta)\!\left(\lsp a_t^{(0)}\gamma_{\Phi^2}^{(\contact)}+ a_s^{(0)} \gamma_{[\Phi, \Phi^*]_0}^{(\contact)}\frac{(-1)^J}{ J^{\Delta}} \frac{\Gamma(2\Delta)}{\Gamma(\Delta)} \right)}{\Gamma(\Delta)\left( 1+a_t^{(0)}\frac{(-1)^J}{J^\Delta }\frac{\Gamma(2\Delta)}{\Gamma(\Delta)}\right)}\, ,
}[largespinLO]
where we have defined
\eqna{
a_t^{(0)}&=c_{\Phi\Phi(\Phi^*)^2}^{(0)}c_{\Phi^*\Phi^*\Phi^2}^{(0)}=2\, , \\
a_s^{(0)}&=c_{\Phi\Phi^* [\Phi, \Phi^*]_0}^{(0)}c^{(0)}_{\Phi^2 (\Phi^{*})^2[\Phi, \Phi^*]_0}=2\, .
}[]
Recalling that the anomalous dimensions of the neutral double-trace operators, due to a $|\phi|^4$ interaction, are simply related to the charged one by $\gamma_{[\Phi, \Phi^*]_0}^{(\contact)}=2 \gamma_{\Phi^2}^{(\contact)}$, then~\eqref{largespinLO} becomes
\eqna{
\tau_{[\Phi, \Phi^2]_J}\approx 3\Delta+\lambda\lsp \gamma^{(\contact)}_{\Phi^2} \left( 1+\frac{(-1)^J}{J^\Delta} \frac{2\Gamma(2\Delta)}{\Gamma(\Delta)}\right)\, .
}[largespinLambda]
\subsubsection*{\hspace{-3pt}Spectrum from Hamiltonian}
When evaluated in the $Q=3$ subsector, the $2\to 2$ Hamiltonian in~\eqref{genSpin0potential} 
has  only one non-vanishing eigenvalue, and the corresponding  eigenstate is $|[\Phi, \Phi^2]_J\>$ from~\eqref{tTrace}.  This behavior is not entirely surprising, given that,  for such an interaction, at $Q=2$ only the spin-0 double-particle state acquires an anomalous dimension. 
This means that $H_2$ acting on the $Q=3$ sector can be concisely expressed as the following projector:
\begin{equation}
\Big( H_2 \Big)_{Q=3} = \gamma_{[\Phi, \Phi^2]_J}^{(\contact)} | [\Phi, \Phi^2]_J \> \< [\Phi, \Phi^2]_J |,
\end{equation}
where $ \gamma_{[\Phi, \Phi^2]_J}^{(\contact)}\equiv \tau_{[\Phi, \Phi^2]_J}^{(\contact)}-3\Delta$ is given by
\eqna{
\gamma_{[\Phi, \Phi^2]_J}^{(\contact)}&=\frac{\lambda}{4}\frac{\pi^{-d}\Delta! \lsp J! \Gamma (J+\Delta )^2 \Gamma (J+3 \Delta -1)}{ (2\Delta)!\Gamma\mleft(\Delta-\frac{d-2}{2}\mright)\Gamma (2 J+3 \Delta -1)}\Bigg( 2 (-1)^J \Gamma (2 \Delta ) \\
&\quad \,  +\Gamma (\Delta )\frac{ \Gamma (J+2 \Delta )}{\Gamma(J+\Delta)} \Bigg) \sum_{\ell=0}^J \frac{f(\ell)}{\Gamma (\ell+1)^2 \Gamma (J-\ell+1) \Gamma (J-\ell+\Delta )}\, .
}[anDimContact]
At large spin,  the sum is a Gaussian peaked around $\ell=\frac{J}{2}$, so it can be performed by saddle-point.  To do so,  we rescale $\ell=J x$, then the sum becomes an integral $ J\int d x $ such that 
\eqna{
\sum_{\ell=0}^J \frac{f(\ell)}{\Gamma (\ell+1)^2 \Gamma (J-\ell+1) \Gamma (J-\ell+\Delta )} &\sim\!\! \int\! \frac{dx}{4\pi^2} \frac{f(J x )\lsp e^{2J(1-(1-x)\log(J(1-x))-x \log(J x))}}{x (1-x)^\Delta J^{\Delta}} \\
&=\frac{2^{2J+\Delta}}{4\pi^{\frac{3}{2}}}e^{2J}J^{-\Delta -2 J-\frac{1}{2}}f\mleft( \frac{J}{2}\mright)\, .
}[]
If we further expand the rest of~\eqref{anDimContact} at $J\gg 1$, paying attention to distinguish between the contribution with and without  $(-1)^J$, we get
\eqna{
\gamma_{[\Phi, \Phi^2]_J}^{(\contact)}& \stackrel{J\gg 1}{\approx}  \frac{\lambda 4^{-(\Delta+1)}\Gamma(\Delta)^2}{\pi^d \Gamma(2\Delta)\Gamma\mleft(\Delta-\frac{d-2}{2}\mright)^2}f\mleft( \frac{J}{2}\mright)\left( J^{2\Delta-1}+ 2(-1)^J \frac{\Gamma(2\Delta)}{\Gamma(\Delta)}J^{\Delta-1}\right)\,.
}[]
To be consistent with the large spin expectation in~\eqref{largespinLambda},  the first term in parenthesis should reproduce $\gamma_{\Phi^2}^{(\contact)}$, this implies that 
\eqna{
f\mleft( \frac{J}{2}\mright) \stackrel{J\gg 1}{\sim}\frac{1}{J^{2\Delta-1}} .
}[]
If we now specify to the actual $|\phi|^4$  interaction with
\eqna{
f(\ell)=\frac{\pi^{\frac{d}{2}}\Gamma(\ell+1)\Gamma\mleft(2\Delta-\frac{d}{2}\mright)}{\Gamma(2\Delta+\ell)}\, ,
}[]
we can either perform the sum over $\ell$ analytically and get
\eqna{
\gamma_{[\Phi, \Phi^2]_J}^{(\contact)}&=\gamma_{\Phi^2}^{(\contact)} \left(1+\frac{2 (-1)^J \Gamma (2 \Delta ) \Gamma (J+\Delta )}{\Gamma (\Delta ) \Gamma (J+2 \Delta )}\right)\, ,
}[Q3contact]
or,  if we are interested only on the large spin behavior,  we can directly use the saddle-point approximation to get
\eqna{
\gamma_{[\Phi, \Phi^2]_J}^{(\contact)}\stackrel{J\gg 1}{\approx}\gamma_{\Phi^2}^{(\contact)} \left( 1+\frac{(-1)^J}{J^\Delta}\frac{2\Gamma(2\Delta)}{\Gamma(\Delta)}\right)\, .
}[Q3contactLargeSpin]
Notice that  in both expressions the first contribution corresponds exactly to $\gamma_{\Phi^2}^{(\contact)}$, 
which is consistent with the expectation 
\eqna{
\tau_{[\Phi, \Phi^2]_J}\xrightarrow[]{J\to \infty} \tau_{\Phi}+\tau_{\Phi^2}+O(J^{-\Delta})=3\Delta+\gamma_{\Phi^2}+O(J^{-\Delta})\,.
}[]
As a consistency check, in appendix~\ref{app:4ptComputation}, we derive the same result starting from the four-point function $\langle \Phi^2(x_1) \Phi(x_2)\Phi^* (x_3) (\Phi^*)^2 (x_4)\rangle$.

\section{$O(\lambda^2)$ Bubble Diagram Contributions to $H$} \label{sec:NLOBubbles}
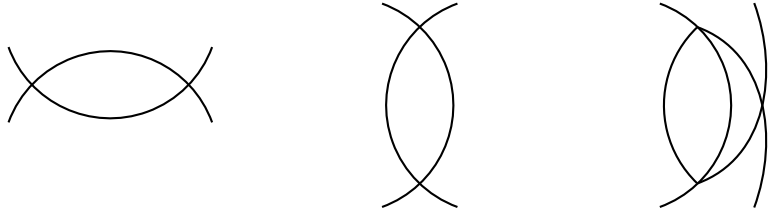
\begin{figure}[t]\center
\begin{tikzpicture}
 \begin{scope}[scale=0.8, transform shape]
\draw[thick] ([shift={(20:1.8cm)}]-5,0) arc (20:160:1.8cm);
\draw[thick] ([shift={(200:1.8cm)}]-5,2.48404) arc (200:340:1.8cm);
\draw[thick] ([shift={(110:1.8cm)}]1.38404,0.9) arc (110:250:1.8cm);
\draw[thick] ([shift={(-70:1.8cm)}]-1.1,0.9) arc (-70:70:1.8cm);
\draw[thick] ([shift={(134:1.8cm)}]6,0.9) arc (134:226:1.8cm);
\draw[thick] ([shift={(-70:1.8cm)}]3.51596,0.9) arc (-70:70:1.8cm);
\draw[thick] (4.755, 2.2) to[out=-20,in=70]  (5.7, -0.8);
\draw[thick] (4.755, -0.4) to[out=20,in=-70]  (5.7, 2.6);
\end{scope}
\end{tikzpicture}\caption{Bubble diagrams contributing to the $Q=2$ spectrum at order $\lambda^2$.} \label{Fig:Q2bubble} 
\end{figure}
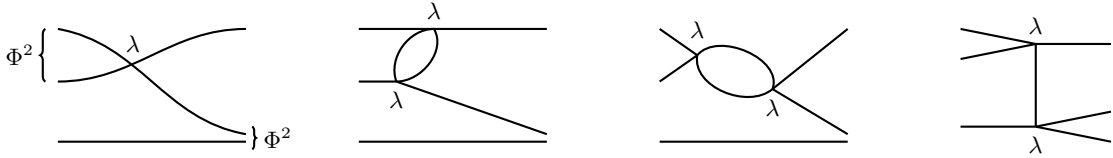
\begin{figure}[t]\center
\begin{tikzpicture}
\draw[thick] (0,0) -- (2.5,0);
\draw[thick] (0,1.5) to[out=-10,in=170] (2.5,0.1);
\draw[thick] (0,0.8) to[out=0,in=180] (2.5,1.5);
\node at (1, 1.3) {\footnotesize $\lambda$};
\draw [thick, decorate,decoration={brace,amplitude=2pt,raise=1ex}]
  (0,0.8) -- (0,1.5);
\node[left=0.18cm] at (0,1.15) {\footnotesize $\Phi^2$};
\draw [thick, decorate,decoration={brace,amplitude=1.5pt, raise=0.5ex}]
  (2.5,0.2) -- (2.5,-0.1);
  \node[right=0.1cm] at (2.5,0.05) {\footnotesize $\Phi^2$};

\draw[thick] (4,0) -- (6.5,0);
\draw[thick] (4,1.5) -- (6.5,1.5);
\draw[thick] (4,0.8) -- (4.5,0.8);
\draw[thick] (4.5,0.8)  to[out=0,in=-60]  (5,1.5);
\draw[thick] (4.5,0.8)  to[out=110,in=-180]  (5,1.5);
\draw[thick] (4.5,0.8) -- (6.5,0.1);
\node[below=0cm] at (4.5,0.8) {\footnotesize $\lambda$};
\node[above=0cm] at (5,1.5) {\footnotesize $\lambda$};

\draw[thick] (8,0) -- (10.5,0);
\draw[thick] (8,1.5) -- (8.5,1.15);
\draw[thick] (8,0.8) -- (8.5,1.15);
\draw[thick] (9.5,0.7) -- (10.5,0.1);
\draw[thick] (9.5,0.7) -- (10.5,1.5);
\draw[thick] (8.5,1.15) to[out=50,in=70]  (9.5,0.7);
\draw[thick] (8.5,1.15) to[out=-100,in=220]  (9.5,0.7);
\node[below=0cm] at(9.5,0.7){\footnotesize $\lambda$};
\node[above=0.1cm] at (8.5,1.15) {\footnotesize $\lambda$};

\draw[thick] (12,0.2) -- (13,0.2);
\draw[thick] (13,0.2) -- (14,0);
\draw[thick] (13,0.2) -- (14,0.4);
\draw[thick] (13,0.2) -- (13,1.3);
\draw[thick] (13,1.3) -- (12,1.5);
\draw[thick] (13,1.3) -- (12,1.1);
\draw[thick] (13,1.3) -- (14,1.3);
\node[below=0cm] at (13,0.2) {\footnotesize $\lambda$};
\node[above=0cm] at  (13,1.3) {\footnotesize $\lambda$};
\end{tikzpicture}  \caption{Diagrams contributing to the $Q=3$ spectrum  at order $\lambda$ and $\lambda^2$.  }
\label{Fig:Q3diagrams}
\end{figure}
Up to this point, all our computations have been at $O(\lambda)$, and the effective lowest-twist Hamiltonian has simply been the bare Hamiltonian.  At $O(\lambda^2)$, however, we start to see the effects from higher-twist states on the lowest-twist Hamiltonian.  In this section, we will see how to treat the effects generated in the two-body Hamiltonian $H_2$, through the bubble loop diagrams shown in Figure~\ref{Fig:Q2bubble} and \ref{Fig:Q3diagrams}.

Following our goal of formulating an EFT based on the separation of scales at large $J$, we do not need to fully evaluate these loop diagrams (though these specific diagrams are simple enough that it is not too hard to do so).  Instead, we want to separate them into long-distance and short-distance contributions to the lowest-twist two-body  Hamiltonian $H_2$.  In practice, this involves performing a $t$-channel spectral decomposition of these contribution, so we can separate out its low- and high-twist components.  

Conveniently, the bubble diagrams in question have a simple closed-form expression for their spectral decomposition~\cite{Fitzpatrick:2011hu},
\begin{equation}
\begin{aligned}
 G_\Delta(X,Y)  &= \CC_\Delta z^{\frac{\Delta}{2}} {}_2F_1(\Delta, d/2, \Delta+1-d/2,z)\, ,  \qquad 
 G^2_\Delta = \sum_{n=0}^\infty b_n^2 G_{\Delta_n}\ , \\
  b_n^2 & =
    \frac{\CC_\Delta^2}{\CC_{\Delta_n}} \frac{\Gamma \mleft(\Delta -\frac{d-2}{2}\mright)^2 \Gamma \mleft(\frac{d}{2}+n\mright)
   \Gamma \mleft(\frac{\Delta_n}{2} \mright)^2 \Gamma \mleft(2 \Delta+n-\frac{d}{2} \mright) \Gamma (\Delta_n -d+1)}{n! \Gamma \mleft(\frac{d}{2}\mright) \Gamma (\Delta )^2 \Gamma
   \mleft(\frac{\Delta_n}{2}-\frac{d-2}{2} \mright)^2 \Gamma (2 \Delta +n-d+1) \Gamma \mleft(\Delta_n-\frac{d}{2}\mright)}\, ,
   \label{eq:BubbleSpecDecomp}
   \end{aligned}
\end{equation}
where $ z =e^{-2\sigma(X,Y)}$ and $\Delta_n = 2 \Delta+2n$.
At the level of the Lagrangian, this decomposition means  that the bubble diagram contributions are completely equivalent to the tree-level exchanges of an infinite tower of scalar fields, 
\begin{equation}
\CL_{\rm eff} \supset  \frac{1}{2}\sum_n \lambda^2  b_n^2 \left( \phi \phi^* \frac{1}{\Box + m_n^2}  \phi \phi^* + \phi^* \phi^* \frac{1}{\Box + m_n^2}  \phi \phi \right),
\end{equation}
 where $m_n^2 = \Delta_n (\Delta_n-d)$.

\subsection{$s$-channel Bubble}

The $s$-channel bubble diagram acting on the lowest-twist sector does not generate long-range interactions, and can be completely absorbed into local counterterms.  Intuitively, this is because the $s$-channel bubble diagrams in flat space depend only on the Mandelstam invariant $s$, which in terms of kinematics is the center-of-mass energy, and therefore related to the AdS twist \cite{Fitzpatrick:2010zm}.  More technically,   the $s$-channel bubble diagrams only contribute to $Q=2$ scalar primary states, and so can be reproduced with spin-0 contact terms.  In fact, {\it they can be completely absorbed into the coefficient of the $|\phi|^4$ contact term in AdS}. 
 Here, we will show this by series expanding the propagator in powers of $1/m_n$.  

The fact that the $s$-channel bubbles can be absorbed into some combination of contact terms is immediately clear from the following expansion:
\begin{equation}
\phi^* \phi^* \frac{1}{\Box + m_n^2} \phi \phi = \phi^* \phi^* \frac{1}{m_n^2} \sum_{j=0}^\infty \left( -\frac{\Box}{m_n^2} \right)^j \phi \phi.
\end{equation}
The next step is to note that when we evaluate the matrix elements of the Hamiltonian from this interaction on the lowest-twist, highest-weight states, the fields $\phi^*$ take on the values of the wavefunctions of the modes they contract with:
\begin{equation}
\phi^2 (x) | \ell_1 , \ell_2 \> = \phi_{\ell_1}(x) \phi_{\ell_2}(x) \propto \frac{\bar{X}_+^{\ell_1 + \ell_2}}{X_-^{2\Delta+\ell_1 + \ell_2}},
\end{equation}
which is the wavefunction for a lowest-twist, highest-weight wavefunction of a spin-$(\ell_1+\ell_2)$ mode of a bulk scalar field with dimension $2\Delta$.  Therefore,
\begin{equation}
\Box \phi_{\ell_1}(x) \phi_{\ell_2}(x) = - 2\Delta (2\Delta-d) \phi_{\ell_1}(x) \phi_{\ell_2}(x).
\end{equation}
Consequently, we can replace $\Box$ with the corresponding eigenvalue, to obtain\footnote{Note that, because $\phi_{\ell_1} \phi_{\ell_2}$ is an eigenfunction of $\Box$, we do not technically even need to series expand the propagator in powers of $\Box$.  }
\begin{equation}
\phi^* \phi^* \frac{1}{\Box + m_n^2} \phi \phi = \phi^* \phi^* \frac{1}{ m_n^2- 2\Delta(2\Delta-d)} \phi \phi ,
\end{equation}
demonstrating that each exchange just shifts the value of the coefficient of $|\phi|^4$.  

The sum $n$ over the spectral decomposition of the bubble diagram can be done in closed form, but in $d\ge 3$ it is divergent due to the familiar UV divergence  of the diagram in flat space.  Therefore, in the lowest-twist Hamiltonian, the $s$-channel bubble diagrams just shift the value of a scheme-dependent counterterm, and have no observable effect.

\subsection{$t$-channel Bubble Spectral Decomposition}

By contrast, the bubble diagrams in the $t$- and $u$-channels generate long-distance effects that affect all spins, and  so we do not replace them completely with local terms.  Instead, we start by separating all contributions  into long-distance and short-distance effects based on the twist cutoff $\Lambda_\tau$.  In fact we have already done the first step in this division, which is to find the spectral decomposition of the bubble diagrams from (\ref{eq:BubbleSpecDecomp}).  For ease of notation, let us write the twist cutoff as 
\begin{equation}
\Lambda_\tau = 2\Delta+2N\, ,
\end{equation}
for some integer $N$, so that all double-trace scalar primaries with $n\le N$ are below the twist cutoff, whereas those with $n > N$ are above the twist cutoff.  Integrating out the primaries above the twist cutoff, we obtain the following effective Lagrangian:
\begin{equation}
\CL_{\rm eff} \supset \frac{\lambda^2}{2} \Big( \sum_{n=0}^N  b_n^2 \phi \phi^* \frac{1}{\Box + m_n^2} \phi \phi^* \Big) + \Big( \sum_m c_m  \phi \phi^* \Box^m  \phi \phi^* \Big) .
\label{eq:TChannelAllEFTTerms}
\end{equation}
The Wilson coefficients at this order in $\lambda$ can be computed explicitly
\begin{equation}
 c_m = \sum_{n>N}  \lambda^2 \frac{b_n^2}{m_n^{2+2m}} \sim O\Big( \frac{\Lambda_\tau^{2(d-3)}}{M^{2(d-3)}}\frac{1}{ \Lambda_\tau^{2m+d-3}}\Big),
 \label{eq:WilsonCoeffsTChannel}
 \end{equation}
 where, as defined in (\ref{eq:MDef}), $\lambda \equiv M^{3-d}$.  As usual, at each order in $\lambda$ some finite number of these coefficients are divergent and cannot be computed from the starting Lagrangian (\ref{eq:BareLag}) and instead must be fixed by renormalization conditions.  In practice, the sum on $m$ should be truncated based on the maximum dimension of irrelevant operators that we wish to keep in the effective theory. For simplicity, we will start by keeping only the counterterm for $|\phi|^4$ itself. 
  Thus, the contribution from bubble diagrams to our effective Lagrangian reduces to
 \begin{equation}
 \CL_{\rm eff} \supset \frac{\lambda^2}{2} \sum_{n=0}^N b_n^2 \phi \phi^* \frac{1}{\Box + m_n^2} \phi \phi^* + \frac{\delta \lambda^{(2)}(N)}{4} (\phi \phi^*)^2.
 \label{eq:LagBubble}
 \end{equation}
The coupling $\lambda$ in the $O(\lambda)$ interaction $\sim \lambda |\phi^4|$ should be treated as a renormalized coupling, which we will fix in terms of the anomalous dimension $\gamma_{\Phi^2}$, and the second-order coefficient $\delta \lambda^{(2)}$ for the counterterm will be chosen to satisfy the renormalization condition that the dimension of $\Phi^2$ is $2\Delta+\gamma_{\Phi^2}$.

We wish to obtain the two-body matrix elements of the Hamiltonian $H_2$ coming from (\ref{eq:LagBubble}). We have already determined the matrix elements for an interaction $(\phi \phi^*)^2$, so all that remains is to find the matrix elements coming from the bulk exchange terms. Individually, each of them is equivalent to the tree-level exchange of a scalar $\chi$ with mass $m_\chi^2 = \Delta_\chi(\Delta_\chi -d)$ and cubic bulk coupling $g_\chi \chi \phi \phi^*$:
\begin{equation}
\CL \supset -g_\chi \chi \phi \phi^* \rightarrow \CL_{\rm eff} \supset \frac{g_\chi^2}{2} \phi \phi^* \frac{1}{\Box + m_\chi^2} \phi \phi^*
\end{equation}
where $\Delta_\chi = 2\Delta+2n$ and the coupling $g_\chi$ can be read off from the spectral  decomposition of the bubble diagram.
The simplest way to compute the resulting two-body matrix elements is to use the fact that the propagator is a Green's function, with a variation on a well-known method from~\cite{DHoker:1999mqo}. We put the details of the computation in appendix \ref{app:scalarmat}, generalizing the result in \cite{Fardelli:2024heb} in $d=3$.  The final result for the two-body matrix elements, in the convention (\ref{eq:PhiFourV2}) for $V(\ell_1, \ell_2, \ell_3, \ell_4)$, is
\begin{align}
&V_{\text{scalar}}(\ell_1,\ell_2,\ell_3,\ell_4)= \sum_{k=\frac{\Delta_\chi}{2}}^\infty V(\ell_1,\ell_2,\ell_3,\ell_4 ; k) - \sum_{k=\Delta}^\infty V(\ell_1,\ell_2,\ell_3,\ell_4; k)\, ,\\
&\begin{aligned}
V(\ell_i; k) & = -\frac{g_\chi^2 \pi^{d/2}  }{\prod_{i=1}^4 \tilde{N}_{\Delta, \ell_i}} a_k  \frac{ \Gamma(\Delta)^2}{\Gamma(k)^2}\frac{2 \ell _2! \ell _4! \Gamma \mleft(\Delta+k-\frac{d}{2}\mright)}{ \Gamma
   \left(\Delta-k\right) \Gamma \left(\ell _2+\Delta\right) \Gamma
   \left(\ell _4+\Delta\right)} \\
&\quad\, \times \sum_{m=0}^{\ell_2}  \frac{\Gamma (k+m)\Gamma \left(k+m-\ell _2+\ell
   _4\right)   \Gamma \left(m+\ell _1+1\right)  \Gamma \left(\ell _2+\Delta-k-m\right)}{\Gamma (m+1) \Gamma
   \left(\ell _2-m+1\right) \Gamma \left(m-\ell _2+\ell _4+1\right) \Gamma
   \left(k+m+\ell _1+\Delta\right)} \, , \\
\end{aligned}\\
&\qquad \, \, a_k  = \frac{\Gamma (k)^2 \Gamma \mleft(\Delta-\frac{\Delta_\chi}{2}\mright) \Gamma \mleft(\Delta+\frac{\Delta_\chi}{2}-\frac{d}{2}\mright)}{4 \Gamma \left(\Delta\right){}^2 \Gamma \mleft(k-\frac{\Delta_\chi}{2}+1\mright) \Gamma
   \mleft(k+\frac{\Delta_\chi}{2}-\frac{d-2}{2}\mright)}\,  .
\label{eq:TChannelV}
\end{align}

\subsubsection{$Q=2$ Spectrum}

In the $Q=2$ spectrum, only the two-body matrix elements contribute. As discussed above, at $O(\lambda^2)$ the only new contributions are the $t$($u$)-channel bulk exchanges, as well as a finite number of local counterterms.  Consider the exchange contributions.  Evaluating our two-body matrix elements from (\ref{eq:TChannelV}) on the two-particle primary states, we find the contribution to the anomalous dimension of the $Q=2$ state at spin $J$ is 
\begin{align} \nonumber 
&\gamma^{(\bubble)}_{[\Phi, \Phi]_J}=\frac{1+(-1)^J}{2} \sum_{n=0}^{\infty}\lambda^2 b_n^2  \gamma^{(\exch)}_{[\Phi, \Phi]_J} (2\Delta+2n)\, , \quad   \gamma^{(\exch)}_{[\Phi, \Phi]_J} (\Delta_\chi) =\sum_{k=\frac{\Delta_\chi}{2}}^\infty \gamma^{(\Delta_\chi)}_{ k}-\sum_{k=\Delta}^\infty \gamma^{(\Delta_\chi)}_{ k} \, , \\
&\gamma^{(\Delta_\chi)}_{ k}= \frac{-\pi ^{-\frac{\mathit{d}}{2}} (k-\Delta)! \Gamma (k)^2 \Gamma \mleft(\Delta -\frac{\Delta_\chi
   }{2}\mright)  \Gamma \mleft(\Delta +\frac{\Delta_\chi-d
   }{2}\mright) \Gamma \mleft(k-\frac{\mathit{d}}{2}+\Delta \mright)}{8  (k-\Delta-J )!  \Gamma
   \mleft(\Delta -\frac{\mathit{d}-2}{2}\mright)^2 \Gamma \mleft(k-\frac{\Delta_\chi }{2}+1\mright) \Gamma (J+k+\Delta ) \Gamma \mleft(k+\frac{\Delta_\chi
  -d+2 }{2}\mright)}\, . \label{scalarGeneric}
  \end{align}
This expression is equivalent to the one obtained from dispersion relations in~\cite{Carmi:2020ekr}.\footnote{In ~\cite{Carmi:2020ekr}, the authors write the anomalous dimension in the form
\eqna{
\gamma^{\left (\bubble\right)}_{[\Phi, \Phi]_J} &=(1+(-1)^J)\sum_{n=0}^\infty \frac{2(\Delta)_n^2 \left(\Delta-\frac{d-2}{2} \right)_n^2}{n!\left( \frac{d}{2}\right)_n (2\Delta+n-d+1)_n \left( 2\Delta+n-\frac{d}{2}\right)_n}\left( \gamma^{(\contact)}_{\Phi \square^n \Phi}\right)^2 \CI_{J, 2\Delta+2n}\, ,
}[]
where in their convention $ \CI{\gamma}_{J, 2\Delta+2n}$ is related to the  anomalous dimension due to a $\chi$-exchange in the bulk written in terms of its OPE coefficient rather than the bulk coupling
\eqna{
 \CI_{J, 2\Delta+2n}&=\lim_{\Delta_\chi\to2\Delta+2n}\frac{g^2 \gamma_{J, \Delta_\chi }}{\lambda_{\Delta\Delta \chi}^2 (\Delta-{\Delta_{\chi}}/{2})^2}\, ,\\
 g&=\frac{2 \sqrt{2} \pi ^{\mathit{d}/4} \Gamma (\Delta ) \sqrt{\Gamma (\Delta \chi )} \Gamma
   \left(-\frac{\mathit{d}}{2}+\Delta +1\right) \sqrt{\Gamma \left(-\frac{\mathit{d}}{2}+\Delta \chi
   +1\right)}}{\Gamma \left(\frac{\Delta \chi }{2}\right)^2 \Gamma \left(\Delta -\frac{\Delta \chi
   }{2}\right) \Gamma \left(\Delta +\frac{\Delta \chi -\mathit{d}}{2}\right)}\,.
}[ZeroOverZero]
}
Although it is not completely manifest from the expression written in this way, we know in general that the contribution from exchange of a scalar with dimension $\Delta_\chi$ should decay at large spin like
\eqna{
\gamma_{[\Phi, \Phi]_J}^{(\exch)}(\Delta_\chi) \sim \frac{1}{J^{\Delta_\chi}}\, ,
}[]
and it is straightforward to check numerically that \eqref{scalarGeneric} does indeed behave this way at large $J$.  This means that if we are interested in the leading large spin behavior of $\gamma^{(\bubble)}_{[\Phi, \Phi]_J}$ it is sufficient to consider only the $n=0$ term in the sum, corresponding to the exchange of the lowest-twist double-trace operator in the bubble diagram.

At finite $J$, we also have to consider the contribution from the local counterterms.  For concreteness, let us take the case $d=4$, so that at $O(\lambda^2)$ only the $|\phi|^4$ counterterm is divergent.  If we also take $\Delta=2$, the expression simplifies significantly
and we find the exchange from a scalar exchange with $\Delta_\chi=2\Delta=4$ reads
\eqna{
\gamma^{(\exch)}_{[\Phi, \Phi]_J}(\Delta_\chi=4)&=-\frac{g_0^2}{(4\pi)^2}\frac{1+(-1)^J}{2}\left(\psi^{(2)}(J+1)+\frac{1}{\CJ^4} \frac{(J+2)(J^2+4J+5)}{J+1}\right)\, , 
}[]
where we have introduced the conformal spin $\CJ^2=(J+1)(J+2)$. The factor $g_n$ is the bulk coupling to the double-trace scalar, which in our application is $g_n = \lambda b_n$, but we will leave it general to emphasize the structure of the diagram. Moreover, in this particular example, we can also find closed-form expressions at small values of $J$ for the contributions from all the higher-twist exchanges $\Delta_\chi=4+2n$. The first few ($J=0,2,4)$ of these are, consistently with~\cite{Aharony:2016dwx},
\threeseqn{
&\gamma^{(\exch)}_{[\Phi, \Phi]_0}=-\frac{g_n^2}{(4\pi)^2}\left( (n+1)^4\psi^{(2)}(n+1)+\frac{2 n^2+6 n+5}{2}\right)\, ,
}[spin0]
{
&\begin{aligned}
\gamma^{(\exch)}_{[\Phi, \Phi]_2}&=-\frac{g_n^2}{(24\pi)^2}\Big(\left(5 n^4+20 n^3+50 n^2+60 n+36\right) (n+1)^4 \psi ^{(2)}(n+1)\\
&\quad\, +\frac{1}{6} \left(30 n^6+210 n^5+735
   n^4+1560 n^3+2041 n^2+1538 n+520\right)\Big)\, , 
\end{aligned}
}[]
{
&\begin{aligned}
\gamma^{(\exch)}_{[\Phi, \Phi]_4}&=-\frac{g_n^2}{2(240 \pi)^2}\Big( (21 n^8+168 n^7+1008 n^6+3696 n^5+10003 n^4+18508 n^3\\
&\quad\, +23716 n^2+18480 n+7200) (n+1)^4 \psi
   ^{(2)}(n+1)+\frac{1}{30}(630 n^{10}\\
   &\quad\, +6930 n^9+46935 n^8+214200 n^7+708225 n^6+1732080 n^5\\
   &\quad\, +3123855
   n^4+4066860 n^3+3635101 n^2+2006882 n+519406)\Big)\, .
   \end{aligned}
}[][gammaBubbleQ2] 
 Notice that at large exchanged twist $m_n\equiv 4+2n$ the anomalous dimensions goes as
\eqna{
\gamma^{(\exch)}_{[\Phi, \Phi]_J} \sim \frac{1}{m_n^{2+2J}}\, .
}[scalarSpecial]
We can compare this result with the expression  (\ref{eq:WilsonCoeffsTChannel}) for the Wilson coefficients of the local contact terms. The key point is that a local contact term must have at least $J$ derivatives in order to contribute to the $Q=2$ spin $J$ anomalous dimensions.  Setting $m=J$ in (\ref{eq:WilsonCoeffsTChannel}), we see that the scaling of the coefficient with $m_n$ exactly matches (\ref{scalarSpecial}).

Finally, given the expression in~\eqref{spin0} we can also compute the contributions to the spin-0 anomalous dimensions from the one-loop $t$- and $u$-channel bubble diagrams
\eqna{
\gamma^{(\bubble)}_{\Phi^2}
&=-\sum_{n=0}^N \frac{\lambda^2 (n+1)^4}{64\pi^2(2n+3)(2n+1)}\left(2 (n+1)^4 \psi ^{(2)}(n+1)+2 n^2+6 n+5\right)\,.
}[]
Following our renormalization condition, we choose the $|\phi|^4$ counterterm to cancel the loop corrections to the spin-0 anomalous dimensions.  In the limit of large twist cutoff $N$, the contribution from bubble diagrams simplifies (still at $d=4, \Delta=2$) to 
\begin{equation}
\gamma^{(\bubble)}_{\Phi^2}
= - \frac{\lambda^2}{(8\pi^2)^2} \left( \frac{N}{12} - \left( \frac{\pi^2}{105} - \frac{1}{8}\right) + O(1/N)\right),
\end{equation}
which explicitly exhibits the expected linear divergence with $N$.  For $J\geq 2$, the anomalous dimensions do not exhibit any divergence, so we can resum them over $n$\footnote{These sums can be performed using the following identities, together with derivatives of the first one
\eqna{
\sum_{n=0}^\infty \psi^{(2)}(n+1)e^{n\epsilon}= \frac{2\left(\text{Li}_3(e^\epsilon)-\zeta_3\right)}{1-e^\epsilon}\,, \qquad \qquad \sum_{n=0}^\infty \frac{\psi^{(2)}(n+1)}{(2n+1)(2n+3)}=8-\frac{\pi ^2}{3}-8 \log (2)\, .
}[polygamma2sums] }
\eqna{
\gamma^{(\bubble)}_{[\Phi, \Phi]_2}&=-\frac{\lambda^2}{(8\pi^2)^2}\left(\frac{5}{162}-\frac{29 \pi ^2}{10395} \right)\, , \\
\gamma^{(\bubble)}_{[\Phi, \Phi]_4}&=-\frac{\lambda^2}{(8\pi^2)^2}\left( \frac{1519}{81000}-\frac{251 \pi ^2}{135135}\right)\, , \\
\gamma^{(\bubble)}_{[\Phi, \Phi]_6}&=-\frac{\lambda^2}{(8\pi^2)^2}\left(\frac{6037}{441000}-\frac{953 \pi ^2}{692835} \right)\, .
}[]
For generic $\Delta$ and $d$, we were not able to find analytic expressions, therefore we will just consider the leading large $J$ behavior coming from the $n=0$ term in the spectral decomposition~\eqref{scalarGeneric}. For $\Delta$ integer, it is possible to write
\eqna{
\gamma_{[\Phi, \Phi]_{J \text{ even}}}^{(\exch)}&\!=\frac{ g_0^2\Gamma\mleft(2\Delta-\frac{d}{2} \mright)(-1)^\Delta }{8\pi^\frac{d}{2}\Gamma\mleft(\Delta-\frac{d-2}{2}\mright)^2}\Bigg[ \frac{ \Gamma\mleft(\frac{d}{2}-\Delta \mright)\Gamma\mleft(2\Delta+J-\frac{d}{2} \mright)}{2\Gamma\mleft(J+\frac{d}{2} \mright)\Gamma\mleft(\Delta-\frac{d-2}{2} \mright)}\!\!\lsp \left(\psi^{(1)}\left(\!\lsp  2\Delta-\frac{d}{2}\right)-\frac{\pi^2}{6}\right)\\
&\quad- \sum_{j=0}^{\Delta+J-1} \frac{(-1)^j \Gamma (j+J +\Delta )}{(j!)^2 (\mathit{d}-2 \Delta +2 j) \Gamma (-j+J +\Delta )}\left( \psi ^{(1)}(j+\Delta )-\frac{\pi ^2}{6}\right)\Bigg]\, , 
}[]
which can be evaluated at large $J$ (for example using a  contour integral trick described below \eqref{contour}) with the result:
\eqna{
\gamma_{[\Phi, \Phi]_J}^{(\bubble)}\approx\gamma_{[\Phi, \Phi]_{J}}^{(\exch)}&\stackrel{J\gg 1}{\approx} -\frac{\lambda^2}{J^{2\Delta}} \frac{\Gamma(\Delta)^4\Gamma\mleft(2\Delta-\frac{d}{2}\mright)^2}{2^5\pi^d\Gamma(2\Delta)\Gamma\mleft(\Delta-\frac{d-2}{2}\mright)^4}\\& \stackrel{\phantom{J\gg 1}}{=}-\frac{2\lambda^2(\gamma_{\Phi^2}^{\contact} )^2}{J^{2\Delta}}\Gamma(2\Delta)\Gamma\mleft( \Delta-\frac{d-2}{2}\mright)\, .
}[bubbleQ2Large]
Although derived for integer $\Delta$, we expect this result to hold for generic values.  
\subsubsection{$Q=3$ Spectrum} \label{subsec:Q=3Spect}
Our next task is to evaluate the contribution of the two-body terms in~\eqref{eq:LagBubble} on $Q=3$ states.  At order $\lambda^2$ is no longer true that only $[\Phi, \Phi^2]_J$ acquires an anomalous dimension; all other degenerate states also receive corrections. In this section, we  detail the computation of the anomalous dimensions for the state $[\Phi, \Phi^2]_J$.  In appendix~\ref{app:fullSpectrumD2d4}, we numerically compute the full spectrum in the specific example $\Delta=2$ in $4d$.

As in the $Q=2$ case,  we  begin by determining the energy corrections arising from a generic bulk exchange.  The general matrix element at $Q=3$ takes the form
\eqna{
\langle \ell_1 \ell_2 \ell_3 |g_\chi \chi \phi \phi^*|\ell_4\ell_5\ell_6 \rangle=\delta_{\ell_1 \ell_4} V_{\rm scalar}(\ell_2, \ell_3, \ell_5, \ell_6)+ \text{permutations}\, ,
}[]
where the permutations account for all possible choices of two $\ell_i$’s acting as spectators and  the  two-two matrix elements are given in \eqref{eq:TChannelV}.

Next, one should sum over all values of the $\ell_i$ to compute the action  on the primary $[\Phi, \Phi^2]_J$ as in~\eqref{tTrace}. However, due to the complexity of the combinatorics, it becomes rather difficult to derive a compact analytic expression for the resulting anomalous dimension as a function of $J$. To proceed concretely, we will therefore restrict to the case $d = 4$ and $\Delta = 2$
\eqna{
\gamma_{[\Phi,\Phi^2]_J}^{(\exch)}(\Delta_\chi)&=g_\chi^2 \left(\sum_{k=\frac{\Delta_\chi}{2}}^\infty \alpha_k \lsp \gamma_{[\Phi,\Phi^2]_J}^{(\exch)}(\Delta_\chi,k) -\sum_{k=2}^\infty \alpha_k \lsp \gamma_{[\Phi,\Phi^2]_J}^{(\exch)}(\Delta_\chi,k) \right)\, ,\\
\alpha_k&=-\frac{\Gamma(k)^2\Gamma\mleft(2-\frac{\Delta_\chi}{2}\mright)\Gamma\mleft(\frac{\Delta_\chi}{2}\mright)}{16\pi^2 k \Gamma\mleft(k+1-\frac{\Delta_\chi}{2}\mright)\Gamma\mleft(k-1+\frac{\Delta_\chi}{2}\mright)}\, ,
}[]
where we have denoted
\eqna{
&\gamma_{[\Phi,\Phi^2]_J}^{(\exch)}(\Delta_\chi,k)\equiv \frac{(J+2)(J+3)}{8(k+1)}-\!\!\sum_{m=2}^{J+1+(-1)^J}\!\!\frac{3(-1)^m \Gamma(J+m+4)p_m}{(k+m) \Gamma (m) \Gamma (m+4) \Gamma (J-m+2)}\, , \\
&p_m=\begin{cases}
\frac{-2 m^3+(J-1) (J+6) m^2+\left(5 J^2+25 J+8\right) m+(J+1) (J+4) \left(J^2+5 J+12\right)}{(J+1)_4 (J^2+5J+18)} & J\text{ even}\, ,\\
\frac{\left(J^2+5 J-2 m\right) (J-m+1) (J+m+4)}{(J+1)_4 (J-1)(J+6)} & J\text{ odd}\, . \\
\end{cases}
}[]
Setting $\Delta_\chi=4+2n$,   partial cancellations between the two sums occur\footnote{Concretely, we can expand arounf $\Delta_\chi=4+2n+\delta$.  Upon summing over $k$, the order $\delta^{-1}$ cancels and we need  to extract the order $\delta^{0}$. } and the final expression drastically simplifies 
\threeseqn{
&\begin{aligned}
\gamma_{[\Phi,\Phi^2]_J}^{(\exch)}(\Delta_\chi=4+2n)&=\frac{3g_n^2 }{4\pi^2}\Bigg\lbrace \frac{(J+2)_2}{24}\left((n+1)^4\psi^{(2)}(n+1)+\frac{2 n^2+6 n+5}{2}\right)\\
&\quad\, \, +\!\sum_{m=2}^{J+1+(-1)^J}\frac{(-1)^m p_m  (m+4)_J }{ \Gamma (m)  (J-m+1)!} \left(f_1(m)+f_2(m)\right)\Bigg\rbrace \, , 
\end{aligned}}[mSumScalar]
{
&
\begin{aligned} f_1(m)&=-2\frac{n+1}{m} \Bigg((1+n)(H_n+\zeta_3)+\frac{n-2 m n-m+1}{2 m}\\
&\quad\, +\frac{(-1)^n \Gamma(m-n)\Gamma(m+n+2)}{2 \Gamma(m+1)^2} \left(\psi ^{(1)}(m+2)-\frac{\pi ^2}{6}\right)\Bigg)\, , 
\end{aligned}}[]
{&f_2(m)=-\sum_{i=1}^n \frac{(-1)^i (n+1) \Gamma (i+n+2)}{i^3 (i-m) \Gamma (i)^2 \Gamma (-i+n+1)}\left( \psi ^{(1)}(i+2)-\frac{\pi ^2}{6}\right)\, .
}[][]
To extract the large $J$-dependence, it is convenient to rewrite the alternating sum as a complex contour integral
\eqna{
&\sum_{m=2}\frac{(-1)^m p_m  (m+4)_J }{ \Gamma (m)  (J-m+1)!} \left(f_1(m)+f_2(m)\right)=\\
&\left(\frac{(J+1)_4}{24}p_1(f_1(1)+f_2(1))\right)+\oint_{\CC} \frac{dm}{2\pi i} \, \frac{\pi}{\sin(\pi  m)} \! \lsp  \frac{p_m  (m+4)_J }{ \Gamma (m)  (J-m+1)!} \left(f_1(m)+f_2(m)\right)\, ,
}[contour]
where $\CC$ is a contour enclosing all the positive integers $m=0,1, \ldots$ and we have used the fact that the $m=0$ term vanishes in the original sum. We can then deform the contour to encircle the negative integers. This has the advantage that, in the large-$J$ limit, the residues at the poles $m = -M$ (with $M = 2, 3, 4, \ldots$, with the residue at $M = -1$ vanishing) fall off as $\frac{1}{J^{2M}}$.  However, one must be careful that there is an addition pole at $m=+\infty$ that one picks up when deforming the contour to enclose the poles at negative $m$.  Schematically,
\eqna{
&\sum_{m=2}\frac{(-1)^m p_m  (m+4)_J }{ \Gamma (m)  (J-m+1)!} \left(f_1(m)+f_2(m)\right)+\left(-\frac{(J+1)_4}{24}p_1(f_1(1)+f_2(1))\right)\stackrel{J\gg 1}{\approx} \\
&  \left(\mathop{\text{Res}}_{m\to +\infty}- \sum_{M=2}^{\tilde{M}}\mathop{\text{Res}}_{m=-M}\right)\left[ \frac{ \pi \! \lsp \csc(\pi m) p_m  (m+4)_J }{ \Gamma (m)  (J-m+1)!} \left(f_1(m)+f_2(m)\right)\right]+O\left( \frac{1}{J^{2\tilde{M}+1}}\right)\, .
}[]
For $n=0$ and $n=1$,  the first few orders at large spin are
\twoseqn{
&\begin{aligned}
&\gamma_{[\Phi,\Phi^2]_J}^{(\exch)}(\Delta_\chi=4)=\gamma_{\Phi^2}^{(\exch)}(\Delta_\chi=4)\left( 1+\frac{12(-1)^J}{\CJ^2}\right)\\
&-\frac{3g_0^2}{4\pi^2 \CJ^4}\left( \log \CJ+\gamma_E-\frac{\pi^2}{6}+24\zeta_3-\frac{117}{4}\right)-\frac{3g_0^2}{\pi^2\CJ^6}\Bigg(\frac{2-9(-1)^J}{3}(\log\CJ+\gamma_E)\\
&  -\frac{\pi^2(1-3(-1)^J)}{6}-\frac{ (-1)^J (576 \zeta _3-703))}{8}+\frac{10}{9} \Bigg)\, ,
\end{aligned}
}[]
{
&\begin{aligned}
&\gamma_{[\Phi,\Phi^2]_J}^{(\exch)}(\Delta_\chi=6)=\gamma_{\Phi^2}^{(\exch)}(\Delta_\chi=6)\left( 1+\frac{12(-1)^J}{\CJ^2}\right)-\frac{4g_1^2}{\pi^2\CJ^4}\left( 72 \zeta_3+\frac{\pi^2}{4}-89\right)\\
&\quad+ \frac{g_1^2}{\pi^2\CJ^6}\left( \log\CJ+\gamma_E-\pi^2(7-12(-1)^J)+48 (-1)^J (72 \zeta_3-89)+\frac{1627}{24}\right)\, ,
\end{aligned}
}[][]
where we have used the ``conformal spin'' $\CJ^2=(J+2)(J+3)$. Notice that the leading piece is proportional to the $O(\lambda)$ $Q=3$ anomalous dimension in~\eqref{Q3contact}.

More generally, for any $\Delta_\chi=4+2n$ and $n\geq 1$,\footnote{For completeness, we should mention that we explicitly checked that for $ n\geq 2$,  no $\log J$ appears at any order in the $1/J$ expansion.} we find the following closed form up to $\CJ^{-4}$:
\eqna{
&\gamma_{[\Phi,\Phi^2]_J}^{(\exch)}(\Delta_\chi\!=4+2n)=\gamma_{\Phi^2}^{(\exch)}(\Delta_\chi\!=4+2n)\left( 1+\frac{12(-1)^J}{\CJ^2}\right)+\frac{3g_n^2}{4\pi^2 \CJ^4}\Bigg\lbrace \frac{-1}{n(n+2)}\\
& +12 n^2+35 n+28+12(n+1)^4 \left( \psi^{(2)}(n+1)+\frac{\psi^{(1)}(n+1)}{{6(n+1)}}-\frac{\psi^{(1)}\left(\frac{n}{2}+1\right)}{12(n+1)}\right)\, .
}[J4bubbles]
This representation is an explicit example of our more general expectation that at leading order in the large-$J$ limit, the $J$-dependence of an exchange is equivalent to a sum over contact terms, where each contact term is weighted by the corresponding $Q=2$ anomalous dimension, as follows:
\eqna{
\gamma_{[\Phi, \Phi^2]_J}^{(\bubble)}&=\lambda^2\sum_{n} \gamma_{[\Phi,\Phi^2]_J}^{(\exch)}(\Delta_\chi=2\Delta+2n)\\
&\stackrel{J\gg 1}{\approx}  \lambda^2 \left( \sum_n  \gamma_{\Phi^2}^{(\exch)}(\Delta_\chi=2\Delta+2n)  \right) \frac{ \gamma_{[\Phi,\Phi^2]_J}^{(\contact)}}{\gamma_{\Phi^2}^{(\contact)}}+O\left( \frac{1}{J^{2\Delta}}\right)\,,
}[]
where we expect that all corrections of the form $J^{-(\Delta + i)}$ are fully captured by the spin-0 contact interaction.
The renormalization procedure described in~\eqref{eq:LagBubble} can then be straightforwardly applied, yielding
\eqna{
\gamma_{[\Phi, \Phi^2]_J}^{(\bubble \text{, eff})}
&\stackrel{J\gg 1}{\approx}   \left(\delta\lambda^{(2)}(N)+\lambda^2 \sum_{n=0}^N  \gamma_{\Phi^2}^{(\exch)}(\Delta_\chi=2\Delta+2n)  \right) \frac{ \gamma_{[\Phi,\Phi^2]_J}^{(\contact)}}{\gamma_{\Phi^2}^{(\contact)}}+\cdots \,.
}[bubbleLargeSpin]
Suppose that we have fixed $\delta \lambda^{(2)}$ to fix the dimension of the $J=0, Q=2$ lowest-twist double-trace state $\Phi^2$.  In the specific example $d=4$ and $\Delta=2$, we can then proceed to completely fix the remaining $J^{-4}$ term in~\eqref{J4bubbles}\footnote{The sums over $n$ can be done using the identities in~\eqref{polygamma2sums} together with
\eqna{
\sum_{n=0}^\infty \psi^{(1)}(n+1)e^{n \epsilon}&=\frac{ \text{Li}_2\left(e^{\epsilon}\right)-\frac{\pi ^2}{6}}{ \left(e^{\epsilon-1 }\right)}\, , \qquad && \sum_{n=0}^\infty \frac{(n+1)\psi^{(1)}(n+1)}{(2n+1)(2n+3)}=\frac{7 \zeta_3}{8}-1+\log (2)\, , \\
\sum_{n=0}^\infty \psi^{(1)}\left(\frac{n}{2}+1\right)e^{n \epsilon}&=\frac{ 4\text{Li}_2\left(e^{\epsilon}\right)-\frac{\pi ^2}{6}(1+3e^\epsilon)}{ \left(e^{2\epsilon-1 }\right)}\, , \qquad &&\sum_{n=0}^\infty \frac{(n+1)\psi^{(1)}\left(\frac{n}{2}+1\right)}{(2n+1)(2n+3)}=\frac{7 \zeta_3}{4}-\frac{\pi ^2}{8}\, .\\
}[]}
\eqna{
\gamma_{[\Phi, \Phi^2]_J}^{(\bubble \text{, eff})}
&\stackrel{J\gg 1}{\approx}
 \gamma^{(2)}_{\Phi^2} \left( 1+\frac{12(-1)^J}{(J+2)(J+3)}\right) -\frac{2\lambda^2}{(4\pi^2)^2}\frac{\left( \log J+\gamma_E+\frac{87\pi^2}{1120}-\frac{29}{12}\right)}{J^4} + \cdots\,.
}[d4del2bubble]

\subsection{Matrix Elements of Local Contact Terms}

Finally, we consider the local quartic contact terms generated by integrating out the higher-twist states.  In  \eqref{eq:TChannelAllEFTTerms}, we wrote them in a standard basis of operators each with a fixed number of derivatives.  However, it is more convenient to reorganize them into terms that each affect only a single value of the spin $\ell$ for the $Q=2$ states.  For $\ell=0$, this operator is the familiar $|\phi|^4$ contact term, but for $\ell>0$ it is a linear combination of the terms in (\ref{eq:TChannelAllEFTTerms}).  The advantage of this basis is that each such operator contributes to the Hamiltonian as a projector onto a single $Q=3$ state, whose  $Q=2$ substate is spin $\ell$:
\begin{equation}
\begin{aligned}
\CL \supset \lambda_\ell [ \phi^4]_\ell \rightarrow \quad  & H_2 \supset \gamma_{[ \Phi, \Phi]_\ell}^{(\contactell{\ell})} |[ \Phi, \Phi]_\ell \> \< [ \Phi, \Phi]_\ell| , \\
& H_3 \supset \gamma_{[\Phi,[\Phi, \Phi]_\ell]_{J-\ell}}^{(\contactell{\ell})} |[ \Phi,[ \Phi, \Phi]_\ell ]_{J-\ell} \> \< [\Phi,  [ \Phi, \Phi]_\ell ]_{J-\ell}| \, ,
\end{aligned}
\end{equation}
where  it is assumed $J\geq \ell$.  For the $Q=2$ anomalous dimensions we were able to find a closed form expression, while for $Q=3$ we report only its leading large-spin behavior\footnote{As an example the full solution for a spin-2 contact interaction is
\eqna{
\gamma_{[\Phi, [\Phi, \Phi]_2]_{J-2}}^{(\contactell{2})}&=\gamma_{ [\Phi, \Phi]_2}^{(\contactell{2})}\Bigg[1+(-1)^J \frac{\Gamma (J+\Delta )}{\Gamma (J+2 \Delta )}\frac{(2 \Delta +3) \Gamma (2 \Delta +1)}{\Gamma (\Delta )} \Bigg( 1-\frac{2 (\Delta +1) (\Delta +2) (2 \Delta +1)}{\Delta  (\Delta +J-1) (2 \Delta +J)}\\
&\quad\, +\frac{2 (\Delta +1) (\Delta +2) (\Delta +3) (2 \Delta +1)}{(\Delta +J-2) (\Delta +J-1) (2 \Delta +J) (2
   \Delta +J+1)}\Bigg)\Bigg]\, .
}[ContactQ3Spin2]}
\twoseqn{
\gamma_{[\Phi, \Phi]_\ell}^{(\contactell{\ell})}&=\frac{\lambda_{\ell}}{4}\frac{ \ell ! \Gamma (\Delta+\ell ) \Gamma (\Delta +\ell  +1) \Gamma
   \mleft(2 \Delta+\ell-\frac{\mathit{d}}{2} \mright)}{\pi^\frac{d}{2} \Gamma (2 \Delta +2 \ell +1)\Gamma \mleft(\Delta-\frac{\mathit{d}-2}{2}
   \mright)^2}\, .
}[]
{
\gamma_{[\Phi,[\Phi, \Phi]_\ell]_{J-\ell}}^{(\contactell{\ell})}&\stackrel{J\gg 1}{\approx} \gamma_{[\Phi, \Phi]_\ell}^{(\contactell{\ell})}\left( 1+\frac{(-1)^J}{J^\Delta}\frac{2 (2 \Delta +2 \ell -1) \Gamma (2 \Delta+\ell -1)}{\Gamma(\ell+1) \Gamma (\Delta )}\right)+\cdots
}[][]
These expressions can also be used to determine the anomalous dimensions of three-particle states at large spin due to a generic scalar exchange. As discussed earlier, any scalar bulk exchange can be replaced by an infinite sum of local quartic contact interactions. In this language, the two-body Hamiltonian at fixed charge $Q=3$ can be rewritten as 
\eqna{
\left(H_2^{(\exch)}\right)_{Q=3}=\sum_{\ell} \frac{\gamma_{[\Phi, \Phi]_\ell}^{(\exch)}(\Delta_\chi)}{\gamma_{[\Phi, \Phi]_\ell}^{(\contactell{\ell})}}\gamma_{[\Phi,[\Phi, \Phi]_\ell]_{J-\ell}}^{(\contactell{\ell})} |[ \Phi,[ \Phi, \Phi]_\ell ]_{J-\ell} \> \< [\Phi,  [ \Phi, \Phi]_\ell ]_{J-\ell}|\, .
}[Q3HasContact]
We emphasize that, although this expression is valid at finite $J$, the states $|[ \Phi,[ \Phi, \Phi]_\ell ]_{J-\ell} \> $ are not orthogonal to each other at finite $J$, and consequently the eigenvectors of $\left(H_2^{(\exch)}\right)_{Q=3}$ are not simply the $|[ \Phi,[ \Phi, \Phi]_\ell ]_{J-\ell} \>$ basis states. 

Using the overlaps computed in~\eqref{ovs} and~\eqref{overlapWithZero}, we can  extract the corrections to the anomalous dimensions of specific states. However, we stress that these states are not, in general, eigenstates of the full Hamiltonian and may mix with each other.  For example
\eqna{
&\langle [\Phi, \Phi^2]_J | \left(H_2^{(\exch)}\right)_{Q=3}| [\Phi, \Phi^2]_J\rangle \approx  \gamma_{\Phi^2}^{(\exch)}\left(1+\frac{(-1)^J}{J^\Delta}\frac{2\Gamma(2\Delta)}{\Gamma(\Delta)}\right)\\
&\quad +\sum_{\ell\geq 2} \gamma_{[\Phi, \Phi]_\ell}^{(\exch)}\frac{4}{J^{2\Delta}}\frac{(2\Delta+2\ell-1)\Gamma(2\Delta)\Gamma(2\Delta+\ell-1)}{\Gamma(\ell+1)\Gamma(\Delta)^2}\left(1-\frac{(-1)^J}{J^\Delta}\frac{2\Gamma(2\Delta)}{\Gamma(\Delta)}\right)\, .
}[Q3contactell]
Notice that this  reproduces exactly our previous results  in~\eqref{J4bubbles}.  

This expression is also particularly useful for predicting the $\frac{\log J}{J^{2\Delta}}$ term from bubble diagrams for generic $\Delta$ and $d$, as in~\eqref{d4del2bubble}. For $\Delta=2$, $d=4$, the leading $\log J$ originates from the exchange of the lowest-twist scalar with $\Delta_\chi=2\Delta$. By analogy, the $\frac{\log J}{J^{2\Delta}}$ term can be extracted from~\eqref{Q3contactell} substituting  $\gamma_{[\Phi, \Phi]_\ell}^{(\exch)}(\Delta_\chi=2\Delta)$ as in~\eqref{bubbleQ2Large} 
\eqna{
\sum_{\ell\geq 2} \gamma_{[\Phi, \Phi]_\ell}^{(\exch)}\frac{4}{J^{2\Delta}}\frac{(2\Delta+2\ell-1)\Gamma(2\Delta)\Gamma(2\Delta+\ell-1)}{\Gamma(\ell+1)\Gamma(\Delta)^2}&\approx -\frac{\lambda^2(\gamma_{\Phi^2}^{(\contact)})^2}{J^{2\Delta}} \frac{16 \Gamma(2\Delta)^2 }{\Gamma(\Delta)^2}\sum_{\substack{\ell=2\\\text{even}}}^J \frac{1}{\ell} \\
&\approx -{\lambda^2(\gamma_{\Phi^2}^{(\contact)})^2}\frac{8\Gamma(2\Delta)^2}{\Gamma(\Delta)^2}\frac{\log J+\gamma_E}{J^{2\Delta}}\, .
}[Q3bubbleResummed]
The emergence of the $1/\ell$ term can be traced to the interplay between the anomalous dimensions and the overlaps: $\gamma_{[\Phi, \Phi]_{\ell}}^{(\contact)}\sim \ell^{-2\Delta}$ combines with $\langle[\Phi, [\Phi, \Phi]_{\ell}]_J | [\Phi, \Phi^2]_{J}\rangle^2\sim \ell^{2\Delta-1}$, yielding precisely the logarithmic enhancement.  Roughly speaking, the extra $\Phi$ in $[\Phi, \Phi^2]_J$ distorts the scalar $\Phi^2$ state (in order for the total state to be an eigenstate of spin), so that the wavefunction of the $\Phi^2$ blob `leaks' into the other spherical harmonics at $Q=2$, and gets a contribution from a sum over their anomalous dimensions $ \sim \ell^{-2\Delta}$ weighted by this leakage $\sim \ell^{2\Delta-1}$. 

Finally, the  effect of bubble diagram for the anomalous dimension of $[\Phi, \Phi^2]_J$ can be summarized as
\eqna{
\gamma_{[\Phi, \Phi^2]_J}^{(\bubble)}\stackrel{J\gg 1}{\approx}  \lambda^2\gamma_{\Phi^2}^{(\bubble)}\left( 1+\frac{(-1)^J}{J^\Delta}\frac{2\Gamma(2\Delta)}{\Gamma(\Delta)}\right)-{\lambda^2(\gamma_{\Phi^2}^{(\contact)})^2}\frac{8\Gamma(2\Delta)^2}{\Gamma(\Delta)^2}\frac{\log J+\gamma_E}{J^{2\Delta}}\, .
}[bubbleTT]
Notice the difference with the respect to the $Q=2$ case in~\eqref{bubbleQ2Large}, in particular the appearance of a $\log J$.\footnote{This implies that, when exchanged in a four-point function, these operators give rise to specific powers of $\log v$ that differ from those generated by double-twist exchanges.}

\section{$O(\lambda^2)$ Three-Body Terms}\label{sec:NLO3To3}

When we integrate out the higher-twist states at $O(\lambda^2)$,  the second class of diagram that gets generated is a three-body interaction where a single scalar is exchanged in the $t$-channel, as depicted in the right-most diagram in Figure~\ref{Fig:Q3diagrams}.  To evaluate this contribution to the Hamiltonian, we will have to follow the same prescription we used in section \ref{sec:OPEcoeff} for evaluating OPE coefficients, in order to correctly capture its large $J$ behavior.  Recall that the key point is that to get the correct large-distance decay of such contributions, we must include the infinite set of intermediate states that combine to make up the bulk propagator for the exchange, {\it minus} the lowest-twist intermediate state contribution.  

In order to perform this computation, we consider deforming the dimension of the exchanged $\phi$ from $\Delta$ to $\Delta_\chi$, to obtain the  diagram in Figure~\ref{fig:3to3}. This step is necessary because both the contribution from the $\chi$ propagator, and the subtraction of the lowest-twist intermediate state, are singular at $\Delta_\chi= \Delta$, and only their combination is regular.  
\tikzset{middlearrow/.style={
        decoration={markings,
            mark= at position 0.5 with {\arrow{#1}} ,
        },
        postaction={decorate}
    }
}
\begin{figure}
\centering
\begin{tikzpicture}
\draw[thick,middlearrow={stealth}] (2,0) -- (3,-0.3);
\draw[thick,middlearrow={stealth}] (2,-0.6) -- (3,-0.3);
\node[left=0cm] at (2,0) {\small $\Phi$};
\node[left=0cm] at (2,-0.6) {\small $\Phi$};
\draw[line width=0.18mm] (2.98, -0.3) -- (2.98, -2);
\draw[line width=0.18mm] (3.02, -0.3) -- (3.02, -2);
\node[right=0.05cm] at (3, -1.15) {\small $\chi$};
\draw[thick,middlearrow={stealth}] (2, -2.3)--(3, -2);
\node[left=0cm] at (2,-2.3) {\small $\Phi$};
\draw[thick,middlearrow={stealth reversed}] (3, -2)--(4, -2.3);
\draw[thick,middlearrow={stealth reversed}] (3, -2)--(4, -1.7);
\draw[thick,middlearrow={stealth reversed}] (3, -0.3)--(4, 0);
\node[right=0cm] at (4,0) {\small $\Phi^\dagger$};
\node[right=0cm] at (4,-1.7) {\small $\Phi^\dagger$};
\node[right=0cm] at (4,-2.3) {\small $\Phi^\dagger$};
\draw[thick] (6, -0.4)--(8.5,0);
\draw[thick] (6, -2.3)--(8.5,-1.9);
\draw[thick,middlearrow={stealth}] (6, 0)--(7,-0.24);
\node[left=0cm] at (6,0) {\footnotesize {$\ell_1$}};
\draw[thick,middlearrow={stealth}] (6, -0.4)--(7,-0.24);
\node[left=0cm] at (6,-0.4) {\footnotesize {$\ell_2$}};
\node[left=0cm] at (6,-2.3) {\footnotesize {$\ell_3$}};
\draw[thick,middlearrow={stealth reversed}] (8.5, 0)--(7,-0.24);
\node[right=0cm] at (8.5,0) {\footnotesize {$\ell_4$}};
\draw[thick,middlearrow={stealth}] (7.5, -2.06)--(8.5,-2.3);
\node[right=0cm] at (8.5,-2.3) {\footnotesize {$\ell_6$}};
\draw[thick,middlearrow={stealth reversed}] (7.5, -2.06)--(6,-2.3);
\node[right=0cm] at (8.5,-1.7) {\footnotesize {$\ell_5$}};
\draw[thick,middlearrow={stealth}] (7.5, -2.06)--(8.5,-1.9);
\draw[thick,middlearrow={stealth }] (7, -0.24)--(7.5,-2.06);
\draw[thick,  red, dashed] (7.25, 0.05)--(7.25, -2.35);
\node[red, right=0cm] at (7.25,-1.15) {\footnotesize $k, n$};
\draw[thick] (10, -0.4)--(12.5,0);
\draw[thick] (10, -2.3)--(12.5,-1.9);
\draw[thick,middlearrow={stealth}] (10, 0)--(11.5,-0.16);
\draw[thick,middlearrow={stealth}] (10, -0.4)--(11.5,-0.16);
\draw[thick,middlearrow={stealth reversed}] (12.5, 0)--(11.5,-0.16);
\draw[thick,middlearrow={stealth}] (11, -2.14)--(12.5,-2.3);
\draw[thick,middlearrow={stealth}] (11, -2.14)--(10,-2.3);
\draw[thick,middlearrow={stealth reversed}] (11, -2.14)--(12.5,-1.9);
\draw[thick] (11, -2.14)--(11.5,-0.16);
\draw[thick,  red, dashed] (11.25, 0.05)--(11.25, -2.35);
\end{tikzpicture}\caption{``Three-to-three'' interaction due to the exchange of $\phi$.  Once properly removed the pole, we can set $\Delta_\chi=\Delta$.} \label{fig:3to3}
\end{figure}
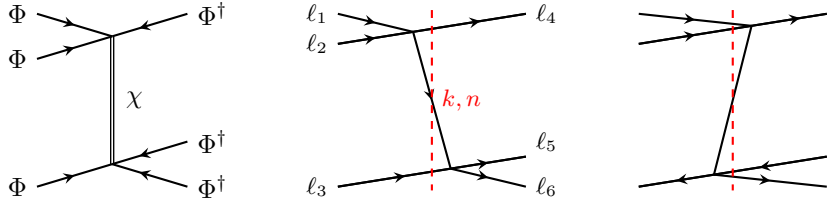
Consider the propagator contribution first.  As before, the easiest way to evaluate its contribution is to use the fact that the propagator is a Green's function.  Therefore, the matrix element takes the form
\begin{equation}
\langle \ell_1 \ell_2 \ell_3 |\chi \phi \phi^\dagger\phi^\dagger |\ell_4\ell_5\ell_6\rangle \equiv \langle \ell_1 \ell_2 |\chi[\phi, \phi, \phi^\dagger]|\ell_4 \rangle \langle \ell_3| \phi \phi^\dagger\phi^\dagger |\ell_5\ell_6\rangle ,
\end{equation}
where $\chi[\phi, \phi, \phi^\dagger]$ is the solution to the bulk equation of motion for $\chi$,
\begin{equation}
\CL \supset -g_\chi \chi \phi \phi^\dagger \phi^\dagger \rightarrow (\square + m_\chi^2) \chi = -g_\chi \phi \phi^\dagger \phi^\dagger.
\end{equation}
In this equation, $\phi$ is a bulk operator, and when we evaluate the matrix elements $\phi$ contracts with an external state and turns into the corresponding wavefunction.  This differential equation can be solved using standard methods, with the following result
\twoseqn{
&\begin{aligned}
\langle \ell_1 \ell_2 \ell_3 |\chi \phi \phi^\dagger\phi^\dagger |\ell_4\ell_5\ell_6\rangle&\equiv \langle \ell_1 \ell_2 |\chi[\phi, \phi, \phi^\dagger]|\ell_4 \rangle \langle \ell_3| \phi \phi^\dagger\phi^\dagger |\ell_5\ell_6\rangle\\
&=\sum_{k=\frac{\Delta_\chi-\Delta}{2}}^\infty f_{3\to 3}(k)-\sum_{k=\Delta}^\infty f_{3\to 3}(k)\, ,
\end{aligned}
}[]
{
&\begin{aligned}
f_{3\to 3}&=\prod_{i=1}^6 \frac{1}{\tilde{N}_{\Delta, \ell_i}}\frac{\Gamma(1-k)\Gamma(k)\Gamma(1-k-\Delta)\Gamma(1+k-\Delta)\Gamma(k+\Delta)\Gamma\mleft(\frac{3\Delta-\Delta_\chi}{2}\mright)}{\Gamma(\Delta+\ell_4)\Gamma(2\Delta+\ell_{12}^+)\Gamma\mleft(k+1+\frac{\Delta-\Delta_\chi}{2}\mright)\Gamma\mleft(1+k+\frac{\Delta+\Delta_\chi-d}{2}\mright)}
\\
&\times \sum_{m=0}^{\ell_{12}^+}\pi ^{\frac{d}{2}} 4\lambda^2 (-1)^{-m-\ell_4-1}\binom{\ell_{12}^+}{m}\binom{\ell_4}{\ell_{12}^+-m}\frac{\Gamma(1-m+\ell_{12}^+)}{\Gamma(1-k-m-\Delta)}\\
&\times \frac{\Gamma(m+1+\ell_3)\Gamma\mleft(k+2\Delta-\frac{d}{2}\mright)\Gamma\mleft(\frac{3\Delta+\Delta_\chi-d}{2}\mright)}{\Gamma(1+\ell_{12}^+-\ell_4-k-m)\Gamma(1+k+m-\ell_{12}^+-\Delta)\Gamma(k+m+\ell_3+2\Delta)}\, ,
\end{aligned}
}[][]
where we have defined $\ell_{12}^+\equiv \ell_1+\ell_2$.  The pole at $\Delta=\Delta_\chi$ is due to the factor of $\Gamma(k)$ evaluated at $k=0$.  \\
Next, we remove the contribution from the lowest-twist intermediate state.  Recall that the propagator contribution can independently be derived in second-order TIPT by summing over all descendants of $\chi$ as well as all `time-reversed' diagrams:
\begin{align}
\langle \ell_1 \ell_2 \ell_3 |\chi \phi \phi^\dagger\phi^\dagger |\ell_4\ell_5\ell_6\rangle & =\sum_{i=1}^2 \frac{V_{\ell_1,\ell_2;\ell_4,k}V^*_{\ell_4,\ell_5;\ell_3,k}}{E_{\rm in}-E_{\mathrm{out},i}}\\ \nonumber
&=\sum_{n=0}^\infty \frac{V_{\ell_1,\ell_2;\ell_4,k}V^*_{\ell_4,\ell_5;\ell_3,k}}{\Delta-\Delta_\chi-2n-k+\ell_{12}^{+}-\ell_4}-\frac{V_{\ell_1,\ell_2;\ell_4,k}V^*_{\ell_4,\ell_5;\ell_3,k}}{\Delta+\Delta_\chi+2n+k+\ell_{12}^{+}-\ell_4}\, ,
\end{align}
where $k>0$ is the total angular momentum of the $\chi$-mode, whose energy is $\Delta_\chi+2n+k$ and we have used $\ell_1+\ell_2+\ell_3=\ell_4+\ell_5+\ell_6$. We have defined the following interaction matrix elements entering TIPT, along the same lines as (\ref{eq:OPEpoleSubtraction}):
\eqna{
V_{\ell_1, \ell_2;\ell_4,k}&=\lambda \int d^d x \sqrt{g} \lsp  \phi_{\ell_1}(X)\phi_{\ell_2}(X)\phi^*_{\ell_4}(X)\chi_{n, k,K}(X)^*\, ,
}[]
with $\phi_{\ell_i}$ as in~\eqref{eq:phiell} and $\chi_{n, k,K}(X)$ is the same as~\eqref{eq:BulkFieldDecomp} with $\Delta\to \Delta_{\chi}$.
Finally,  notice that two external operators are in the highest-weight (HW) spin representation, one in the lowest-weight (LW) spin representation, and the $\chi$ state in a generic one.\footnote{For instance, in $d=3$ the SO$(3)$ representations are labeled by $J^2$ eigenvalue $\ell$ and $J_z$ one $m$.  So in that case $m_1=\ell_1, \, m_2=\ell_2, \, m_4=-\ell_4$,  so $m_\chi=\ell_1+\ell_2-\ell_4$ but $k= |\ell_1+\ell_2-\ell_4|, \cdots, \ell_1+\ell_2+\ell_4$.}

The point of writing the propagator in TIPT is, as before, that the sum over intermediate states is not supposed to include the term where the intermediate state and external state are the same. In the present context, this means that 
we are not supposed to sum over the $n=0$ $\chi$-state in the HW representation for $k=\ell_1+\ell_2-\ell_4\geq 0$, where the denominator indeed diverges.  Subtracting this term from the calculation with the bulk propagator, 
the correct matrix element is
\eqna{
\lim\limits_{\Delta_\chi\to \Delta} \left( f_{3\to3}\left(k=\frac{\Delta_\chi-\Delta}{2}\right)-\frac{|V|_{\rm pole}}{\Delta-\Delta_\chi}\right) + \sum\limits_{k=1}^\infty f_{3\to 3}(k)-\sum\limits_{k=\Delta}^\infty f_{3\to 3}(k) \, , 
}[]
where we have stressed the fact the only pole comes from the $k=\frac{\Delta_\chi-\Delta}{2}\to 0$ contribution in $f_{3\to 3}(k)$ and that the subtraction is necessary only for $ \ell_1+\ell_2-\ell_4\geq 0$ 
\eqna{
|V|_{\text{pole}}\equiv \begin{cases}  V_{\ell_1,\ell_2;\ell_4, \ell_1+\ell_2-\ell_4}V^*_{\ell_5,\ell_6;\ell_3, \ell_1+\ell_2-\ell_4}   &\,\,  \ell_{12}^+-\ell_4\geq 0\, \\ 0 &  \,\,\text{otherwise }\,   \end{cases}.
}[]
When evaluated between $[\Phi, \Phi^2]_J$ primaries the subtracted pole has a  compact formula: 
\eqna{
\frac{|V|_{\rm pole}}{\Delta-\Delta_\chi} &=\frac{(-1)^J  \left(2 (-1)^J \Gamma (2 \Delta ) \Gamma (J+\Delta )+\Gamma (\Delta ) \Gamma (J+2 \Delta )\right)}{2\pi^d (\Delta-\Delta_\chi)}\\
&\quad\, \times \frac{\Gamma(\Delta)\Gamma(J+\Delta)\Gamma(\Delta_\chi)\Gamma\mleft( \frac{3\Delta+\Delta_\chi-d}{2}\mright)^2}{\Gamma(2\Delta)\Gamma\mleft(\Delta-\frac{d-2}{2}\mright)^3\Gamma\mleft(\Delta_\chi-\frac{d-2}{2}\mright)\Gamma\mleft(J+\frac{3\Delta+\Delta_\chi}{2}\mright)^2}\,.
}[d4del23to3]
For the particular case $\Delta=2$ in $d=4$
\eqna{
{\gamma}_{[\Phi, \Phi^2]_J}^{(\three)}
=\frac{\lambda^2}{16}\frac{12+(-1)^J(J+2)(J+3)}{24\pi^4 (J+2)^3(J+3)^3}\Bigg\lbrace (J(J+3)+1)-\frac{4 (J+2) (J+3)}{(J+1) (J+4)}\Bigg\rbrace\, ,
}[]
where the first contribution comes from canceling the pole and the rest from the two sums ($k=1, \cdots, \infty$ and $k=\Delta, \cdots,  \infty$).  Notice that the leading large $J$ behavior always come from the contribution removing the pole and this can be derived for every $\Delta$ and $d$
\eqna{
{\gamma}_{[\Phi, \Phi^2]_J}^{(\three), k=0 {\textrm{pole} \atop \textrm{removed}} }
&=\frac{(-1)^J\lambda^2\Gamma (\Delta )^2  \Gamma (J+\Delta )\Gamma \mleft(2 \Delta -\frac{\mathit{d}}{2}\mright)^2}{64 \pi^d\Gamma (2 \Delta ) \Gamma (J+2 \Delta )^2 \Gamma \mleft(\Delta -\frac{\mathit{d}-2}{2}\mright)^4}\left( H_{\Delta -1}+H_{J+\Delta -1}-H_{J+2 \Delta -1}\right)\\
&\quad\, \times \left(2 (-1)^J \Gamma (2 \Delta ) \Gamma (J+\Delta )+\Gamma (\Delta ) \Gamma (J+2 \Delta ) \right)\, ,\\
&  \stackrel{J\gg 1}{\approx}  {\lambda^2} \frac{(-1)^J}{J^\Delta} \frac{\Gamma(\Delta)^3 \Gamma\mleft(2\Delta-\frac{d}{2}\mright)^2H_{\Delta-1}}{64\pi^d\Gamma(2\Delta) \Gamma\mleft(\Delta-\frac{d-2}{2}\mright)^4}\,.  
}[resfrompoleSpin0]
Note that this diagram only affects the anomalous dimension of $[\Phi, \Phi^2]_J$, keeping the other possible triple-particle operators untouched.\\
Putting everything together, we get at large $J$
\eqna{
\gamma_{[\Phi, \Phi^2]_J}^{(\three)}&\approx \frac{(-1)^J}{J^\Delta}\left(\gamma_{\Phi^2}^{(\contact)}\right)^2 \frac{\Gamma(2\Delta)H_{\Delta-1}}{\Gamma(\Delta)}\left(1+\frac{(-1)^J}{J^\Delta} \frac{2\Gamma(2\Delta)}{\Gamma(\Delta)}\right)\\
&\quad\, -\frac{(-1)^J}{J^{\Delta+2}}\left(\gamma_{\Phi^2}^{(\contact)}\right)^2\frac{\Delta  (\mathit{d}-4 \Delta ) \Gamma (2 \Delta )}{(\mathit{d}-2 (\Delta +1)) \Gamma (\Delta -1)}\left(1+\frac{(-1)^J}{J^\Delta} \frac{2\Gamma(2\Delta)}{\Gamma(\Delta)}\right)\, ,
}[largeSpin3To3]
where the first line corresponds to the contribution from $k=0$ and the pole and the second line is from the rest of the $k$ sum.\footnote{
Shamefully, we derived (\ref{largeSpin3To3}) by assuming that $\Delta$ is an integer, and then assuming the result is true for all real $\Delta$.}

\section{$[\Phi, \Phi^2]_J$ anomalous dimension  } \label{sec:Results}
\subsection{Results at Large Spin}
Having determined all diagrams up to $O(\lambda^2)$, we now combine them to determine the large-spin behavior of the anomalous dimension of $[\Phi, \Phi^2]_J$.
\eqna{
\gamma_{[\Phi, \Phi^2]_J}&\approx \left(\lambda \gamma_{\Phi^2}^{(\contact)}+\lambda^2 \gamma_{\Phi^2}^{(2)}\right) \left( 1+\frac{(-1)^J}{J^\Delta } \frac{2\Gamma(2\Delta)}{\Gamma(\Delta)}\right)+\lambda^2 \left(\gamma_{\Phi^2}^{(\contact)}\right)^2 \frac{\Gamma(2\Delta)}{\Gamma(\Delta)}\Bigg\lbrace \frac{(-1)^J}{J^\Delta} H_{\Delta-1}\\
&\quad\, +\frac{1}{J^{2\Delta}}\frac{2\Gamma(2\Delta)}{\Gamma(\Delta)} \left( H_{\Delta-1}-4(\log J+\gamma_E)\right)\Bigg\rbrace\, , 
}[resQ3Full]
where $\gamma_{\Phi^2}^{(2)}=\gamma_{\Phi^2}^{(\bubble)}+$(counterterms/$s$-channel bubbles). We can compare this result against the expectation from the large spin expansion in~\eqref{largeSpinDT}.  To extract the leading $1/J$ corrections,  we consider the exchange $\Phi$ itself in the $t$-channel OPE and $[\Phi, \Phi^*]_0$ in the $s$-channel one,  neglecting contributions from double-twist neutral operators with $\ell \geq 2 $: 
\begin{align} \nonumber
\tau_{[\Phi, \Phi^2]_J}& \approx \Delta+\Delta_{\Phi^2}+\lambda \gamma_{\Phi^2}^{(\contact)} \frac{(-1)^J}{J^\Delta}\frac{2\Gamma(2\Delta)}{\Gamma(\Delta)}+\lambda^2 \Bigg\lbrace \frac{(-1)^J}{J^\Delta}\frac{2\Gamma(2\Delta)}{\Gamma(\Delta)}\Bigg( \frac{ H_{\Delta-1}(\gamma_{\Phi^2}^{(\contact)})^2}{2}+\gamma_{\Phi^2}^{(2)}\Bigg)\\
&\quad \,+ \frac{2}{J^{2\Delta}}\frac{\Gamma(2\Delta)^2}{\Gamma(\Delta)^2}(\gamma_{\Phi^2}^{(\contact)})^2 \left(5H_{\Delta-1}-4\log J-4\gamma_E \right)\Bigg\rbrace+ O\left(\frac{\lambda^2 \gamma_{[\Phi, \Phi^*]_{\ell \geq 2}}^{(2)}} {J^{2\Delta}}\right) \,,
\end{align}
where we have used
\eqna{
\gamma_{[\Phi, \Phi^*]_0}^{(\contact)}&=2\gamma_{\Phi^2}^{(\contact)}\, , \qquad a_t=2+\lambda a_t^{(\contact)}\, ,  \qquad a_s =2+3\lambda a_t^{(\contact)}\, , \\
a_t^{(\contact)}&=2\gamma_{\Phi^2}^{(\contact)}\left( 2H_{\Delta-1}-H_{2\Delta-1}\right)\,, 
}[]
and for the $O(\lambda^2)$ anomalous dimensions $\gamma^{(2)}_{[\Phi, \Phi^*]_0}=2\gamma^{(2)}_{\Phi^2}$.  
It is more suggestive to write the anomalous dimension $\gamma_{[\Phi, \Phi^2]_J}\equiv \tau_{[\Phi, \Phi^2]_J}-3\Delta$:
\eqna{
\gamma_{[\Phi, \Phi^2]_J} &\approx \gamma_{\Phi^2}\left(1+\frac{(-1)^J}{J^\Delta} \frac{2\Gamma(2\Delta)}{\Gamma(\Delta)}\right)+\lambda^2 (\gamma_{\Phi^2}^{(\contact)})^2\frac{\Gamma(2\Delta)}{\Gamma(\Delta)}\Bigg\lbrace\frac{(-1)^JH_{\Delta-1}}{J^\Delta}\\
&\quad\, -\frac{8}{J^{2\Delta}}\frac{\Gamma(2\Delta)}{\Gamma(\Delta)}\left(\log J+\gamma_E-\frac{5}{4} H_{\Delta-1}\right)\Bigg\rbrace+O\left(\frac{\lambda^2 \gamma_{[\Phi, \Phi^*]_{\ell \geq 2}}^{(2)}} {J^{2\Delta}}\right)\, ,
}[largeSpinExpectation]
where we have denoted with $\gamma_{\Phi^2}= \left(\lambda \gamma_{\Phi^2}^{(\contact)}+\lambda^2\gamma_{\Phi^2}^{(2)} \right)$. At order $O(\lambda^2)$,  this result exactly reproduces the bubble-diagram contribution in~\eqref{bubbleTT}  to the  $\frac{1}{J^\Delta}$ and $\frac{\log J}{J^{2\Delta}}$ terms and the leading part of~\eqref{largeSpin3To3}.  Interestingly,  the $\log J$ term originates from different mechanisms in the two approaches: in the Hamiltonian computation it arises from averaging over $\gamma_{[\Phi, [\Phi, \Phi]_{\ell}]_{J-\ell}}$, while in the large-spin formula it is tied to the anomalous dimension that  $[\Phi, \Phi^*]_0$ gets at order $\lambda$.  Clarifying this relation would be an interesting direction for future work.

\subsection{Prediction for $d=4$ and $ \Delta=2$}
The calculation becomes technically simpler for the special choice of parameters $\Delta = 2$ and $d = 4$, for which we have obtained the anomalous dimensions analytically up to $O(1/J^4)$
\eqna{
\gamma_{[\Phi, \Phi^2]_J}^{(\rm full)}&\approx\left( \frac{\lambda}{48\pi^2}+\lambda^2 \gamma_{\Phi^2}^{(2)}\right)\left[ 1+12 \frac{(-1)^J}{J^2}\left( 1-\frac{5}{J}+\frac{19}{J^4}\right)\right]\\
&\quad\, +\frac{\lambda^2}{(4\pi^2)^2} \left [ \frac{(-1)^J}{24J^2}\left( 1-\frac{7}{J}+\frac{30}{J^2}\right)-\frac{2}{J^4}\left( \log J+\gamma_E+\frac{87 \pi ^2}{1120} -\frac{8}{3}\right) \right]\, .
}[fullSpecial]
We would like to compare this expression with what one expects from combining the large-spin analysis in~\eqref{largeSpinDT} and  the Lorentzian inversion formula~\cite{Caron-Huot:2017vep}.  A detailed derivation is given in appendices~\ref{app:inverison1} and~\ref{app:inversion2}, here we summarize the main results. 

To fix the $[\Phi, \Phi^2]_J$ anomalous dimensions up to $J^{-4}$, one needs to consider, apart from the identity,  the exchange of $\Phi$ in the $t$-channel OPE and, in the $s$-channel one, the exchange of $S\equiv [\Phi, \Phi^*]_0$ as well as the remaining twist-2 charged operators $J_{\ell}\equiv [\Phi, \Phi^*]_{\ell}$ with $\ell \geq 2$:
\eqna{
\gamma_{[\Phi, \Phi^2]_J}^{(\rm LIF)}&=\gamma_{\Phi^2}+\gamma_{[\Phi, \Phi^2]_J}^{(\Phi), \text{LIF}}+\gamma_{[\Phi, \Phi^2]_J}^{(S), \text{LIF}}+\gamma_{[\Phi, \Phi^2]_J}^{(J_\ell), \text{LIF}}+ \dots \, ,
}[eqn:d4del2invA]
where $\dots$ indicates contributions beyond $1/J^4$.  The first two terms, up to $O(1/J^4)$, are
\eqna{
\gamma_{[\Phi, \Phi^2]_J}^{(\Phi), \text{LIF}}&=\left(\frac{\lambda}{4\pi^2}+12\lambda^2 \gamma_{\Phi^2}^{(2)}\right)\frac{(-1)^J}{ J^2}\left({1}-\frac{5}{J}+\frac{19-12(-1)^J}{J^2}\right)\\
&\quad\, +\frac{\lambda^2}{(4\pi^2)^2} \frac{(-1)^J}{J^2}\left[\frac{1}{24}\left( 1-\frac{7}{J}+\frac{30}{J^2}\right)-\frac{(-1)^J}{J^2}(3-\log 2) \right]+O(J^{-5})\, , \\
\gamma_{[\Phi, \Phi^2]_J}^{(S), \text{LIF}}&=\frac{3\lambda}{\pi^2}\frac{1}{J^2}+  \frac{144 \lambda^2 \gamma_{\Phi^2}^{(2)}}{J^4}-\frac{2\lambda^2}{(4\pi^2)^2 J^4}\left( \log J+\gamma_E-2\right)+O(J^{-5})\,  ,
}[eqn:d4del2inv]
where we have used 
\begin{align}
\begin{aligned}
\gamma_{\Phi^2}&=\frac{\lambda}{48\pi^2}+\lambda^2\gamma_{\Phi^2}^{(2)}\, ,  &&\quad  \gamma_{S}=\frac{\lambda}{24\pi^2}\, , \\
a_t&=c_{\Phi\Phi[\Phi^*,\Phi^*]_0}^2=2+\frac{\lambda}{144\pi^2} \, , 
&&\quad a_s=c_{\Phi\Phi^*[\Phi,\Phi^*]_0}c_{\Phi^2(\Phi^*)^2[\Phi,\Phi^*]_0}=2+\frac{\lambda}{48\pi^2}\, .
\end{aligned}\label{eq:datad2D4}
\end{align}
These two contributions reproduce the result in~\eqref{fullSpecial} exactly at order $\lambda$, but at $O(\lambda^2)$ we start to see that  the twist-two operators $J_\ell$ need to be included in the inversion formula at $O(1/J^4)$:
\eqna{
\gamma_{[\Phi, \Phi^2]_J}^{(\rm full)}- \gamma_{[\Phi, \Phi^2]_J}^{(\rm LIF)}=\frac{\lambda^2}{(4\pi^2)^2}\frac{1}{J^4}\left( \frac{17}{6}-\frac{87 \pi ^2}{560}-\frac{\log 2}{2}\right)-\gamma_{[\Phi, \Phi^2]_J}^{(J_\ell), \text{LIF}}+O(J^{-5})\, .
}[eq:SpecialCaseComparison]
 Unfortunately, we were not able to independently obtain a closed form analytic expression for the contribution $\gamma_{[\Phi, \Phi^2]_J}^{(J_\ell), \text{LIF}}$ from double twists to the inversion formula at $O(\lambda^2/J^4)$ by computing the double-discontinuity of the four-point correlator; in appendix \ref{app:inversion2} we describe a numeric approximation of $\gamma_{[\Phi, \Phi^2]_J}^{(J_\ell), \text{LIF}}$ along such lines.  We emphasize that we think our calculation \eqref{fullSpecial}, obtained using the  bulk Hamiltonian, is correct and should ultimately be reproduced by the inversion formula.  Accordingly,  the expression~\eqref{eq:SpecialCaseComparison} should be viewed as a prediction for the contributions from $\gamma_{[\Phi, \Phi^2]_J}^{(J_\ell), \text{LIF}}$. 
The precise mechanism by which the Hamiltonian formalism resums the effect of the twist-accumulation point remains somewhat obscure, and we plan to investigate this aspect further in the future.
\section{Future Directions}\label{sec:Generalization}

In this section we will make a few comments about how to generalize the procedure we have used here for the leading-twist effective Hamiltonian to more realistic models.  In particular, we will more often be presented with a set of OPE data and CFT correlators, rather than a bulk Lagrangian.  And even in the case where a bulk Lagrangian does exist, it may be more efficient to construct the effective Hamiltonian starting directly with CFT data, which is physical and free of gauge or field reparameterization redundancies.  

The starting point of the construction is to match the $t$-channel exchanges of low twist operators.  For two-particle states when the leading $t$-channel exchanges are isolated in twist, this matching is  well-understood, and guarantees that the effective Hamiltonian will correctly reproduce the anomalous dimensions of two-particle states at large spin $J$. It often works surprisingly well at small $J$ as well, but even if it did not it would always be possible to add in a finite set of bulk quartic  terms in order to match the CFT data at all values of $J$ up until some value $J_*$ chosen around where the large spin approximation eventually becomes accurate. In our simple holographic model with a twist gap set to $\Lambda_\tau = 2\Delta$ and $J_*=0$, this would correspond to including just the $t$-channel exchange of the neutral $\Phi \Phi^*$ operator, which is the lowest-dimension mode inside the bubble diagram, and then fixing the coefficient of the contact term from the dimension of the charged scalar operator $\Phi^2$. Proceeding to states with three particles, we can think of them as double-trace states made from $\Phi$ and a two-particle state $[\Phi, \Phi]_\ell$.  By considering the four-point function $\< \Phi \Phi^* [\Phi, \Phi]_\ell [\Phi^* , \Phi^*]_\ell\> $, one can again use results about anomalous dimensions of double-twist states from $t$-channel exchanges.  However, some of these contributions will already be present due to the two-body terms in the Hamiltonian introduced by matching the two-particle states, and others, such as our ``three-to-three'' diagram, can be thought of as being generated by our two-body terms at second order.  Since the CFT data involved in the $t$-channel exchanges in $\< \Phi \Phi^* [\Phi, \Phi]_\ell [\Phi^* , \Phi^*]_\ell\> $ depends on OPE coefficients involving the two-particle operator $ [\Phi, \Phi]_\ell$, the more accurately our two-body and three-body Hamiltonian terms capture these OPE coefficients, the more accurately it will predict the triple-twist anomalous dimensions $[\Phi, [\Phi, \Phi]_\ell]_{J-\ell}$.  For this reason, it will typically not be sufficient to choose quartic terms for the two-particle states in order to match the anomalous dimensions at $J\le J_*$. Rather, it will also be necessary to choose additional quartic terms (two for each $J\le J_*$) to fix the OPE coefficients as well; we have checked explicitly that it is possible to add a linear combination of $\partial_\mu \phi \partial^\mu \phi^* \phi \phi^* $ and $|\phi|^4$ whose only effect on the lowest twist $Q=2$ states is to modify the OPE coefficient  $c_{\Phi \Phi \Phi^2}$.  

 Depending on the level of accuracy desired at large $J$, it may also be necessary to add more terms of the form, say,  $\sim s |\phi|^4$, in order to fix the OPE coefficient $c_{\Phi^2 \Phi^{*2} s}$ of two double-twists $\Phi^2$, $\Phi^{*2}$ and a low-spin exchanged operator $S$. Consider for instance the effect of a bulk cubic coupling $\sim s^3$ for $s$, with coefficient fixed by the $c_{SSS}$ OPE coefficient.  Because of the cubic interaction $\sim s |\phi|^2$, there is a bulk Witten diagram for the six-point function $\< \Phi \Phi \Phi \Phi^* \Phi^* \Phi^*\>$, with three $\phi \phi^*$ pairs each producing a bulk $s$ that then connect to each other at a bulk $s^3$ vertex, in a `snowflake' formation.\footnote{We thank Gabriel Cuomo for emphasizing this diagram.}  For a typical large $J$ configuration where all the $\phi$s have spin $\approx J/3$, the bulk $s^3$ vertex will be near the center of AdS, and each of the $s$ bulk-to-bulk propagators will decay like $J^{-\Delta_S/2}$ for a combined scaling $J^{-3\Delta_S/2}$ of the total contribution to the energy of the state.  In the EFT approach, we approximate non-local contributions with local contributions if they decay faster than $1/J^{\Lambda_\tau}$ for our choice of the twist cut-off $\Lambda_\tau$. So, if $\Lambda_\tau < 3 \Delta_S/2$ then we would replace the  snowflake diagram with a diagram containing the $s |\phi|^2$ vertex and a $s |\phi|^4$ vertex (plus additional irrelevant couplings depending on the desired degree of accuracy), connected by a single bulk $s$ propagator.

A potential obstacle is that correlators will generally include the $t$-channel exchange of twist-families with accumulation points in twist, and so the multi-twist families themselves become the source of large $J$ corrections \cite{Fitzpatrick:2015qma,Simmons-Duffin:2016wlq}.  Naively, this means that at some sufficiently large value of $\Lambda_\tau$ (usually not very large, and at most $2 \Delta$), including exchanges of all states with twist below $\Lambda_\tau$ would require an infinite set of bulk fields.  On the other hand, though, these multi-twist exchanges at large spin are approximately just weakly-coupled multi-particle states in AdS, which are already exchanged in the $t$-channel through the existing bulk diagrams, and so one might hope that their contributions can be modeled through the same kind of interaction terms in the Hamiltonian that we have already been using. For instance, in our holographic model, most of the double- and triple-twist $t$-channel exchanges in the correlators $\< \Phi \Phi \Phi^* \Phi^*\>$ and $\< \Phi \Phi^* [\Phi, \Phi]_\ell [\Phi^* , \Phi^*]_\ell\> $ did not require us to introduce any nonlocal bulk potential at $O(\lambda^2)$.

A more technical remark is that describing  more realistic models may require more complicated interactions, leading to operator dimensions that cannot be obtained analytically. In these situations, it may be necessary to study very large values of $J$ numerically in order to extract the asymptotic large-spin behavior. For this purpose, it could be useful to employ the method for constructing states used in~\cite{Kravchuk:2024wmv,Harris:2024nmr}.

Beyond these prescriptive comments, it is still unclear how well or widely such an effective Hamiltonian should work.  The fact that the $O(\lambda)$ effects from the quartic term $|\phi|^4$ are connected on general grounds to $O(\lambda^2)$ effects that must be included through the three-to-three exchange diagram would seem to imply that at least the anomalous dimensions computed through bulk interactions should be small.  By contrast, if the dimension of, say, $\Phi^2$ is very far from $2\Delta$, then it is not clear that it is even very meaningful to call it `$\Phi^2$', and at that point it may be necessary to simply introduce it as a new fundamental bulk field.  An extreme version of this is the $O(2)$ model operator $\Phi^2 \Phi^*$.  Due to the equations of motion, it is actually a descendant of $\Phi$ rather than a primary, so in a sense it is completely removed from the theory by strong coupling effects.  However, the free holographic effective theory does contain the primary state $\Phi^2 \Phi^*$, which must therefore be removed.  One possible way to do this is to introduce a bulk ghost field, which simply cancels the unphysical $\Phi^2 \Phi^*$ degree of freedom \cite{CompanionPaper2}. Nevertheless, the conformal bootstrap shows \cite{Fitzpatrick:2012yx,Komargodski:2012ek} that every CFT in $d >2$ contains an infinite amount of data at large $J$ that is perturbatively close to that of a GFF, and we hope that the methods we have described in this paper can be used to model it quantitatively.   
\vspace{0.75cm}

\headline{\large \textbf{Summary of Key Technical Results}} 
\vspace{0.5cm}
 \noindent 
 $Q=3$ normalized state built from $\Phi^2$
\eqna{
[\Phi, \Phi^2]_J&=\frac{1}{\sqrt{\CN}}\sum_{m=0}^J\sum_{i=0}^{J-m}\sqrt{\frac{2^J\Gamma (i+\Delta ) \Gamma (J-m-i+\Delta )}{i! m! \Gamma (\Delta )^7 \Gamma (m+\Delta ) (J-m-i)!}} \frac{(-1)^m \ket{i, m,J-m-i}}{\Gamma(2\Delta+J-m)}\, , \\
\CN&=\frac{2^{J+1} \left(2 (-1)^J \Gamma (2 \Delta ) \Gamma (J+\Delta )+\Gamma (\Delta ) \Gamma (J+2 \Delta
   )\right) \Gamma (2 J+3 \Delta -1)}{\Gamma (\Delta )^6 \Gamma (2 \Delta ) \Gamma (J+1) \Gamma (J+\Delta
   ) \Gamma (J+2 \Delta )^2 \Gamma (J+3 \Delta -1)}\,.
}
Overlaps between different $Q=3$ states at large $J$
\eqna{
\langle [\Phi,[\Phi, \Phi]_{\ell^\prime}]_{J-\ell^\prime}| [\Phi,[\Phi, \Phi]_\ell]_{J-\ell}\rangle &=\begin{cases}
1 &\quad \ell^\prime =\ell\, ,\\
 \frac{2}{J^\Delta}\sqrt{\frac{(2\Delta+2\ell-1)(2\Delta+2\ell^\prime-1)\Gamma(2\Delta+\ell-1)\Gamma(2\Delta+\ell^\prime-1)}{\Gamma(\ell+1)\Gamma(\ell^\prime+1)\Gamma(\Delta)^2}} & \quad \ell^\prime \neq \ell\, .
\end{cases}
}
$Q=3$ anomalous dimension due to a $|\phi|^4$-like contact term
\eqna{
\gamma_{[\Phi, \Phi^2]_J}^{(\contact)}&=\frac{\lambda}{4}\frac{\pi^{-d}\Delta! \lsp J! \Gamma (J+\Delta )^2 \Gamma (J+3 \Delta -1)}{ (2\Delta)!\Gamma\mleft(\Delta-\frac{d-2}{2}\mright)\Gamma (2 J+3 \Delta -1)}\Bigg( 2 (-1)^J \Gamma (2 \Delta ) \\
&\quad \,  +\Gamma (\Delta )\frac{ \Gamma (J+2 \Delta )}{\Gamma(J+\Delta)} \Bigg) \sum_{\ell=0}^J \frac{f(\ell)}{\Gamma (\ell+1)^2 \Gamma (J-\ell+1) \Gamma (J-\ell+\Delta )}\, .
}
$Q=2$ anomalous dimension due to a $\Delta_\chi$-scalar exchange
\begin{align*} \nonumber 
 \gamma^{(\exch)}_{[\Phi, \Phi]_J} (\Delta_\chi) &=\sum_{k=\frac{\Delta_\chi}{2}}^\infty \gamma^{(\Delta_\chi)}_{ k}-\sum_{k=\Delta}^\infty \gamma^{(\Delta_\chi)}_{ k} \, , \\
\gamma^{(\Delta_\chi)}_{ k}&= \frac{-\pi ^{-\frac{\mathit{d}}{2}} (k-\Delta)! \Gamma (k)^2 \Gamma \mleft(\Delta -\frac{\Delta_\chi
   }{2}\mright)  \Gamma \mleft(\Delta +\frac{\Delta_\chi-d
   }{2}\mright) \Gamma \mleft(k-\frac{\mathit{d}}{2}+\Delta \mright)}{8  (k-\Delta-J )!  \Gamma
   \mleft(\Delta -\frac{\mathit{d}-2}{2}\mright)^2 \Gamma \mleft(k-\frac{\Delta_\chi }{2}+1\mright) \Gamma (J+k+\Delta ) \Gamma \mleft(k+\frac{\Delta_\chi
  -d+2 }{2}\mright)}\, . 
  \end{align*}
$Q=2$ Hamiltonian and anomalous dimension due to a $[\phi^4]_{\ell}$ interaction
\eqna{
H_2 &\supset \gamma_{[ \Phi, \Phi]_\ell}^{(\contactell{\ell})} |[ \Phi, \Phi]_\ell \> \< [ \Phi, \Phi]_\ell| \, , \\
\gamma_{[\Phi, \Phi]_\ell}^{(\contactell{\ell})}&=\frac{\lambda_{\ell}}{4}\frac{ \ell ! \Gamma (\Delta+\ell ) \Gamma (\Delta +\ell  +1) \Gamma
   \mleft(2 \Delta+\ell-\frac{\mathit{d}}{2} \mright)}{\pi^\frac{d}{2} \Gamma (2 \Delta +2 \ell +1)\Gamma \mleft(\Delta-\frac{\mathit{d}-2}{2}
   \mright)^2}\, .
}
$Q=3$ Hamiltonian and anomalous dimension due to a $[\phi^4]_{\ell}$ interaction
\eqna{
H_3 &\supset \gamma_{[\Phi,[\Phi, \Phi]_\ell]_{J-\ell}}^{(\contactell{\ell})} |[ \Phi,[ \Phi, \Phi]_\ell ]_{J-\ell} \> \< [\Phi,  [ \Phi, \Phi]_\ell ]_{J-\ell}| \, ,\\
\gamma_{[\Phi,[\Phi, \Phi]_\ell]_{J-\ell}}^{(\contactell{\ell})}&\stackrel{J\gg 1}{\approx} \gamma_{[\Phi, \Phi]_\ell}^{(\contactell{\ell})}\left( 1+\frac{(-1)^J}{J^\Delta}\frac{2 (2 \Delta +2 \ell -1) \Gamma (2 \Delta+\ell -1)}{\Gamma(\ell+1) \Gamma (\Delta )}\right)\, .
}
$Q=3$ anomalous dimension due to bubble diagrams
\eqna{
\gamma_{[\Phi, \Phi^2]_J}^{(\bubble)}\stackrel{J\gg 1}{\approx}  \lambda^2\gamma_{\Phi^2}^{(\bubble)}\left( 1+\frac{(-1)^J}{J^\Delta}\frac{2\Gamma(2\Delta)}{\Gamma(\Delta)}\right)-{\lambda^2(\gamma_{\Phi^2}^{(\contact)})^2}\frac{8\Gamma(2\Delta)^2}{\Gamma(\Delta)^2}\frac{\log J+\gamma_E}{J^{2\Delta}}\, .
}
$Q=3$ anomalous dimension due to a three-body interaction at large $J$
\eqna{
\gamma_{[\Phi, \Phi^2]_J}^{(\three)}&\approx \frac{(-1)^J}{J^\Delta}\left(\gamma_{\Phi^2}^{(\contact)}\right)^2 \frac{\Gamma(2\Delta)H_{\Delta-1}}{\Gamma(\Delta)}\left(1+\frac{(-1)^J}{J^\Delta} \frac{2\Gamma(2\Delta)}{\Gamma(\Delta)}\right)\\
&\quad\, -\frac{(-1)^J}{J^{\Delta+2}}\left(\gamma_{\Phi^2}^{(\contact)}\right)^2\frac{\Delta  (\mathit{d}-4 \Delta ) \Gamma (2 \Delta )}{(\mathit{d}-2 (\Delta +1)) \Gamma (\Delta -1)}\left(1+\frac{(-1)^J}{J^\Delta} \frac{2\Gamma(2\Delta)}{\Gamma(\Delta)}\right)\, .
}
\hrule

\acknowledgments{We are grateful to Agnese Bissi, Gabriel Cuomo,  Ami Katz,  Petr Kravchuk, and Jeremy Mann  for useful discussions and we thank Agnese Bissi for comments on a draft. GF,  ALF, and WL  are supported by the US Department of Energy Office of Science under Award Number DE-SC0015845, and GF was partially supported by the Simons Collaboration on the Non-perturbative Bootstrap. }

\newpage 
\appendix

\section{$Q=2$ Scalar Matrix Element}
\label{app:scalarmat}
In this appendix we provide details about the derivation of the $Q=2$ monomial matrix element $V_{\text{scalar}}(\ell_1,\ell_2,\ell_3,\ell_4)$ corresponding to the exchange of a bulk field $\chi(x)$
\eqna{
H^{(\exch)}&=g_{\chi} \int d^{d}x \sqrt{g} \chi(x) \phi(x) \phi^*(x)\, , \\
 V_{\text{scalar}}(\ell_1,\ell_2,\ell_3,\ell_4)&=\langle\ell_1,\ell_2|H^{(\exch)}|\ell_3,\ell_4\rangle\, .
}[]
Following the construction in~\cite{Fitzpatrick:2011hh} and summarized in~\cite{Fardelli:2024heb}, this matrix element can be evaluated as  
\begin{equation}\label{eqn:scalarV}
V_{\text{scalar}}(\ell_1,\ell_2,\ell_3,\ell_4)=g_\chi \int {d}^dx\sqrt{g}\phi_{\ell_1}^*(x)\phi_{\ell_3}(x)\langle\ell_2|\chi(x)|\ell_4\rangle+(\ell_1\leftrightarrow\ell_2,\ell_3\leftrightarrow\ell_4)\, ,
\end{equation}
where $\langle\ell|\chi(x)|k\rangle$ depends on the bulk-to-bulk propagator  as
\begin{equation}\label{eqn:3pteps}
\langle\ell|\chi(x)|k\rangle=\int{d}^dx\sqrt{g}\phi^*_{\ell}(x)G_{\chi}(x,y)\phi_k(y)\, .
\end{equation}
Direct evaluation of this integral is quite challenging. Instead, we can proceed by first expressing the scalar background in terms of the boundary CFT operators in embedding space, and then selecting the desired external states. Concretely,
\eqna{
\langle \ell| \chi (x)| k\rangle \sim  \left( Z_1 \cdot \frac{\partial}{\partial P_1}\right)^\ell \left( Z_3 \cdot \frac{\partial}{\partial P_3}\right)^k \langle \Phi^\dagger (P_1) \chi(X) \Phi(P_3)\rangle \, , 
}[extractMatrixEl]
where $P_i$ are $(d+2)$-dimensional embedding space coordinates for the external boundary points,  $X$ is a bulk point and $Z_i$ are null polarization vectors. The advantage of this approach is that it reduces the problem to computing derivatives of a three-point function, whose form is fixed by conformal invariance to be
\eqna{
 \langle \Phi^\dagger (P_1) \chi(X) \Phi(P_3)\rangle = \frac{f_{\text{scalar}}(t)}{(-2P_1\cdot P_3)^{\Delta}}\, , \qquad t=\frac{-2P_1\cdot P_3}{(-2X\cdot P_1)(-2X\cdot P_3)}\, .
}[eqn:3ptbulk]
Imposing the bulk equation of motion for the field $\chi$   then translates into an ordinary differential equation for $f_{\text{scalar}}(t)$~\cite{DHoker:1999mqo}
\begin{equation}\label{eqn:scalarODE}
4t^2(t-1)f^{''}_{\text{scalar}}(t)+4t\Big(t+\frac{d}{2}-1\Big)f^{'}_{\text{scalar}}(t)+\Delta_\chi(\Delta_\chi-d)f_{\text{scalar}}(t)=g_\chi t^{\Delta}\, .
\end{equation}
Requiring $f_{\rm scalar}$ to be finite at $t\to 0$ and to vanish at $t\to \frac{\pi}{2}$, this equation has a unique solution
\begin{align}\label{eqn:ftseries}
\begin{aligned}
f_{\text{scalar}}(t)&=\sum_{k=0}^\infty \alpha \left(\frac{\Delta_\chi}{2}+k\right)t^{\frac{\Delta_\chi}{2}+k}-\sum_{k=0}^\infty \alpha (\Delta+k)t^{\Delta+k}\, ,\\
\alpha(k)&=\frac{\Gamma(k)^2\Gamma\mleft(\Delta-\frac{\Delta_\chi-d}{2}\mright)\Gamma\mleft(\Delta-\frac{\Delta_\chi}{2}\mright)}{4\Gamma(\Delta)^2\Gamma\mleft(1+k-\frac{\Delta_\chi}{2}\mright)\Gamma\mleft(k+\frac{\Delta_\chi-d+2}{2}\mright)}\,.
\end{aligned}
\end{align}
To effectively extract the lowest-twist matrix element in~\eqref{extractMatrixEl},  it is convenient to use the following coordinates  
\begin{align}
\begin{aligned}
X^M&=\Big(\frac{e^{-T}}{\cos\rho},\frac{e^{T}}{\cos\rho},\tan\rho\lsp \vec{\Omega}\Big)\, ,\\
P_1^M&=\Big(0,1,\vec{0}_{d-3},\frac{z}{2},\frac{z}{2i}\Big)\, ,\\
P_3^M&=\Big(1,0,\vec{0}_{d-3},\frac{\bar{z}}{2},-\frac{\bar{z}}{2i}\Big)\, , 
\end{aligned}
\end{align}
such that scalar products reduce to 
\begin{align}
\begin{aligned}
X\cdot P_1&=\frac{1}{2\cos\rho}\left[z\left (\prod_{i=1}^{d-2}\sin\theta_i\right)e^{-i\varphi}\sin\rho-e^{-T}\right]\,  , \\
X\cdot P_3&=\frac{1}{2\cos\rho}\left[\bar{z}\left(\prod_{i=1}^{d-2}\sin\theta_i\right)e^{i\varphi}\sin\rho-e^{T}\right]\, , \\
P_1\cdot P_3&=\frac{1}{2}(z\bar{z}-1)\, , 
\end{aligned}
\end{align}
and computing derivatives simplify to 
\begin{equation}\label{eqn:scalar3ptlksol}
\langle\ell |\chi(x) |k\rangle=\frac{-g_{\chi}\pi^{-\frac{d}{2}}}{\Gamma\mleft(\Delta-\frac{d-2}{2}\mright)}\sqrt{\frac{\Gamma(\Delta+\ell)\Gamma(\Delta+k)}{\Gamma(\ell+1)\Gamma(k+1)}}\frac{1}{(\Delta)_k(\Delta)_\ell}\partial_{\bar{z}}^{k}\partial_{z}^{\ell}\left[\frac{f_{\text{scalar}}(t)}{(-2 P_1\cdot P_3)^{\Delta}}\right]\Bigg|_{z=\bar{z}=0}\, .
\end{equation}
Integrating this expression against the $\phi_{\ell_1}^* \phi_{\ell_2}$ current as in~\eqref{eqn:scalarV} we get 
\begin{align}
\begin{aligned}
V_{\text{scalar}}(\ell_1,\ell_2,\ell_3,\ell_4)&=\frac{-g_\chi^2 \pi^{-\frac{d}{2}}}{\Gamma\mleft(\Delta-\frac{d-2}{2}\mright)^2}\sqrt{\frac{\Gamma(\Delta+\ell_1)\Gamma(\Delta+\ell_2)\Gamma(\Delta+\ell_3)\Gamma(\Delta+\ell_4)}{\Gamma(\ell_1+1)\Gamma(\ell_2+1)\Gamma(\ell_3+1)\Gamma(\ell_4+1)}}\\
&\quad\, \times\Big[\frac{\ell_2!\ell_4!}{(\Delta)_{\ell_2}(\Delta)_{\ell_4}}\Theta(\ell_2,\ell_4)+\frac{\ell_1!\ell_3!}{(\Delta)_{\ell_1}(\Delta)_{\ell_3}}\Theta(\ell_1,\ell_3)\Big]\, ,
\end{aligned}
\end{align}
with
\begin{equation}
\Theta(\ell_i,\ell_j)=\sum_k \alpha(k)\!\!\sum_{n=0}^{\text{min}(\ell_i,\ell_j)}\frac{(\Delta-k)_n(k)_{\ell_i-n}(k)_{\ell_j-n}}{n!(\ell_i-n)!(\ell_j-n)!}\frac{\Gamma(1+\ell_1+\ell_2-n)\Gamma\mleft(\Delta+k-\frac{d}{2}\mright)}{\Gamma(\Delta+\ell_1+\ell_2-n+k)}\, .
\end{equation}
Finally,  the expression for $H^{(\exch)}$  between $Q=2$ primaries  reads
\eqna{
\gamma_{[\Phi, \Phi]_J}^{(\exch)}(\Delta_\chi)&=\frac{g_\chi^2}{4\Gamma\mleft( \Delta-\frac{d-2}{2}\mright)^2}\left(\sum_{k=\frac{\Delta_\chi}{2}}^\infty \gamma_{[\Phi, \Phi]_J}^{(\exch)}(\Delta_\chi, k)- \sum_{k=\Delta}^\infty \gamma_{[\Phi, \Phi]_J}^{(\exch)}(\Delta_\chi, k)\right)\, ,\\
\gamma_{[\Phi, \Phi]_J}^{(\exch)}(\Delta_\chi, k) &=\frac{ \pi^{-\frac{d}{2}}\Gamma(k)^2 \Gamma(k-\Delta-1)\Gamma\mleft( k+\Delta-\frac{d}{2}\mright)\Gamma\mleft(\Delta-\frac{\Delta_\chi}{2}\mright)\Gamma\mleft(\Delta+\frac{\Delta_\chi-d}{2}\mright)}{\Gamma(k-\Delta-J+1) \Gamma(k+\Delta+J) \Gamma\mleft(k-\frac{\Delta_\chi-2}{2} \mright)\Gamma\mleft(k+1+\frac{\Delta_\chi-d}{2} \mright)}\,.
}[]
We can generalize the same procedure to pairwise identical operators using 
\begin{align}
 &\prod_{i=1}^4 \tilde{N}_{\Delta_i, \ell_i} \!\int\! d^d x \sqrt{g} \phi_{\Delta_3, \ell_2}(x)\phi_{\Delta_1, \ell_4}^*(x) \langle \ell_1, \Delta_1| \chi(x)| \ell_3, \Delta_3\rangle= \!\!\!\sum_{k=\frac{\Delta_\chi-\Delta_{13}}{2}}^\infty\!\!\!\! \! \tilde{V}(k) -\sum_{k=\Delta_3}^\infty\tilde{V}(k)\, ,\\ \nonumber 
 &\tilde{V}(k)= \!\!\sum_{m=0}^{\ell_1}\! \frac{  \pi^{\frac{d}{2}}\Gamma(k)\Gamma(1-k)\Gamma(k+\Delta_{13})\Gamma(1-k-\Delta_{13}) \Gamma\mleft(k+\Delta_1-\frac{d}{2}\mright)\Gamma(k+1-\Delta_3)}{\Gamma(\Delta_1+\ell_1)\Gamma(\Delta_3+\ell_3) \Gamma\mleft(k+1+\frac{\Delta_{13}-\Delta_\chi}{2}\mright)\Gamma\mleft(k+1+\frac{\Delta_{13}+\Delta_\chi-d}{2}\mright)(\ell_{13}-m-k)!} \\ \nonumber 
&\qquad \quad \,  \times\begin{pmatrix}
\ell_1 \\ m
\end{pmatrix}\!\!
\begin{pmatrix}\!
\ell_3 \\ \ell_1-m\!
\end{pmatrix}\!\frac{(-1)^{m+\ell_3+1} (\ell_1-m)!(\ell_2+m)!\Gamma\mleft(\frac{\Delta_1+\Delta_3-\Delta_\chi}{2}\mright)\Gamma\mleft(\frac{\Delta_1+\Delta_3+\Delta_\chi-d}{2}\mright) }{2(\ell_2+m+k+\Delta_1-1)!(m-\ell_1+k-\Delta_3)!(-m-k-\Delta_{13})!}\, ,
\end{align}
with  $\ell_{ij}\equiv \ell_i-\ell_j$ and similarly for $\Delta_{ij}$.
\section{Tree-level Spectrum from Four-point Functions}\label{app:4ptComputation}
In this appendix,  we describe how to extract the anomalous dimension of the operator $[\Phi, \Phi^2]_{\ell}$ from a four-point function at leading order in $\lambda$. To do so, we consider the correlator with two insertions of the scalar operator  $T=[\Phi, \Phi]_0$
\eqna{
\langle T(x_1) \Phi (x_2) \Phi^* (x_3) T^* (x_4)\rangle &=\frac{1}{(x_{12}^2 x_{34}^2)^{\frac{\Delta+\Delta_T}{2}}}\left( \frac{x_{13}^2x_{24}^2}{(x_{14}^2)^2}\right)^{\frac{\Delta_T-\Delta}{2}} \CG_{\Delta_T \Delta \Delta \Delta_T}(u,v)\, , \\
\CG_{\Delta_T \Delta \Delta \Delta_T}&=\sum_{\Delta_{\rm ex}, \ell} a_{\Delta_{\rm ex}, \ell}\lsp  g_{\Delta_{\rm ex}, \ell}^{\scriptscriptstyle (\Delta_T-\Delta,\Delta-\Delta_T)} (u,v)\, ,
}[eq:4ptT]
where in the second line we have expanded the correlator in terms of conformal blocks $g_{\Delta_{\rm ex}, \ell}^{(\Delta_{12},\Delta_{34})}(u,v)$, with $a_{\Delta_{\rm ex}, \ell}$ denoting the squared OPE coefficients. The cross-ratios $u$ and $v$ are defined as
\eqna{
u=\frac{x_{12}^2x_{34}^2}{x_{13}^2 x_{24}^2}\, ,  \qquad\qquad v=\frac{x_{14}^2x_{23}^2}{x_{13}^2 x_{24}^2}\, .
}[]
The  operators exchanged in the OPE are double-twist operators of the form $[\Phi, T]_{n,\ell}$, with classical dimension $\Delta_{\rm ex} = \Delta + \Delta_T + 2n + \ell$. Since we are interested in extracting the anomalous dimension of the lowest-twist operators, we focus on the limit of small $u$, where the conformal blocks simplify significantly
\eqna{
g_{\Delta_{\rm ex}, \ell}^{\scriptscriptstyle(\Delta_T-\Delta,\Delta-\Delta_T)}(u,v) \approx u^{\frac{\Delta_T+\Delta}{2}}\left( \frac{1-v}{2}\right)^\ell \!{}_2 F_1 \left(\Delta+\ell, \Delta+\ell, \Delta+\Delta_T+2\ell; 1-v\right)\, .
}[]
We can then further consider the expansion at leading order in $\lambda$,  where together with $\Delta_T= 2\Delta+\lambda \gamma_{\Phi^2}^{(\contact)}$,  both the OPE coefficients and the dimensions of the exchanged operators receive small corrections. Specifically, we write  
\eqna{
a_{\Delta_{\rm ex}, \ell}=a^{(0)}_{\ell}+\lambda \lsp a^{(1)}_{\ell} \, , \qquad\quad  \Delta_{\rm ex}=2\Delta+\lambda (\gamma_{\Phi^2}^{(\contact)}+\gamma_{\ell})\, ,
}[]
and correspondingly expand the four-point function in~\eqref{eq:4ptT}
\eqna{
&\langle T\Phi \Phi^* T^* \rangle \approx \frac{1}{(x_{12}^2 x_{34}^2)^{\frac{3\Delta}{2}}}\left( \frac{x_{13}^2x_{24}^2}{(x_{14}^2)^2}\right)^{\frac{\Delta}{2}}u^\frac{3\Delta}{2}\sum_{\ell}\Big( a^{(0)}_{\ell}+\lambda \Big[ -a^{(0)}_{\ell} \gamma_{\Phi^2}^{(\contact)} \log(x_{14}^2)\\
&\qquad\, +\frac{1}{2}a^{(0)}_{\ell}\gamma_{\ell} \log u  +\cdots \Big] \Big) \left( \frac{1-v}{2}\right)^\ell \!{}_2 F_1 \left(\Delta+\ell, \Delta+\ell, 3\Delta +2\ell; 1-v\right)\, , 
}[eq:4ptExpanded]
where the $\cdots$ represent terms without $\log$'s.   This expansion has to be compared with the contribution of the Witten diagrams up to this order.  Starting from $O(\lambda^0)$
\begin{align}
  \begin{tikzpicture}[baseline={([yshift=-.5ex]current bounding box.center)},vertex/.style={anchor=base,
    circle,fill=black!25,minimum size=18pt,inner sep=2pt}]
\draw[thick]  (0,0) circle (1);
\draw[thick] (0,0)-- (0.707107,0.707107);
\draw[thick] (0,0)-- (-0.707107,-0.707107);
\draw[thick, double ] (-0.0707107,0.0707107)-- (-0.707107,0.707107);
\draw[thick, double ] (0.0707107,-0.0707107)-- (0.707107,-0.707107);
\node at (1.5,0) {+};
\draw[thick]  (3,0) circle (1);
\draw[thick] (2.29289,0.707107) --(3.707107,0.707107) ; 
\draw[thick] (2.29289,-0.707107) --(3.707107,-0.707107) ; 
\draw[thick] (2.29289,0.707107) --(3.707107,-0.707107) ; 
\end{tikzpicture}
=\frac{1}{(x_{12}^2 x_{34}^2)^{\frac{3\Delta}{2}}}\left( \frac{x_{13}^2x_{24}^2}{(x_{14}^2)^2}\right)^{\frac{\Delta}{2}}\!u^\frac{3\Delta}{2}\left( 2+\frac{1}{v^\Delta}\right)\, , 
\end{align}
and comparing with the expansion in~\eqref{eq:4ptExpanded} we can extract the free OPE coefficients
\eqna{
a_{\ell}^{(0)}&=\begin{cases}
3 & \quad  \text{if}\quad \ell=0\, ,\\
\frac{2^{\ell } \Gamma (\ell +\Delta ) \Gamma (\ell +3 \Delta -1)}{\ell! \Gamma (\Delta )
   \Gamma (2 \ell +3 \Delta -1)}\left(\frac{2 \Gamma (\ell +\Delta )}{\Gamma (\Delta )}+\frac{(-1)^{\ell } \Gamma (\ell +2 \Delta )}{\Gamma (2
   \Delta )} \right)& \quad  \text{if}\quad \ell\geq 2 \,.
\end{cases}
}[]
At order $\lambda$, we encounter three distinct classes of diagrams contributing to the four-point function. The first type consists of the four-point contact Witten diagram for $\Phi$ multiplied by a free propagator 
\begin{align}
  \begin{tikzpicture}[baseline={([yshift=-.5ex]current bounding box.center)},vertex/.style={anchor=base,
    circle,fill=black!25,minimum size=18pt,inner sep=2pt}]
\draw[thick]  (0,0) circle (1);
\draw[thick] (-0.707107,0.707107)-- (0.707107,-0.707107);
\draw[thick] (0.707107,0.707107)-- (-0.707107,-0.707107);
\node[above=0.1cm, myRed] at (0,0) {\footnotesize $\lambda$};
\draw[thick] (-0.176777,  -0.0353553) -- (-0.813173, 0.601041);
\draw[thick] (-0.0353553, -0.176777) -- (0.601041, -0.813173);
\filldraw[myRed] (0,0) circle (0.03);
\end{tikzpicture}
= \frac{1}{(x_{13}^2x_{14}^2x_{24}^2)^\Delta}\frac{-\lambda\pi^{-\frac{d}{2}}\Gamma\mleft(2\Delta-\frac{d}{2} \mright)}{4\Gamma(\Delta)^2\Gamma\mleft(\Delta-\frac{d-2}{2}\mright)^2}\Db_{\Delta\Delta\Delta\Delta}\, .
\end{align}
The second and third types of diagrams involve a free propagator together with the $T$ two-point function or the three-point function $\langle T \Phi \Phi \rangle$. To efficiently extract the contribution at order 
$\lambda$,  it is convenient to express these diagrams in terms of $\Delta_T$ and  the OPE coefficients involving $T$, such that 
\begin{align}
  \begin{tikzpicture}[baseline={([yshift=-.5ex]current bounding box.center)},vertex/.style={anchor=base,
    circle,fill=black!25,minimum size=18pt,inner sep=2pt}]
\draw[thick]  (0,0) circle (1);
\draw[thick](0,0) to[out=70,in=20]  (-0.707107,0.707107);
\draw[thick](0,0) to[out=200,in=250]  (-0.707107,0.707107);
\draw[thick] (0,0) to[out=20,in=70]  (0.707107,-0.707107);
\draw[thick] (0,0) to[out=250,in=200]  (0.707107,-0.707107);
\node[myRed] at (-0.15,0.18) {\footnotesize $\lambda$};
\draw[thick] (0.106066,0.106066) --  (0.707107,0.707107);
\draw[thick] (-0.106066,-0.106066) --  (-0.707107,-0.707107);
\filldraw[myRed] (0,0) circle (0.03);
\end{tikzpicture}
= \frac{1}{(x_{23}^2)^\Delta(x_{14}^2)^{\Delta_T} }\Big|_{\lambda}=-\lambda \gamma_{\Phi^2}^{(\contact)}\frac{\log(x_{14}^2)}{(x_{23}^2)^\Delta(x_{14}^2)^{2\Delta} }\, ,
\end{align}
\begin{align} \nonumber 
  \begin{tikzpicture}[baseline={([yshift=-.5ex]current bounding box.center)},vertex/.style={anchor=base,
    circle,fill=black!25,minimum size=18pt,inner sep=2pt}]
\draw[thick]  (0,0) circle (1);
\draw[thick](0,0) to[out=70,in=20]  (-0.707107,0.707107);
\draw[thick](0,0) to[out=200,in=250]  (-0.707107,0.707107);
\node[myRed] at (-0.15,0.18) {\footnotesize $\lambda$};
\draw[thick] (0,0) --  (0.707107,0.707107);
\draw[thick] (0,0) --  (0.707107,-0.707107);
\draw[thick]  (-0.707107,-0.707107)--  (0.707107,-0.707107);
\filldraw[myRed] (0,0) circle (0.03);
\node at (1.5,0) {+};
\draw[thick]  (3,0) circle (1);
\draw[thick] (3,0) to[out=20,in=70]  (3.707107,-0.707107);
\draw[thick] (3,0) to[out=250,in=200]  (3.707107,-0.707107);
\node[myRed] at (3.15,-0.18) {\footnotesize $\lambda$};
\draw[thick] (3,0) --  (2.29289,0.707107);
\draw[thick] (3,0) --  (2.29289,-0.707107);
\draw[thick] (3.707107,0.707107) --  (2.29289,0.707107);
\filldraw[myRed] (3,0) circle (0.03);
\end{tikzpicture}
&=\frac{ \sqrt{2}c_{\Phi \Phi \Phi^2}}{(x_{14}^2)^\frac{\Delta_T}{2}}\!\!\left[\frac{(x_{34}^2)^{\frac{\Delta_T-2\Delta}{2}}}{(x_{24}^2)^{\Delta}(x_{13}^2)^\frac{\Delta_T}{2}}+\frac{(x_{12}^2)^{\frac{\Delta_T-2\Delta}{2}}}{(x_{13}^2)^{\Delta}(x_{24}^2)^\frac{\Delta_T}{2}} \right]_{\lambda}\\
&=\frac{\lambda \left[ c_{\Phi \Phi \Phi^2}^{(0)}\gamma_{\Phi^2}^{(\contact)}(\log u-2\log(x_{14}^2))+4 c_{\Phi \Phi \Phi^2}^{(1)} \right]}{\sqrt{2} (x_{13}^2 x_{14}^2 x_{24}^2)^\Delta}\, . \label{eq:thirdDiagram}
\end{align}
Putting everything together and comparing with~\eqref{eq:4ptExpanded}, we first observe that the second and third diagrams correctly combine to reproduce the $\log (x_{14}^2)$ term.  To extract the anomalous dimension  $\gamma_\ell$,  we must isolate the $\log u$  contribution in the correlator. This contribution arises from two sources: the $\Db_{\Delta\Delta\Delta\Delta}$ and the explicit $\log u$ term present in~\eqref{eq:thirdDiagram}.   Expanding these terms and using
\eqna{
\gamma_{\Phi^2}^{(\contact)}=\lambda \frac{\Gamma(\Delta)^2 \Gamma\mleft(2\Delta-\frac{d}{2}\mright)}{8\pi^\frac{d}{2}\Gamma(2\Delta)\Gamma\mleft(\Delta-\frac{d-2}{2}\mright)^2}\, ,
}[]
we obtain
\eqna{
\gamma_\ell= \gamma_{\Phi^2}^{(\contact)}\frac{2 (-1)^{\ell } \Gamma (2 \Delta ) \Gamma (\ell +\Delta )}{\Gamma (\Delta ) \Gamma (\ell +2 \Delta
   )}\, ,
}[]
which exactly reproduces~\eqref{Q3contact}. 
Extending this computation to order  $\lambda^2$ would be significantly more challenging due to the rapid increase in the number and complexity of diagrams.  A possible way forward is to use the Mellin space results in~\cite{Ma:2022ihn}, although we did not pursue this approach in the present work.

\section{Spectrum from Bubble Diagrams} \label{app:fullSpectrumD2d4}
\begin{figure}[t]
    \centering
    \begin{minipage}{0.48\textwidth}
        \centering
        \includegraphics[width=\linewidth]{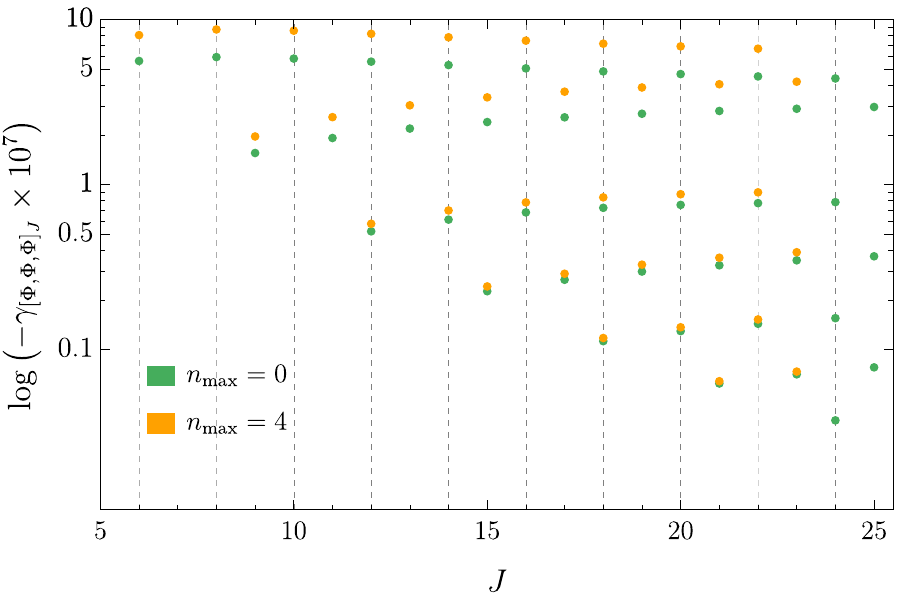}
    \end{minipage}
    \hfill
    \begin{minipage}{0.48\textwidth}
        \centering
        \includegraphics[width=\linewidth]{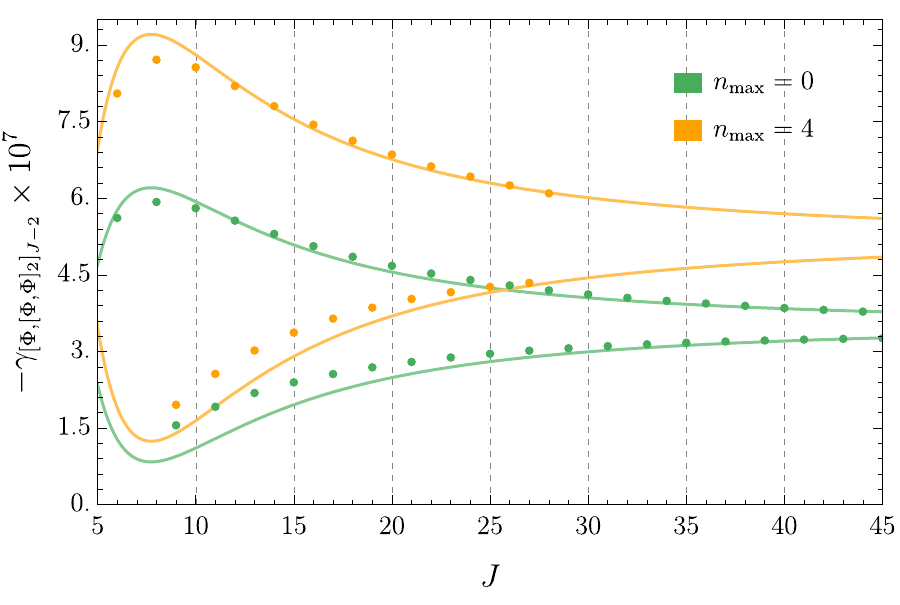}
    \end{minipage}
    \caption{\textit{Left}: $Q=3$ spectrum at $O(\lambda^2)$ from bubble diagrams for different truncations $n_{\rm max}$. \textit{Right}:  upper trajectories, interpreted as the anomalous dimensions of  $[\Phi, [\Phi, \Phi]_2]_{J-2}$. Solid lines correspond to the large spin approximation in~\eqref{lspinB2}. }\label{fig:bubbles}
\end{figure}
As anticipated in section~\ref{subsec:Q=3Spect}, the $t$- and $u$-channel bubble diagrams affect not only the dimension of $[\Phi, \Phi^2]_J$ but also that of all $Q=3$ operators.
In Figure~\ref{fig:bubbles}  we compute numerically the full spectrum  for the case $\Delta=2$ in $d=4$, using the spectral decomposition in~\eqref{eq:BubbleSpecDecomp}.
Concretely, within the $Q=3$ subsector orthogonal to $[\Phi, \Phi^2]_J$, we construct and diagonalize the Hamiltonian obtained by including the exchange of bulk scalars with dimensions $\Delta_\chi = 4+2n$, $n=0,\dots,n_{\rm max}$.  Even for the relatively small values of $J$ accessible in our computation, two pairs of upper trajectories are clearly visible, approaching the anomalous dimensions of $[\Phi, \Phi]_{\ell}$ in~\eqref{gammaBubbleQ2} with $\ell = 2$ and $\ell = 4$ as $J$ increases. These trajectories can be identified at large spin with operators of the form $[\Phi, [\Phi, \Phi]_{\ell}]_{J-\ell}$. 
The slower convergence in $n$ observed for these states follows directly from this interpretation, since $\gamma_{[\Phi, \Phi]_{\ell}} \sim 1/n^{2+2\ell}$, as shown in~\eqref{scalarSpecial}. 

In the right panel of Figure~\ref{fig:bubbles}, we focus on the $[\Phi, [\Phi, \Phi]_2]_{J-2}$ trajectories.  For even and odd $J$, these trajectories approach  $\sum_{n=0}^{n_{\rm max}}\gamma_{[\Phi, \Phi]_2}^{(\exch)}(4+2n)$ from above and below, respectively.  
Although the eigenstate matches  $[\Phi, [\Phi, \Phi]_2]_{J-2}$ only at large $J$,  to leading order we may treat it as such and consider
\eqna{
\langle [\Phi, [\Phi, \Phi]_2]_{J-2} | \sum_{n=0}^{n_{\rm max}}\left(H_2^{(\exch)}\right)|[\Phi, [\Phi, \Phi]_2]_{J-2} \rangle \approx  \sum_{n=0}^{n_{\rm max}} \frac{\gamma_{[\Phi, \Phi]_2}^{(\exch)}(4+2n)}{\gamma_{[\Phi, \Phi]_2}^{(\contactell{2})}}\gamma_{[\Phi,[\Phi, \Phi]_2]_{J-2}}^{(\contactell{2})} \, , 
}[lspinB2]
where we have used~\eqref{Q3HasContact} with $\gamma_{[\Phi, \Phi]_2}^{(\contactell{2})}$ in ~\eqref{ContactQ3Spin2} and the  behavior of the overlaps in~\eqref{ovs}.  The solid lines in the figure correspond to this approximation  and we see that its agreement with the numerical data gets better and better  as $J$ increases.

\section{Inversion Integrals}\label{app:inverison1}
In this appendix we summarize the technical ingredients needed to apply the perturbative inversion formula and to evaluate the relevant integrals used in appendix~\ref{app:inversion2}. For a comprehensive treatment we refer the reader to the original references~\cite{Alday:2017vkk,Alday:2017zzv,Henriksson:2020jwk},  our presentation will largely follow the discussion in~\cite[Sec.~2.3, 5.3]{Bertucci:2022ptt}.

The perturbative inversion formula can be viewed as a one-dimensional reduction of the Lorentzian inversion formula~\cite{Caron-Huot:2017vep}, designed to extract OPE data for a fixed twist family of twist $\tau_0$. It is formulated in terms of the generating function $\mathbb{T}_{\tau_0}(\log z,\bar{h})$, defined as
\begin{equation}
\mathbb{T}_{\tau_0}(\log z,\bar{h})=\kappa_{\bar{h}}^{1234}\int_0^1\frac{{d}\bar{z}}{\bar{z}^2}k_{\bar{h}}^{1234}(\bar{z})(1-\bar{z})^{\frac{\Delta_{21}+\Delta_{34}}{2}}\mathrm{dDisc}^{1234}[\mathcal{G}_{1234}(z,\bar{z})]\Big|_{z^{\frac{\tau_0}{2}}}+(-1)^\ell(1\leftrightarrow 2)\, ,
\end{equation}
where $\mathcal{G}_{1234}(z,\bar{z})$ denotes the four-point function of four scalar operators $\CO_i$, $k_{\bar{h}}^{1234}(\bar{z})$  the SL$(2,\mathbb{R})$ block and $\mathrm{dDisc}$ is the double-discontinuity,  defined as two different analytic continuation around $\zb=1$ 
\begin{align}\label{eqn:dDisc}
\begin{aligned}
k_{\beta}^{1234}(x)&=x^\beta{}_2F_1\Big(\beta+\frac{\Delta_{12}}{2},\beta-\frac{\Delta_{34}}{2};2\beta;x\Big)\, , \\
\kappa_\beta^{1234}&=\frac{\Gamma\mleft(\beta-\frac{\Delta_{12}}{2}\mright)\Gamma\mleft(\beta+\frac{\Delta_{12}}{2}\mright)\Gamma\mleft(\beta-\frac{\Delta_{34}}{2}\mright)\Gamma\mleft(\beta+\frac{\Delta_{34}}{2}\mright)}{2\pi^2\Gamma(2\beta)\Gamma(2\beta-1)}\, , \\
\mathrm{dDisc}^{1234}[f(\bar{z})]&=\cos\!\left(\pi\frac{\Delta_{21}+\Delta_{34}}{2}\right)f(\bar{z})-\frac{1}{2}e^{i\pi\frac{\Delta_{21}+\Delta_{34}}{2}}f^{\circlearrowleft}(\bar{z})-\frac{1}{2}e^{-i\pi\frac{\Delta_{21}+\Delta_{34}}{2}}f^{\circlearrowright}(\bar{z})\, .
\end{aligned}
\end{align}
The generating function encodes OPE coefficient $a_{\tau_0, \ell}$ and anomalous dimensions $\gamma_{\tau_0, \ell}$ of the operators exchanged in the $s$-channel OPE as
\eqna{
\mathbb{T}_{\tau_0}(\log z,\bar{h})&=T_{\bar{h}}^{(0)}+\frac{1}{2}T_{\bar{h}}^{(1)}\log z+\frac{1}{8}T_{\bar{h}}^{(2)}\log^2z+\dots\, , \\
a_{\tau_0,\ell}(\gamma_{\tau_0,\ell})^p&=T_{\bar{h}}^{(p)}+\frac{1}{2}\partial_{\bar{h}}T_{\bar{h}}^{(p+1)}+\frac{1}{8}\partial_{\bar{h}}^2T_{\bar{h}}^{(p+2)}+\dots\Big|_{\bar{h}=\frac{\tau_0}{2}+\ell}\, .
}[]
In the following we restrict to pairwise-identical operators. Our main goal will be to compute the generating functions associated with $\langle 1212\rangle$- and $\langle 2112\rangle$-type correlators, which will be useful  in appendix~\ref{app:inversion2}.
\subsection*{$\boldsymbol{\langle}\mathbf{ 1212}\boldsymbol{\rangle}$}
Starting with the case $\CO_3=\CO_1$ and $\CO_4=\CO_1$, the most general integral we need to evaluate (for integers $a_1,a_2\geq0$) is
\begin{equation}
H^{1212}(\Delta_{21},\alpha,\beta,a_1,a_2)=\kappa_{\bar{h}}^{1212}\int_0^1\frac{\mathrm{d}\bar{z}}{\bar{z}^2}k^{1212}_{\bar{h}}(\bar{z})\mathrm{dDisc}\Big[\frac{\bar{z}^\alpha}{(1-\bar{z})^\beta}(\log\bar{z})^{a_1}(\log(1-\bar{z}))^{a_2}\Big]\, .
\end{equation}
For $H^{1212}(\Delta_{21},\alpha,\beta,0,0)$, the integral admits a closed form in terms of a regularized hypergeometric function,
\begin{align}
&H^{1212}(\Delta_{21},\alpha,\beta,0,0)=\frac{\Gamma(\bar{h}-1+\alpha)\Gamma\mleft(\bar{h}-\frac{\Delta_{21}}{2}\mright)^2\Gamma\mleft(\bar{h}+\frac{\Delta_{21}}{2}\mright)^2}{\Gamma(2\bar{h}-1)\Gamma(1+\bar{h}-\alpha)\Gamma(\beta)^2}\notag\\
&\times{}_3\tilde{F}_2\Big[\{\bar{h}-1+\alpha,\alpha-\beta-\frac{\Delta_{21}}{2},\alpha-\beta+\frac{\Delta_{21}}{2}\};\{\bar{h}+\alpha-\beta,\bar{h}+\alpha-\beta\};1\Big]\, .
\end{align}
Then by using 
\begin{equation}
\mathrm{dDisc}^{1212}\Big[\frac{\bar{z}^\alpha}{(1-\bar{z})^\beta}(\log\bar{z})^{a_1}(\log(1-\bar{z}))^{a_2}\Big]=(-1)^{a_2}\partial_\alpha^{a_1}\partial_\beta^{a_2}\Big[2\frac{\bar{z}^\alpha}{(1-\bar{z})^\beta}\sin(\pi\beta)^2\Big]\, ,
\end{equation}
we obtain
\begin{equation}
H^{1212}(\Delta_{21},\alpha,\beta,a_1,a_2)=(-1)^{a_2}\partial_{\alpha^\prime}^{a_1}\partial_{\beta^\prime}^{a_2}\Big[H^{1212}(\Delta_{21},\alpha^\prime,\beta^\prime,0,0)\Big]\Big|_{\alpha^\prime=\alpha,\beta^\prime=\beta}\,.
\end{equation}
Below we list all the $\langle1212\rangle$-type inversion integrals used in this paper:
\eqna{
H^{1212}(2+\lambda a+\lambda^2 b,3,2,0,0)&=\frac{\Gamma(\bar{h}-1)^2}{2\Gamma(2\bar{h}-1)}(a\lambda(\bar{h}(\bar{h}(2\bar{h}-5)-1)+2)+2(\bar{h}-2)_4)\, , \\
H^{1212}(2,3,2,1,0)&=-\frac{\Gamma(\hb-1)\Gamma(\hb+1)}{\Gamma(2\hb-1)}\, ,\\
H^{1212}(2,3,2,0,1)&=\frac{\Gamma(\hb-1)}{\Gamma(2\hb-1)}\Big[\Gamma(\bar{h}-1)(\bar{h}(\bar{h}(2(\bar{h}-1)\bar{h}-3)-1+2)\\
&\quad\, -2(\bar{h}-2)H_{\bar{h}}\Gamma(\bar{h}+2)\Big]\, .
}[]
\subsection*{$\boldsymbol{\langle}\mathbf{ 2112}\boldsymbol{\rangle}$}
The other relevant case is
\begin{equation}
H^{2112}(\Delta_{21},\alpha,\beta,a_1,a_2)=\kappa_{\bar{h}}^{2112}\!\!\int_0^1\frac{\mathrm{d}\bar{z}}{\bar{z}^2}k^{2112}_{\bar{h}}(\bar{z})\mathrm{dDisc}^{2112}\Big[\frac{\bar{z}^\alpha}{(1-\bar{z})^\beta}(\log\bar{z})^{a_1}(\log(1-\bar{z}))^{a_2}\Big]\, .
\end{equation}
Proceeding as before, we first get a closed form expression for
\begin{align}\label{eqn:2112inv}
&H^{2112}(\Delta_{21},\alpha,\beta,0,0)=\frac{\Gamma(\bar{h}-1+\alpha)\Gamma\mleft(\bar{h}-\frac{\Delta_{21}}{2}\mright)^2\Gamma\mleft(\bar{h}+\frac{\Delta_{21}}{2}\mright)^2}{\Gamma(2\bar{h}-1)\Gamma(1+\bar{h}-\alpha)\Gamma(\beta)\Gamma(\beta+\Delta_{21})}\notag\\
&\times{}_3\tilde{F}_2\Big[\{\bar{h}-1+\alpha,\alpha-\beta-\frac{\Delta_{21}}{2},\alpha-\beta+\frac{\Delta_{21}}{2}\};\{\bar{h}+\alpha-\beta,\bar{h}+\alpha-\beta-\Delta_{21}\};1\Big]\, , 
\end{align}
that we then combine with 
\begin{equation}
\mathrm{dDisc}^{2112}\Big[\frac{\bar{z}^\alpha (\log\bar{z})^{a_1}(\log(1-\bar{z}))^{a_2}}{(1-\bar{z})^\beta}\Big]\!=\partial_\alpha^{a_1}\partial_\beta^{a_2}\Big[\frac{2(-1)^{a_2}\bar{z}^\alpha}{(1-\bar{z})^\beta}\sin(\pi\beta)\sin[\pi(\beta+\Delta_{21})]\Big]\, .
\end{equation}
Below we list all the $\langle2112\rangle$-type inversion integrals used in this paper:
\begin{align}
\begin{aligned}
H^{2112}(2,2,1,0,1)&=\frac{\Gamma(\bar{h}+1)^2}{4\Gamma(2\bar{h}-1)}\bigg[3-4\gamma_E-4\psi^{(0)}(\bar{h}-1)-2\Big(\frac{1}{\bar{h}}+\frac{1}{\bar{h}-1}\Big)\bigg]\, , \\
H^{2112}(2,3,2,1,1)&=\frac{\Gamma(\bar{h}-2)\Gamma(\bar{h}-1)}{12\Gamma(2\bar{h}-1)}\bigg[12(\bar{h}-2)(\bar{h}-1)^2\bar{h}^2H_{\bar{h}-3}\\
&\quad\, +(3\bar{h}(\bar{h}((20-3\bar{h})\bar{h}-39)+30)-28)\bar{h}+8\bigg]\,  ,\\
H^{2112}(2,2,1,2,1)&=\frac{\Gamma(\bar{h}-1)^2}{\Gamma(2\bar{h}-1)}\, ,\\
H^{2112}(2,2,1,1,1)&=\frac{\Gamma(\bar{h}+1)^2}{2\Gamma(2\bar{h}-1)}\Big[1+2\bar{h}(\bar{h}-1)\Big]\, , 
\end{aligned}
\end{align}
with $\psi^{(0)}(z)$  the digamma function. 
\section{$\gamma_{[\Phi,\Phi^2]_J}$ from the Lorentzian inversion formula for $d=4$ and $\Delta=2$}\label{app:inversion2}
In this section we detail the large-spin computation of $\gamma_{[\Phi,\Phi^2]_J}$ at $d=4,\Delta=2$, using both the perturbative inversion formula and the large spin expansion in~\eqref{largeSpinDT}.  Our goal is to compute  $\gamma_{[\Phi,\Phi^2]_J}$ up to $\frac{\lambda^2}{J^4}$, so we can just include the exchange of $\Phi$, $S=[\Phi,\Phi^*]$ and $J_\ell=[\Phi,\Phi^*]_\ell$. 

We start from considering the exchange of $\Phi$, through its conformal block $g_{\Delta,0}$,  and of the identity operator 
\begin{align}
\begin{aligned}
\mathbb{T}_{T\Phi\Phi T}^{(\Phi)}&=c_{\Phi\Phi T}^2(-1)^\ell\kappa_{\bar{h}}^{\Phi T\Phi T}\int_0^1\frac{{d}\bar{z}}{\bar{z}^2}k_{\bar{h}}^{\Phi T\Phi T}(\bar{z})\mathrm{dDisc}^{\Phi T\Phi T}\bigg[\Big(\frac{u}{v}\Big)^{\frac{\Delta_T+\Delta}{2}}g_{\Delta,0}(1-\bar{z},1-z)\bigg] \\
& \quad\,  +\kappa_{\bar{h}}^{T\Phi\Phi T}\int_0^1\frac{{d}\bar{z}}{\bar{z}^2}k_{\bar{h}}^{T\Phi\Phi T}(\bar{z})\mathrm{dDisc}^{T\Phi\Phi T}\left(\frac{u^{\frac{\Delta_T+\Delta}{2}}}{v^{\Delta}}\right)\, ,
\end{aligned}
\end{align}
with $\Delta=2,\Delta_T=4+\lambda \gamma_{\Phi^2}^{(\contact)}+\lambda^2\gamma_{\Phi^2}^{(2)}$. The anomalous dimension $[\Phi, \Phi^2]_J$ can then be extracted from this generating function as 
\twoseqn{
\mathbb{T}_{T\Phi\Phi T}^{(\Phi)}&=\mathbb{T}_\Phi^{(0)}(\bar{h})+\frac{1}{2}\mathbb{T}_\Phi^{(1)}(\bar{h})\log z+\dots
}[]
{
\gamma_{[\Phi,\Phi^2]_J}^{(\Phi),\text{LIF}}&=\frac{\mathbb{T}_\Phi^{(1)}(\bar{h})}{\mathbb{T}_\Phi^{(0)}(\bar{h})+\frac{1}{2}\partial_{\bar{h}}\mathbb{T}_\Phi^{(1)}(\bar{h})}\Big|_{\bar{h}=\frac{\Delta+\Delta_T}{2}+J}\,.
}[mixedIn][]
By using the inversion integrals listed in appendix~\ref{app:inverison1} and  explicit expansion of the conformal block to order $\lambda^2$
\eqna{
g_{\Delta,0}(z,\bar{z})&=u+\lambda\frac{\gamma_{\Phi^2}^{(\contact)}}{2}u\log v+\frac{\lambda^2u}{4}\bigg[-\left(\gamma_{\Phi^2}^{(\contact)}\right)^2+\Big(2\gamma_{\Phi^2}^{(2)}+\frac{\left(\gamma_{\Phi^2}^{(\contact)}\right)^2}{v-1}\Big)\log v\\
&\quad\, +\frac{\left(\gamma_{\Phi^2}^{(\contact)}\right)^2 u(3-4v+v^2+2\log v)}{(v-1)^3}-\left(\gamma_{\Phi^2}^{(\contact)}\right)^2 \mathrm{Li}_2(1-v)\bigg]+\mathcal{O}(\lambda^3)\, ,
}[]
we can then compute
\twoseqn{
&
\begin{aligned}
\mathbb{T}_\Phi^{(0)}(\bar{h})&=\frac{\Gamma(\hb-1)^2(\hb-2)_4}{\Gamma(2\hb-1)}\Bigg\lbrace (-1)^\ell a_t^{(0)}+\frac{\hb(\hb-1)}{6}+\lambda \Bigg [a_t^{(0)} \gamma^{(\contact)}_{\Phi^2} \left( H_{\hb+1}-\frac{\hb-1}{\hb-2}\right) 
 \\
&\, + (-1)^J a_t^{(1)}+\frac{\gamma^{(\contact)}_{\Phi^2}}{6}\!\left( (\hb-1)_2H_{\hb-2}-\frac{11  \hb^4-34  \hb^3+16  \hb^2+31 \hb-12}{6(\hb-2)(\hb+1)}\right) \Bigg] \Bigg\rbrace\, ,
\end{aligned}
}[]
{
&\begin{aligned}
\mathbb{T}_\Phi^{(1)}(\bar{h})&=\lambda (-1)^J \frac{\Gamma(\hb-1)^2(\hb-2)_4}{2\Gamma(2\hb-1)} \Bigg\lbrace {a_t^{(0)}\gamma_{\Phi^2}^{(\contact)}}+\lambda \Bigg[ a_t^{(0)}\gamma_{\Phi^2}^{(2)}+a_t^{(1)}\gamma_{\Phi^2}^{(\contact)}\\
&\, +a_t^{(0)}\left(\gamma_{\Phi^2}^{(\contact)}\right)^2 \left( H_{\hb}-\frac{3\hb^2-3\hb+1}{2(\hb-2)(\hb-1)}\right)\Bigg]\Bigg\rbrace\, ,
\end{aligned}
}[]
where $a_t^{(i)}=(c_{\Phi\Phi T}^{(i)})^2$. Plugging the explicit values in~\eqref{eq:datad2D4} into~\eqref{mixedIn}, finally we get 
\eqna{
&\gamma_{[\Phi,\Phi^2]_J}^{(\Phi), \text{LIF}}\approx\lambda\frac{(-1)^J}{4\pi^2}\Big[\frac{1}{J^2}-\frac{5}{J^3}+\frac{19-12(-1)^J}{J^4}\Big]+\lambda^2(-1)^J\Bigg[\frac{12\gamma_{\Phi^2}^{(2)}}{J^2}\left(1-\frac{5}{J}\right)\\
&+\frac{1}{384 \pi^4 J^2}\!\left(\!1-\frac{7}{J}\right)+\frac{1}{64\pi^4 J^4}\Big(5+14592\gamma_{\Phi^2}^{(2)}\pi^4+(-1)^J(-6-9216\gamma_{\Phi^2}^{(2)}\pi^4+\log4\Big)\Bigg]\,.
}[]

For the exchange of $S$, we only need the leading $\frac{1}{J^4}$ contribution,  which can be more easily obtained from the large spin formula \eqref{largeSpinDT},
\begin{equation}
\gamma_{[\Phi,\Phi^2]_J}^{(S), \text{LIF}}=\frac{3\lambda}{J^4\pi^2}+\frac{\lambda^2}{32J^4\pi^4}\Big(8-4\gamma_E+4608\pi^4\gamma_{\Phi^2}^{(2)}-4\log J\Big)+\mathcal{O}\Big(\frac{1}{J^5}\Big)\, .
\end{equation}

The last contribution comes from the infinite tower of neutral twist-two operators  $J_\ell=[\Phi, \Phi^*]_{\ell}$.  In this case, we again rely on the perturbative inversion formula. Unlike the single-exchange case, however, the prescription here is to first resum over $\ell$ and only then take the double discontinuity. The resummation we need to perform is
\begin{align}
F_2(1-\bar{z},1-z)&=\frac{\log v}{2}\sum_{\ell=2,4,\dots,\infty}c_{\Phi\Phi J_\ell}^{(0)}c_{TT J_\ell}^{(0)}\gamma_{[\Phi,\Phi^*]_\ell}g_{2\Delta+\ell,\ell}(1-\bar{z},1-z)\notag\\
&=\frac{v\log v}{2(\bar{z}-z)}\!\sum_\ell\! c_{\Phi\Phi J_\ell}^{(0)}c_{TT J_\ell}^{(0)}\gamma_{[\Phi,\Phi^*]_\ell}  \left[k_{\Delta-1}(1-\bar{z})k_{\Delta+\ell}(1-z)-(z\leftrightarrow\bar{z})\right],
\end{align}
where 
\begin{equation}
c_{\Phi\Phi J_\ell}^{(0)}c_{TT J_\ell}^{(0)}=\left[1+(-1)^\ell\right]\frac{4^{1-\ell-\Delta}\sqrt{\pi}\Gamma(\ell+\Delta\Gamma(\ell-1+2\Delta)}{\Gamma(1+\ell)\Gamma(\Delta)^2\Gamma(\ell-1/2+\Delta)}\,.
\end{equation}
Unfortunately a closed analytic expression for $\gamma_{[\Phi, \Phi^*]_{\ell}}$ is not known,  so the resummation cannot be carried out explicitly.  Nevertheless, following the procedure of~\cite{Simmons-Duffin:2016wlq}, we can approximate the result by using the large spin expansion of these anomalous dimensions
\begin{equation}
\gamma_{[\Phi,\Phi^*]_\ell}=-\frac{\lambda^2}{192\pi^4(\mathcal{J}^2)^2}+\mathcal{O}\Big(\frac{1}{(\mathcal{J}^2)^3}\Big)\,, \qquad\mathcal{J}^2=(\ell+1)(\ell+2)\, . 
\end{equation}
With the help of the regulator function
\begin{equation}
S_a(h)=\frac{\Gamma(h)^2\Gamma(h-a-1)}{\Gamma(-a)^2\Gamma(2h-1)\Gamma(h+a+1)}
\end{equation}
the sum can then be approximated by
\begin{align}\label{eqn:gammaapprox}
c_{\Phi\Phi J_\ell}^{(0)}c_{TT J_\ell}^{(0)}\gamma_{[\Phi,\Phi^*]_\ell}^{\rm approx}& \approx-\frac{\lambda^2}{96\pi^4}\lim_{a\to0}\Gamma(-a)^2S_a(2+\ell)=-\frac{2^{-7-2\ell}\lambda^2\Gamma(1+\ell)}{3(2+\ell)\pi^{7/2}\Gamma(\ell+3/2)}\, , 
\end{align}
and resummed exactly to
\begin{align}
\begin{aligned}
\tilde{g}(z)&=\sum_{\ell}c_{\Phi\Phi J_\ell}^{(0)}c_{TTJ_\ell}^{(0)}\gamma_{[\Phi,\Phi^*]_\ell}^{\text{approx}}k_{2+\ell}(1-z)=-\frac{\lambda^2}{32\pi^4}\Big(1+\frac{\log z}{2}-\frac{z\log z}{z-1}+\frac{\log^2z}{12}\Big)\, ,\\
F_2^{\rm approx}&=\frac{(1-\bar{z})(1-z)\log[(1-z)(1-\bar{z})]}{2({z}-\zb)}\left( \tilde{g}(z)\log \zb -\tilde{g}(\zb) \log z\right)\, .
\end{aligned}
\end{align}
Plugging this expression into the generating function
\begin{equation}
\mathbb{T}_{T\Phi\Phi T}^{(J_{\ell})}(z,\bar{h})\approx\kappa_{\bar{h}}^{T\Phi\Phi T}\int_0^1\frac{{d}\bar{z}}{\bar{z}^2}k_{\bar{h}}^{T\Phi\Phi T}(\bar{z})\mathrm{dDisc}^{T\Phi\Phi T}\left[\frac{u^{\frac{\Delta_T+\Delta}{2}}}{v^{\Delta}}+F_{2}^{\text{approx}}(1-\bar{z},1-z)\right]\, ,
\end{equation}
we can extract the anomalous dimensions
\twoseqn{
\mathbb{T}_{T\Phi\Phi T}^{(J_{\ell})}(z,\bar{h})&=T_{J_\ell}^{(0)}(\bar{h})+\frac{1}{2}T_{J_\ell}^{(1)}(\bar{h})\log z+\frac{1}{8}T_{J_\ell}^{(2)}(\bar{h})\log^2z+\dots
}[]
{
\gamma_{[\Phi,\Phi^2]_J}^{(J_\ell), \text{LIF}}&=\frac{\mathbb{T}_{J_\ell}^{(1)}(\bar{h})+\frac{1}{2}\partial_{\bar{h}}\mathbb{T}_{J_\ell}^{(2)}(\bar{h})}{\mathbb{T}_{J_\ell}^{(0)}(\bar{h})+\partial_{\bar{h}}\mathbb{T}_{J_\ell}^{(1)}(\bar{h})}\Big|_{\bar{h}=\frac{\Delta+\Delta_{T}}{2}+J}
}[][]
Since we only need to extract the leading $1/J^4$ contribution,  this expression further simplifies
\begin{equation}
\gamma_{[\Phi,\Phi^2]_J}^{(J_\ell), \text{LIF}}\approx\frac{\mathbb{T}_{J_\ell}^{(1)}(\bar{h})+\frac{1}{2}\partial_{\bar{h}}\mathbb{T}_{J_\ell}^{(2)}(\bar{h})}{\mathbb{T}_{\mathbf{1}}^{(0)}(\bar{h})}\Big|_{\bar{h}=\frac{\Delta+\Delta_{T}}{2}+J}\, ,
\end{equation}
with
\begin{align}
\begin{aligned}
&\mathbb{T}^{(0)}_{\mathbf{1}}(\bar{h})=\frac{(\bar{h}-2)\Gamma(\bar{h}+1)\Gamma(\bar{h}+2)}{6\Gamma(2\bar{h}-1)},\qquad \mathbb{T}_{J_\ell}^{(2)}(\bar{h})=\frac{\lambda^2\left(1+2\bar{h}(\bar{h}-1)\right)\Gamma(\bar{h}-1)^2}{192\pi^4\Gamma(2\bar{h}-1)}\\
&\mathbb{T}_{J_\ell}^{(1)}(\bar{h})=\frac{\lambda^2\left(1+2\bar{h}(\bar{h}-1)\right)\Gamma(\bar{h}-1)^2}{128 \pi^{4}\Gamma(2\bar{h}-1)}\, .
\end{aligned}
\end{align}
All in all
\begin{equation}
\gamma_{[\Phi,\Phi^2]_J}^{(J_\ell), \text{LIF}}=-\frac{\lambda^2}{32\pi^4J^4}(\log4-3)+\mathcal{O}\Big(\frac{1}{J^5}\Big)\,.
\end{equation}

\Bibliography[refs.bib]

\end{document}